\documentclass{aa}  

\usepackage{graphicx}
\usepackage{txfonts}
\usepackage[flushleft]{threeparttable}
\usepackage{tabularx}
\def\arraybackslash{\let\\\tabularnewline} % center text in single cell

\usepackage{xcolor}

\usepackage{gensymb}

\usepackage{soul}

\usepackage[version=4]{mhchem}
\usepackage{chemformula}

\usepackage[labelformat=simple,justification=centering]{subcaption}
\renewcommand\thesubfigure{(\alph{subfigure})}

\begin{document}

   \title{A major asymmetric ice trap in a planet-forming disk}
   \subtitle{IV. Nitric oxide gas and a lack of CN tracing sublimating ices and a C/O ratio $< 1$}

   \author{M. Leemker
          \inst{1}
          \and
          A. S. Booth\inst{1}
          \and
          E. F. van Dishoeck\inst{1, 2}
          \and
          N. van der Marel\inst{1}
          \and
          B. Tabone\inst{3}
          \and
          N. F. W. Ligterink\inst{4}
          \and
          N. G. C. Brunken\inst{1} 
          \and
          M. R. Hogerheijde\inst{1,5}         
          }
          
   \institute{Leiden Observatory, Leiden University, P.O. box 9513, 2300 RA Leiden, The Netherlands\\
              \email{leemker@strw.leidenuniv.nl}
            \and
         	 Max-Planck-Institut f\"ur Extraterrestrische Physik, Giessenbachstrasse 1, 85748 Garching, Germany
         	\and 
         	Université Paris-Saclay, CNRS, Institut d’Astrophysique Spatiale, 91405 Orsay, France
         	\and
         	Physics Institute, Space Research and Planetary Sciences, University of Bern, Sidlerstrasse 5, CH-3012 Bern, Switzerland
         	\and
            Anton Pannekoek Institute for Astronomy, University of Amsterdam, Science Park 904, 1090 GE Amsterdam, The Netherlands
             }

   \date{Received XXX; accepted YYY}

  \abstract
{Most well-resolved disks observed with ALMA show signs of dust traps. These dust traps set the chemical composition of the planet forming material in these disks, as the dust grains with their icy mantles are trapped at specific radii and could deplete the gas and dust at smaller radii of volatiles.}
{In this work we analyse the first detection of nitric oxide (NO) in a protoplanetary disk. We aim to constrain the nitrogen chemistry and the gas-phase C/O ratio in the highly asymmetric dust trap in the Oph-IRS~48 disk.}
{We use ALMA observations of NO, CN, \ce{C2H}, and related molecules in the Oph-IRS~48 disk. We model the effect of the increased dust-to-gas ratio in the dust trap on the physical and chemical structure using a dedicated nitrogen chemistry network in the thermochemical code DALI. Furthermore, we explore how ice sublimation contributes to the observed emission lines. Finally, we use the model to put constraints on the nitrogen containing ices.}
{NO is only observed at the location of the dust trap but CN and \ce{C2H} are not detected in the Oph-IRS~48 disk. This results in an CN/NO column density ratio of $< 0.05$ and thus a low C/O ratio at the location of the dust trap. Models show that the dust trap cools the disk midplane down to $\sim$30~K, just above the NO sublimation temperature of $\sim$25~K. The main gas-phase formation pathways to NO though OH and NH in the fiducial model predict NO emission that is an order of magnitude lower than is observed. The gaseous NO column density can be increased by factors ranging from 2.8 to 10 when the \ce{H2O} and \ce{NH3} gas abundances are significantly boosted by ice sublimation. However, these models are inconsistent with the upper limits on the \ce{H2O} and OH column densities derived from \textit{Herschel} PACS observations and the upper limit on CN derived from ALMA observations. As the models require an additional source of NO to explain its detection, the NO seen in the observations is likely the photodissociation product of a larger molecule sublimating from the ices. The non-detection of CN provides a tighter constraint on the disk C/O ratio than the \ce{C2H} upper limit.}
{We propose that the NO emission in the Oph-IRS~48 disk is closely related to the nitrogen containing ices sublimating in the dust trap. The non-detection of CN constrains the C/O ratio both inside and outside the dust trap to be $< 1$ if all nitrogen initially starts as \ce{N2} and $\leq 0.6$, consistent with the Solar value, if (part of) the nitrogen initially starts as N or \ce{NH3}. }

   \maketitle

\section{Introduction}

Almost all bright and well-resolved disks observed with ALMA show dust traps seen as rings or cavities possibly generated by forming planets in the disk \citep[e.g.,][]{vanderMarel2016, Huang2018DSHARP, Long2018, Andrews2020}. 
These dust traps are the result of a local maximum in the gas pressure, causing the dust to drift to these regions \citep{Whipple1972, Pinilla2012}. These pressure maxima can be caused by planet but also by e.g. dead zones, magnetic fields, gravitational instability, Rossby wave instability, or snowlines \citep[e.g.,][]{Stevenson1988, Boss1997, Paardekooper2006, Rice2006, Regaly2012, Zhu2012, Meheut2013, Alexander2014, Suzuki2016, Stammler2017}. Once dust traps are formed, trapped dust pebbles may continue to grow to planetesimals, comets, and planets.

These dust traps can affect the global chemical composition of the planet forming material, as the dust grains beyond the major snowlines are covered in icy mantles \citep{McClure2020}. Therefore, they trap ices at specific radii in the disk making the dust traps act as ice traps. 
The freeze-out and desorption of the major volatiles, \ce{H2O}, \ce{NH3}, and CO, at their respective snowlines changes the chemical composition of the gas and dust \citep[e.g.,][]{Oberg2011, Pinilla2017, Oberg2021, vanderMarel2021C_O}. Inside the water snowline in a smooth disk, the C/O ratio of the gas is expected to be close to the stellar value as all major volatiles are in the gas-phase, whereas outside the \ce{CO2} or CO snowlines, the C/O ratio in the gas is higher because oxygen is locked up in the icy mantles. Similarly, the C/O ratio of the ices changes across the snowlines and is expected to be low compared to the gas. However, if a dust trap is present in the disk, this C/O ratio may change as a result of transport of icy pebbles. One extreme example of such a dust trap is the Oph-IRS~48 disk, hereafter the IRS~48 disk, making it a unique laboratory to study icy chemistry in disks.

The relative volatile carbon to oxygen budget in disks can be traced by the emission of [C {\sc I}], [O {\sc I}] observed by the PACS instrument on \textit{Herschel} \citep{Thi2010, Fedele2013, Howard2013}, and \ce{C2H} as these carbon species only become bright when the C/O ratio exceeds unity \citep{Bergin2016, Kama2016, Bergner2019, Miotello2019, Bosman2021, Guzman2021}. 
Other tracers of the C/O ratio are the ratio of CS to SO \citep{Dutrey2011, Semenov2018, Facchini2021, Booth2021irs48, LeGal2021} and the less explored CN to NO ratio \citep[e.g.,][]{Hily-Blant2010, Daranlot2012, LeGal2014}. The commonly observed CN radical is destroyed by atomic oxygen \citep{Cazzoletti2018}, whereas NO is enhanced if the gas is oxygen-rich because it is mainly formed through the reaction
\begin{align}
\ce{N + OH \to NO + H}
\end{align}
\citep{Millar1991}. NO is destroyed by photodissociation and the reaction
\begin{align}
\ce{N + NO \to N2 + O}. 
\end{align}
Therefore, NO is expected to be bright when the gas-phase C/O ratio is low \citep{Schwarz2014}. 

Ices, which have a low C/O ratio, are nitrogen poor as $70-85$\% of the total available nitrogen is not identified in ices and could thus be in the form of gas-phase N or \ce{N2} \citep{Oberg2011, Boogert2015}. The main carrier of nitrogen is \ce{NH3}, which typically has an abundance of only 5\% compared to \ce{H2O} \citep{Boogert2015}. Other nitrogen bearing ices seen in the comet 67P are HCN (21\% w.r.t. \ce{NH3}), \ce{N2} (13\%) and HNCO, \ce{NH2CHO}, and \ce{CH3CN} (all $< 5$\%), and ammonia salts \citep{Rubin2019, Altwegg2020, Altwegg2022}.  

An extreme example of an asymmetric dust trap is the IRS~48 dust disk \citep{vanderMarel2013}. The large mm-sized dust is concentrated in an arc in the south of this disk, possibly due to a vortex generated by a massive companion. 
Recent work has shown many complex organic molecules (COMs) such as \ce{CH3OH} and \ce{CH3OCH3} but also smaller species like SO and, for the first time in a disk, \ce{SO2} and NO, are seen in the gas-phase at the location of the dust trap (\citealt{vanderMarel2021irs48}, hereafter Paper~I, \citealt[][hereafter Paper~II]{Booth2021irs48, Brunken2022}). The excitation temperature of \ce{CH3OH} and \ce{H2CO} of 100-200~K indicates that these molecules are thermally sublimated just inside the water snowline (Paper~I and II). Furthermore, the detections of SO and \ce{SO2} and the non-detection of CS point to a low C/O ratio of $< 1$ \citep{Booth2021irs48}. Therefore, the dust trap is an ice trap, making this disk a unique laboratory to study the ice content in this disk. The snowlines of these COMs are most likely at the dust cavity edge that is heated by stellar radiation. The detection of NO further points to the presence of nitrogen bearing ices on the dust.

ALMA dust polarization observations together with VLA data show that the dust surface density in the IRS~48 dust trap is very high at $2-8$~g~cm~$^{-2}$ \citep{Ohashi2020}. Together with the low gas surface density of 0.07~g~cm~$^{-2}$ at this location \citep{vanderMarel2016}, the local gas-to-dust ratio in this disk region is $0.04-0.009$ instead of the nominal gas-to-dust ratio of 100 in the ISM. In contrast to the large mm-sized grains, the small micron-sized dust and the gas traced by CO isotopologs are seen throughout the full 360$\degree$ of the disk \citep[e.g.,][]{Geers2007, vanderMarel2013, Bruderer2014, vanderMarel2016}. Modelling of these CO isotopologue observations have constrained the IRS~48 disk mass to be as low as $5.5\times 10^{-4}$~M$_{\odot}$ \citep{vanderMarel2016} with an inclination of 50$\degree$ and a position angle of 100$\degree$ \citep{Bruderer2014} based on \ce{^12CO} observations.

IRS~48 is likely is an A$0 \pm 1$ type star with a stellar luminosity of 14.3~L$_{\odot}$ using a dereddening slope of $R_v = 5.5$ \citep{Brown2012, Follette2015}. The high visual extinction of $A_V = 11.5$ makes a precise measurement uncertain. \citet{Schworer2017} inferred a $\sim$3 times higher stellar luminosity as more flux was attributed to the star instead of a $\sim1$~AU inner disk, and a higher visual extinction of $A_v = 12.9$ together with a dereddening slope of $R_v = 6.5$ were used. Given the high extinction and for consistency with previous work modelling the IRS~48 disk, we use a stellar luminosity of 14.3~L$_{\odot}$. The star is located in the Ophiuchus star forming cloud at a distance of 135~pc \citep{Gaia2018}.

In this paper we analyse the impact of the dust trap on the chemistry in the IRS~48 disk and in particular focus on the detection of NO. To complement this we search for other carbon and nitrogen bearing species and combine these with upper limits on OH, \ce{H2O}, CN, \ce{C2H}, \ce{N2O}, \ce{NO2}, and \ce{NH2OH} in Sect.~\ref{sec:obs}. We model the molecular abundances using a thermochemical model where we explore different scenarios in Sect.~\ref{sec:dali}. Finally, the observational and modelling results are discussed in Sect~\ref{sec:discussion}, and the conclusions are summarized in Sect.~\ref{sec:conclusions}.

\begin{table*}[h]
\begin{threeparttable}
    \centering
        \caption{Molecular line and continuum ALMA observations used in this work }    
     \begin{tabularx}{\linewidth}{p{0.15\columnwidth}p{0.3\columnwidth}p{0.16\columnwidth}p{0.05\columnwidth}p{0.19\columnwidth}p{0.15\columnwidth}p{0.1\columnwidth}p{0.1\columnwidth}p{0.33\columnwidth}p{0.05\columnwidth}}    
    \hline\hline \noalign {\smallskip}
       \centering\arraybackslash Molecule & \centering\arraybackslash Transition$^{(a)}$ & \centering\arraybackslash $A_{ul}$ (s$^{-1}$) & \centering\arraybackslash $E_u$ (K) & \centering\arraybackslash Frequency (GHz) & \centering\arraybackslash Int. flux$^{(b)}$ (mJy km~s$^{-1}$) &  \centering\arraybackslash Channel width (km~s$^{-1}$) & \centering\arraybackslash Channel rms (mJy beam$^{-1}$) & \centering\arraybackslash Beam & \centering\arraybackslash Ref.   \\ \hline
        \\[-0.7em] 
NO$^{(b)}$ & $4_{-1, 7/2, 9/2} - 3_{1, 5/2, 7/2}$ & $5.4\times 10^{-6}$ & 36 & 350.689 & \ce{CH3OH} blend & 1.7 & $\sim1.2$ & $0\farcs55\times 0\farcs44\ (-79.8\degree)$ & (1)  \\  
NO$^{(b)}$ & $4_{-1, 7/2, 7/2} - 3_{1, 5/2, 5/2}$ & $5.0\times 10^{-6}$ & 36 & 350.691 & \ce{CH3OH} blend & 1.7 & $\sim1.2$ & $0\farcs55\times 0\farcs44\ (-79.8\degree)$ & (1)  \\    
NO$^{(b)}$ & $4_{-1, 7/2, 5/2} - 3_{1, 5/2, 3/2}$ & $4.8\times 10^{-6}$ & 36 & 350.695 & \ce{CH3OH} blend & 1.7 & $\sim1.2$ & $0\farcs55\times 0\farcs44\ (-79.8\degree)$ & (1)  \\    
NO$^{(c)}$ & $4_{1, 7/2, 9/2} - 3_{-1, 5/2, 7/2}$ & $5.4\times 10^{-6}$ & 36 & 351.044 & 31 & 1.7 & $\sim1.2$ & $0\farcs55\times 0\farcs44\ (-79.8\degree)$ & (1)  \\    
NO$^{(c)}$ & $4_{1, 7/2, 7/2} - 3_{-1, 5/2, 5/2}$ & $5.0\times 10^{-6}$ & 36 & 351.052 & 23 & 1.7 & $\sim1.2$ & $0\farcs55\times 0\farcs44\ (-79.8\degree)$ & (1)  \\    
NO$^{(c)}$ & $4_{1, 7/2, 5/2} - 3_{-1, 5/2, 3/2}$ & $4.8\times 10^{-6}$ & 36 & 351.052 & 16 & 1.7 & $\sim1.2$ & $0\farcs55\times 0\farcs44\ (-79.8\degree)$ & (1)  \\      
\ce{N2O} & $14_{0, 0} - 13_{0, 0}$ & $6.3\times 10^{-6}$ & 127 & 351.668 & $<45$ & 1.7 & $\sim1.2$ & $0\farcs55\times 0\farcs44\ (-79.8\degree)$ & (2)  \\ 
\ce{NO2}$^{(d)}$ & $5_{1, 5, 11/2, 13/2} - 4_{0, 4, 9/2, 11/2}$ & $1.4\times 10^{-5}$ & 29 & 348.821 & $<65$ & 1.7 &  $\sim 1.2$ & $0\farcs56\times 0\farcs44\ (80.2\degree)$ & (2)  \\               
\ce{NH2OH}$^{(d)}$ & $7_{1, 7} - 6_{1, 6}$  & $8.1\times 10^{-5}$ & 76 & 352.298 & $<46$ & 1.7 & $\sim 1.2$ & $0\farcs55\times 0\farcs44\ (79.8\degree)$ & (2) \\   
CN$^{(e)}$ & $3_{0, 7/2, 7/2} - 2_{0, 5/2, 5/2}$ & $3.8\times 10^{-4}$ & 33 & 340.248 & $<51$ & 1.0 & 1.8 & $0\farcs25\times 0\farcs20\ (80.9\degree)$ & (3) \\
CN$^{(e)}$ & $3_{0, 7/2, 9/2} - 2_{0, 5/2, 7/2}$ & $4.1\times 10^{-4}$ & 33 & 340.248 & $<69$ & 1.0 & 1.8 & $0\farcs25\times 0\farcs20\ (80.9\degree)$ & (3) \\
CN$^{(e)}$ & $3_{0, 7/2, 5/2} - 2_{0, 5/2, 3/2}$ & $3.7\times 10^{-4}$ & 33 & 340.249 & $<37$ & 1.0 & 1.8 & $0\farcs25\times 0\farcs20\ (80.9\degree)$ & (3) \\
\ce{C2H}$^{(f)}$ & $4_{9/2, 5} - 3_{7/2, 4}$ & $1.3\times 10^{-4}$ & 42 & 349.338 & $<26$ & 1.7 & 1.0 & $0\farcs64\times 0\farcs51\ (80.1\degree)$ & (4) \\
\ce{C2H}$^{(f)}$ & $4_{9/2, 4} - 3_{7/2, 3}$ & $1.3\times 10^{-4}$ & 42 & 349.339 & $<26$ & 1.7 & 1.0 & $0\farcs64\times 0\farcs51\ (80.1\degree)$ & (4) \\
\ce{^13CO} & $3-2$ & $2.2\times 10^{-6}$ & 11 & 330.588 & $4.5\times 10^3$ & 0.5 & 2.2 & $0\farcs19\times 0\farcs14\ (64.8\degree)$ & (3)  \\
cont.$^{(g)}$ & 0.89~mm & && 336 & 0.18 & & 0.056 & $0\farcs19\times 0\farcs14\ (65.4\degree)$ & (3) \\
       \hline        
    \end{tabularx}
    \begin{tablenotes}
      \small
      \item \textbf{Notes.} $^{(a)}$ Quantum numbers are formatted for NO as $N_{K, J, F_1}$ and line strengths taken from CDMS and \citealt{Varberg1999}, \ce{NH2OH} as $N_{K_a, K_c}$ \citep[CDMS,][]{Tsunekawa1972}, \ce{N2O} as $N_{K,v}$ \citep[JPL,][]{Burrus1956}, \ce{NO2} as $J_{\Omega, \Lambda, (F1),(F2),(F)}$ \citep[JPL,][]{Bowman1982, SemmoudMonnanteuil1989}, CN as $N_{v, J, F_1}$ \citep[CDMS,][]{Skatrud1983}, \ce{C2H} as $N_{J,	F_1}$ \citep[CDMS,][]{Sastry1981}, and \ce{^13CO} as $J$ \citep[CDMS][]{Klapper2000}. $^{(b)}$ Disk integrated flux within the Keplerian mask. Blended lines are corrected for their expected line strength by scaling the total integrated flux with $A_{ij}\times g_u$ neglecting the dependence on $E_u$, unless denoted otherwise. $^{(b)}$.~Detected but blended with the two other NO lines at 350.7~GHz and a methanol line. $^{(c)}$~Blended with the two other NO lines at $351.04-351.05$~GHz. $^{(d)}$~Only the line that is expected to be brightest is listed. This line is blended with two weak 266~K line, that are neglected for the calculation of the integrated flux. $^{(e)}$~Blended with another CN line at 340.2~GHz. $^{(f)}$~Blended with another \ce{C2H} line at 349.3~GHz. $^{(g)}$ Unit of the integrated continuum flux is Jy.
\item \textbf{References} (1)~ALMA project code: 2017.1.00834.S (Paper~II). (2)~ALMA project code: 2017.1.00834.S (this work). (3)~ALMA project code: 2013.1.00100.S (this work). (4)~ALMA project code: 2017.1.00834.S \citep{vanderMarel2021C_O}.       
    \end{tablenotes}
    \label{tab:obs}
\end{threeparttable}
\end{table*}

\section{Observations and methods} \label{sec:obs}

\subsection{Data}

We use ALMA observations covering transitions of \ce{^13CO}, NO, \ce{N2O}, \ce{NO2}, \ce{NH2OH}, CN, and \ce{C2H}. Only \ce{^13CO} and NO are detected. A summary of the transitions covered with ALMA, details of the observations, and an overview of the line strengths obtained from the CDMS and JPL databases is presented in Table~\ref{tab:obs} \citep{Pickett1998, Muller2001, Muller2005, Endres2016}.

Six emission lines of NO with upper energy levels of 36~K at 350.7~GHz (3 lines), 351.04~GHz (1 line), and 351.05~GHz (2 lines) were observed as part of the ALMA project 2017.1.00834.S (PI: A. Pohl; see Table~\ref{tab:obs}; \citealt{Ohashi2020}).
Additionally, the data cover transitions of \ce{N2O}, \ce{NO2}, and \ce{NH2OH}. This data set has a sensitivity of $\sim$1.2~mJy~beam$^{-1}$~channel$^{-1}$ and a spectral resolution of 1.7~km~s$^{-1}$. The imaging is described in detail in Paper~I and II. 

The CN $3_{0, 7/2, 7/2} - 2_{0, 5/2, 5/2}$, $3_{0, 7/2, 9/2} - 2_{0, 5/2, 7/2}$, and $3_{0, 7/2, 5/2} - 2_{0, 5/2, 3/2}$ transitions with $E_u = 33$~K at 340.248~GHz (2 lines) and 340.249~GHz (1 line), and \ce{^13CO} $3-2$ transition with $E_u = 32$~K at 330.588~GHz in Band~7 have not been published before. These two molecules were targeted as part of the ALMA program 2013.1.00100.S (PI: N. van der Marel). The IRS~48 disk was observed on June 14 2015 for 27 minutes on-source with 41 antennas and baselines ranging from 21.4~m to 783.5~m. The spectral windows covering the CN and the \ce{^13}CO transitions have a resolution of 282.23~kHz (0.25~km~s$^{-1}$). Both were self-calibrated using CASA version 4.2.2 \citep{McMullin2007} with 4 rounds of phase calibration down to a solution interval of 24~s and subsequently 1 round of amplitude calibration with a solution interval of 48~s. This increased the signal-to-noise ratio on the continuum from 217 to 804. The continuum and the \ce{^{13}CO} are both detected and were imaged using the \texttt{clean} function in casa with used-defined masks and Briggs weighing with robust $= 0.5$. The images were \texttt{clean}ed to a noise threshold of 3$\times$ the rms that was measured in the line-free channels. The CN emission is imaged using natural weighing (equivalent to robust = 2.0) to maximize the sensitivity but it is not detected. Furthermore, the CN data were binned to 1~km~s$^{-1}$ and the \ce{^13CO} to 0.5~km~s$^{-1}$ bins. The resulting 0.89~mm continuum image has a spatial resolution of $0\farcs19 \times 0\farcs14\ (65.4\degree)$ with the position angle of the beam indicated within the brackets, and a sensitivity of $56~\mu$Jy~beam$^{-1}$. The \ce{^{13}CO} and CN images have a spatial resolution of $0\farcs19 \times 0\farcs14\ (64.8\degree)$ and $0\farcs25 \times 0\farcs20\ (80.9\degree)$ and a sensitivity of 2.2~mJy~beam$^{-1}$ and 1.8~mJy~beam$^{-1}$, respectively. 

In addition to the ALMA observations, we also use the observations of the IRS~48 disk in the \textit{Herschel} PACS DIGIT survey \citep{Fedele2013}. These observations cover \ce{H2O} and OH far infrared lines in the spectral range of $51-220~\mu$m, but neither of these molecules is detected, with \ce{H2O} fluxes lower than $(1.6-8)\times 10^{-17}$~W~m$^{-2}$ and OH fluxes lower than $(1.2-9.6)\times 10^{-17}$~W~m$^{-2}$. A full description of the data reduction is reported in \citet{Fedele2013}. As the $9\farcs4\times 9\farcs4$ spaxels of the \textit{Herschel} PACS data are much larger than the IRS~48 disk, only a disk integrated analysis is done using these \ce{H2O} and OH data.

\begin{figure*}
\centering
\includegraphics[width=\hsize]{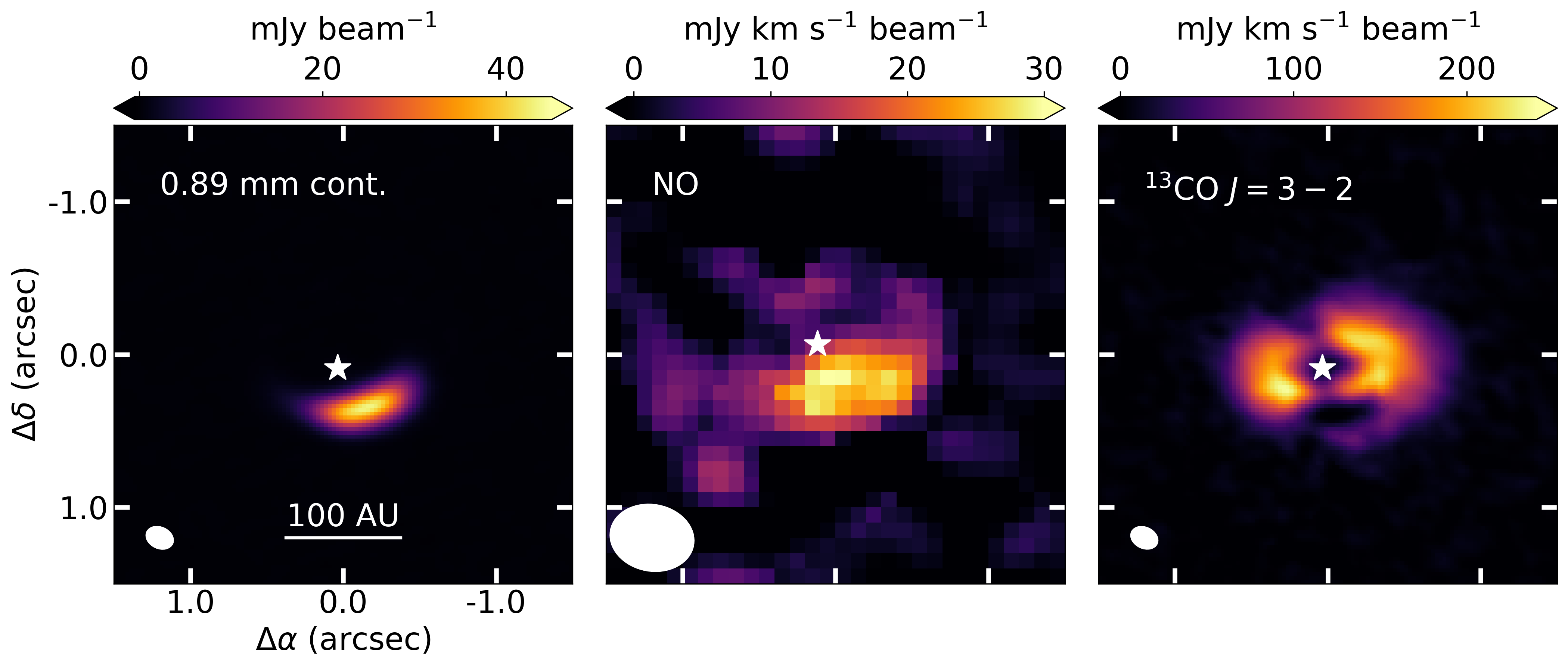}
\caption{Dust and gas observations in the IRS~48 disk. \textit{Left:} 0.89~mm continuum emission showing the major asymmetric dust trap in the south. \textit{Middle:} moment 0 map of the sum of the three NO lines at 351.1~GHz. \textit{Right}: moment 0 map of the \ce{^13CO} $3-2$ transition. The position of the star and the beam of the observations are indicated with the white star and the white ellipse, respectively, in each panel. A 100~AU scalebar is indicated in the bottom of the left panel.  }
\label{fig:mom0}
\end{figure*}

\subsection{Data analysis}

To increase the signal to noise ratio of the NO and \ce{^13CO} integrated intensity (moment 0) maps, as well as that of the ALMA data covering the lines that are not detected, we use a Keplerian mask \citep{Teague2020_software}\footnote{\url{https://github.com/richteague/keplerian_mask}}. The mask for each line is created using a stellar mass of 2~M$_{\odot}$, an position angle of 100$\degree$, an inclination of 50$\degree$, a distance of 135~pc, and a source velocity of 4.55~km~s$^{-1}$ \citep{Bruderer2014}. Furthermore, the mask for the moment 0 maps has a large maximum radius of $4\farcs0$ to avoid any loss of diffuse emission in the outer disk. Additionally, the mask is smoothed with a beam 1.5 times larger than the spatial resolution of the data and assumes all emission is coming from the disk midplane. Finally, the line width in the mask is calculated as $0.5\ \mathrm{km~s^{-1}} \times (r/1\farcs0)^{-0.5}$ to encompass both the \ce{^13CO} $3-2$ (presented in this work) and the previously published $6-5$ line \citep{vanderMarel2016}. For the uncertainty on the integrated fluxes, a Keplerian mask with a smaller outer radius of $0\farcs 9$, fitting the \ce{^13CO} $J=3-2$ and $J=6-5$ transitions, is used. This ensures that we include all pixels that could have emission but as few of those without as possible. Furthermore, in the case of the blended NO lines at 351.05~GHz, we use the union of two Keplerian masks centred at the frequencies of the NO lines at 351.043~GHz and 351.051~GHz.

The error on the integrated intensity maps are calculated using the channel rms $\sigma_{\mathrm{channel\ rms}}$:
\begin{align}
\sigma_{\mathrm{mom0}} = \sigma_{\mathrm{channel\ rms}}\times \Delta V_{\mathrm{channel}}\times \sqrt{N_{\mathrm{channel}}},
\end{align}
with $\Delta V_{\mathrm{channel}}$ the spectral resolution of the image cube in km~s$^{-1}$ and $N_{\mathrm{channel}}$ the number of channels in the Keplerian mask in that pixel. To increase the signal to noise ratio of the radial profile, the average over an 100$\degree$ wedge is taken, following \citet{vanderMarel2021irs48}. This wedge is centred an position angle of 192\degree, close to the disk minor axis in the south to encompass the emission of the dust, NO, \ce{^13CO}, and \ce{CH3OH}. The error on this radial profile is propagated from that on the integrated intensity map:
\begin{align}
\sigma_{\mathrm{radial\ profile}} = \sqrt{\frac{1}{N_{\mathrm{beams\ bin}}N_{\mathrm{pix\ bin}}} \sum_{\mathrm{pix}} \sigma_{\mathrm{mom0}}^2},
\end{align}
with $N_{\mathrm{beams\ bin}} \geq 1$ the number of independent beams in 1 radial bin within the wedge, and $N_{\mathrm{pix\ bin}}$ the number of pixels used in that bin. For the azimuthal profile, the same equation is used but then with the number of beams and pixels in 1 azimuthal bin in a ring. 

The error on the integrated flux, $\sigma_{F_{\nu}\Delta V}$, and the upper limits on the integrated fluxes of the non-detections are calculated as follows:
\begin{align}
\sigma_{F_{\nu}\Delta V} = 1.1 \times \sigma_{\mathrm{channel\ rms}} \times  \Delta V_{\mathrm{channel}} \times \sqrt{\frac{N_{\mathrm{pix\ mask}}}{N_{\mathrm{pix\ beam}}}}
\end{align}
for an extended region, and as follows:
\begin{align}
\sigma_{F_{\nu}\Delta V} = 1.1 \times  \sigma_{\mathrm{mom0}}
\end{align}
for the spectra within 1 beam.
The factor of 1.1 accounts for the 10\% absolute flux calibration error of ALMA which is necessary to include when comparing data from different observing projects, $N_{\mathrm{pix\ mask}}$ is the total number of pixels included in the Keplerian mask and $N_{\mathrm{pix\ beam}}$ is the number of pixels per beam. For the data used in this work $N_{\mathrm{pix\ mask}}/N_{\mathrm{pix\ beam}} = 80-320$. In the case of a non-detection, a 3$\sigma_{F_{\nu}\Delta V}$ upper limit is used.

\begin{table}[]
\begin{threeparttable}
    \centering
        \caption{Derived column densities}    
     \begin{tabularx}{\linewidth}{p{0.2\columnwidth}p{0.1\columnwidth}p{0.25\columnwidth}p{0.25\columnwidth}}    
    \hline\hline \noalign {\smallskip}
       \centering\arraybackslash Molecule & \centering\arraybackslash $T_{\mathrm{ex}}$ (K) & \centering\arraybackslash $N_{\mathrm{ext}}$ (cm$^{-2}$)  & \centering\arraybackslash $N_{\mathrm{peak}}$ (cm$^{-2}$)\\ \hline
        \\[-0.7em]
         NO (north) &  40 & & $<3.0\times 10^{14}$ \\
         NO (north) &  100 & & $<5.0\times 10^{14}$ \\
         NO (south) &  40 & $\sim 6.7\times 10^{14}$ & $7.7\times 10^{14}$ \\
         NO (south) &  100 & $\sim 1.1\times 10^{15}$ & $1.3\times 10^{15}$ \\         
         \ce{N2O} & 40 & $< 1.8\times 10^{15}$ & $<1.7\times 10^{15} $ \\ 
         \ce{N2O} & 100 & $< 6.6\times 10^{14}$ & $< 6.4\times 10^{14}$  \\ 
         \ce{NO2} & 40 & $<2.1 \times 10^{15}$ & $<2.0 \times 10^{15}$ \\ 
         \ce{NO2} & 100 & $<5.6 \times 10^{15}$  & $<5.2 \times 10^{15}$  \\          
         \ce{NH2OH} & 40 & $< 1.3\times 10^{14}$ & $< 1.4\times 10^{14}$ \\         
         \ce{NH2OH} & 100 & $< 1.7\times 10^{14}$ & $< 1.8\times 10^{14}$ \\         
         CN & 40 & $<2.7 \times 10^{12}$ & $< 4.1\times 10^{13}$ \\       
         CN & 100 & $< 4.1\times 10^{12}$ & $< 6.2\times 10^{13}$ \\
         \ce{C2H} &  40 & $< 1.5\times 10^{12}$ & $< 1.1\times 10^{13}$ \\
         \ce{C2H} & 100 & $< 2.1\times 10^{12}$ & $< 1.5\times 10^{13}$ \\
         OH & $<100$ & no constraint & \\
         OH & 150 & $\lesssim 6\times 10^{14}$ & \\
         OH & $>200$ & $\lesssim 10^{14}$ & \\
         \ce{H2O} & $<100$ & no constraint &  \\ 
         \ce{H2O} & 150 & $\lesssim 1\times 10^{14}$ &  \\ 
         \ce{H2O} & $>200$ & $\lesssim 10^{14}$ &  \\        
       \hline        
    \end{tabularx}
    \begin{tablenotes}
      \small
      \item \textbf{Notes.} The upper limits on the column densities of \ce{N2O}, \ce{NH2OH}, \ce{OH}, and \ce{H2O} derived in the extended region covered by the Keplerian mask $N_{\mathrm{ext}}$ are derived assuming an emitting region of $1.4\times 10^{-11}$~sr, similar to NO. This is because these species are expected to be closely related to the emitting region of NO. The emitting regions of CN and \ce{C2H} are assumed to be $8.0\times 10^{-11}$~sr (3.4 square arcseconds) and $1.6\times 10^{-10}$~sr (7.0 square arcseconds), respectively, using the area covered in the Keplerian mask. The peak column densities $N_{\mathrm{peak}}$ are derived using a spectrum within 1 beam at the location where the emission is brightest. 
    \end{tablenotes}
    \label{tab:N}
\end{threeparttable}
\end{table}

\subsection{Column densities} \label{sec:N}

To compare our observations to other disks and disk models, we convert the (upper limits on) the velocity integrated fluxes to a column density following e.g., \citet{Goldsmith1999}:
\begin{align}
N_{u}^{\mathrm{thin}} = \frac{4\pi F_{\nu}\Delta V}{A_{ul}hc\Omega},
\end{align}
with $N_{u}^{\mathrm{thin}}$ the column density of the upper energy level $u$, $F_{\nu}\Delta V$ the integrated flux, $A_{ul}$ the Einstein $A$ coefficient, $c$ the speed of light, $h$ the Planck constant and $\Omega$ the emitting region. 
The column density of the upper level is set by the total column density, $N$, of the molecule assuming local thermodynamical equilibrium (LTE) and optically thin emission. The total column density $N$ then follows from 
\begin{align}
N = \frac{N_u Q(T_{\mathrm{ex}})}{g_u} e^{E_u/kT_{\mathrm{ex}}},
\end{align}
with $Q(T_{\mathrm{ex}})$ the partition function at excitation temperature $T_{\mathrm{ex}}$, $g_u$ the degeneracy of the upper energy level, $E_u$ the energy, and $k$ the Boltzmann constant. The column density for each molecule is computed four times: we compute $N$ for two different assumed excitation temperatures and for two different regions in the disk. The first temperature of 40~K corresponds to the assumed excitation temperature of NO in Paper~II and the second temperature of 100~K corresponds to the excitation temperature of methanol (Paper~I) in the IRS~48 disk. The excitation temperature of NO cannot be derived from the observations as all detected NO lines have an upper energy level energy of $E_u = 36$~K.

For the two regions, the column density within 1 beam ($N_{\mathrm{peak}}$) and within an extended region ($N_{\mathrm{ext}}$) are computed. The column density within 1 beam is computed using the spectrum extracted from the brightest pixel. 
To compute column densities, the physical size of the beam area is used.
For comparison to other work, the column density is also calculated in an extended region, $N_{\mathrm{ext}}$, using the total flux in the small Keplerian mask listed in Table~\ref{tab:obs} and an assumed emitting region. For the molecules that are directly related to the dust trap (\ce{NO, N2O, NO2, NH2OH, H2O}, and OH) an emitting area of $1.4\times 10^{-11}$~sr (0.6 square arcseconds, typically 10 times smaller than that covered by the Keplerian mask) is used as done in Paper~I, II, and \citet[][]{Booth2021irs48}, while for the molecules where this is not necessarily the case (CN and \ce{C2H}) the area covered by the Keplerian mask is used ($8.0\times 10^{-11}$~sr, equivalent to 3.41 square arcseconds, and $1.6\times 10^{-10}$~sr equivalent to 7.0 square arcseconds, respectively). An overview of the resulting column densities at 40 and 100~K is presented in Table~\ref{tab:N}, the column densities at a range of temperatures are presented in Fig.~\ref{fig:T_vs_N}.

\begin{figure}
\centering
\includegraphics[width=\hsize]{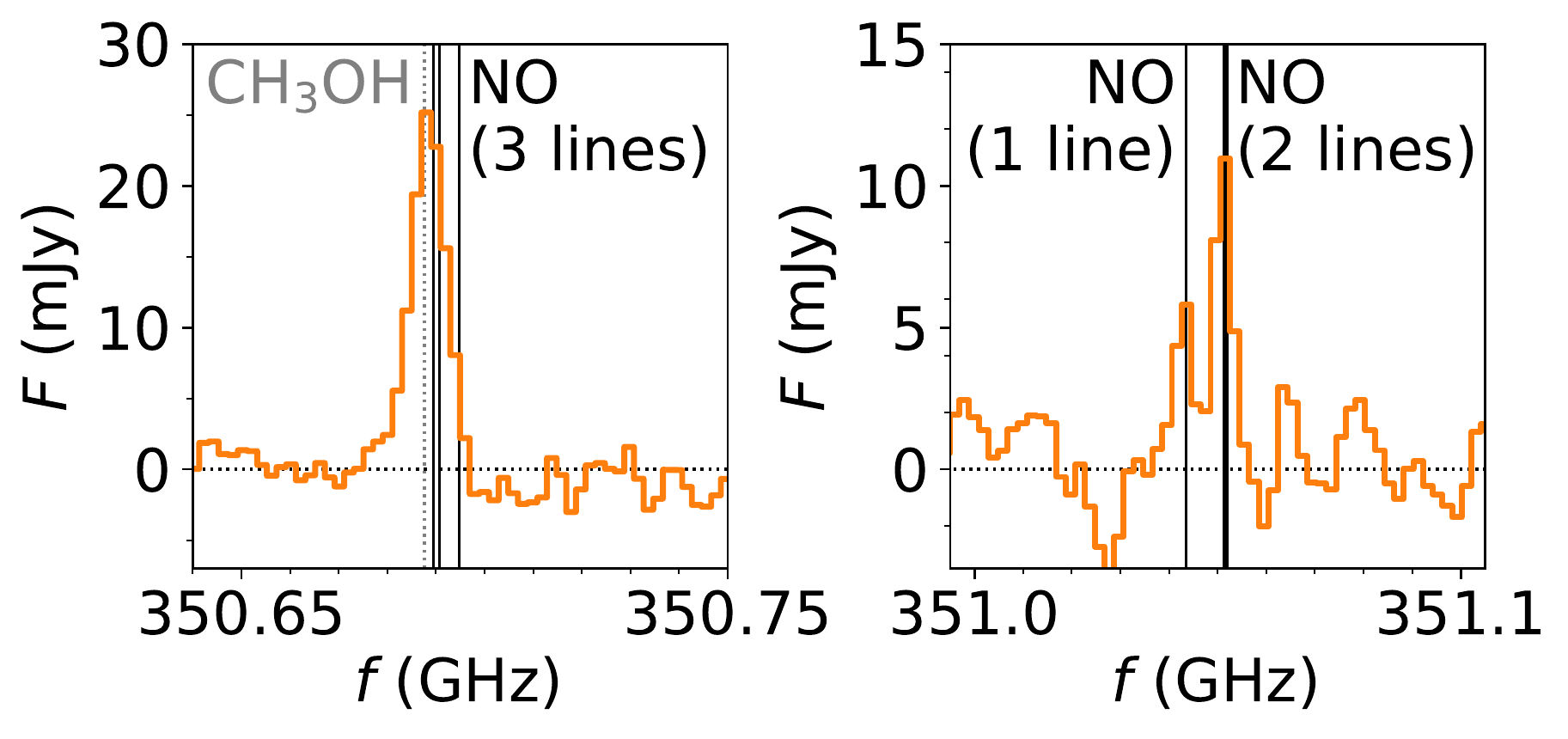}
\caption{Continuum subtracted, stacked spectra of the six NO transitions detected inside 115~AU. Left: the three NO lines at 350.7~GHz, indicated with the thick black vertical line, are blended with methanol (dotted grey line). Right: the two NO lines at 351.05~GHz are slightly blended with the final NO line at 351.04~GHz. }
\label{fig:NO_spec_Jy}
\end{figure}

\section{Observational results} \label{sec:results}
\subsection{Detected lines}

The observational results are presented in Fig.~\ref{fig:mom0} and \ref{fig:NO_spec_Jy}. We detect the 0.89~mm continuum, the \ce{^{13}CO} $3-2$ transition and six transitions of NO and provide upper limits for \ce{CN, N2O, NO2, NH2OH}, and \ce{C2H}. The three NO lines at 350.7~GHz are blended with the $4_{0,4} - 3_{1,3}$ transition of \ce{CH3OH} (Paper~II, see left panel in Fig.~\ref{fig:NO_spec_Jy}). Therefore, we focus on the three NO lines at 351.1~GHz, that are blended with themselves but not with emission from other molecules (see right panel in Fig.~\ref{fig:NO_spec_Jy}). A spectrum in units of K can be found in Fig.~\ref{fig:spec_NO_K}. 

The Keplerian masked integrated intensity maps of these three NO lines, together with the Keplerian masked \ce{^{13}CO} $3-2$ transition and the 0.89~mm continuum are presented in Fig.~\ref{fig:mom0}. Both the NO and the continuum emission are only detected in the south of the IRS~48 disk, where the dust trap is located. 
In contrast, the \ce{^{13}CO} emission does not show this asymmetry. Thus, there is no indication that the IRS~48 gas disk is azimuthally asymmetric apart from cloud absorption along the minor axis \citep{Bruderer2014, vanderMarel2016}.
The lack of \ce{^{13}CO} emission at the location of the dust trap is likely due to \ce{^13CO} $3-2$ and continuum optical depth effects and not due to a lack of \ce{^{13}CO} gas at that location (see Fig.~\ref{fig:obs_13CO32_chans_w_cont}; \citealt{Isella2016, Weaver2018}). Additionally, the 25~AU gas cavity in this disk is also seen in the \ce{^13CO} $J=3-2$ transition \citep{vanderMarel2016}. The emission from COMs and NO is (close to) optically thin as this depression at the location of the dust trap is not seen in those moment 0 maps. The small depression of \ce{^13CO} emission along the disk minor axis is likely due to foreground cloud absorption at $3-4.5$~km~s$^{-1}$ that is seen in the \ce{^13CO} $J=3-2$ channel maps (Fig.~\ref{fig:obs_13CO32_chans}) and in the \ce{^12CO} $6-5$ transition \citep{vanKempen2009, Bruderer2014}. 

\begin{figure}
   \centering
  \includegraphics[width=\columnwidth]{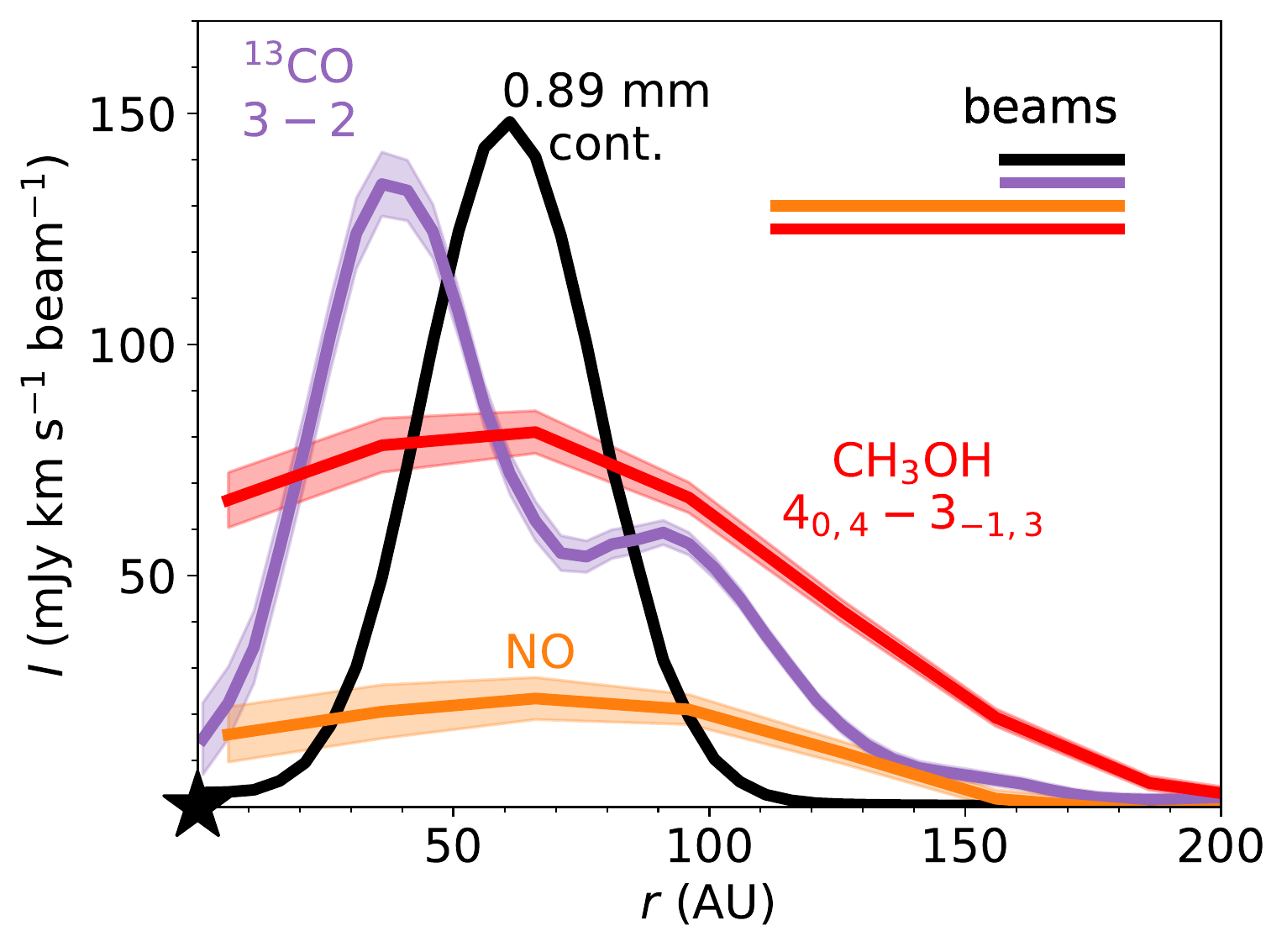}
      \caption{Radial profile of the continuum (black), NO (orange), \ce{^13CO} $3-2$ (purple), and \ce{CH3OH} $4_{0,4}-3_{-1,3}$ (red) emission averaged over an 100$\degree$ angle, centred at 192$\degree$ east of north where the large dust is located.} 
         \label{fig:rad_azi}
   \end{figure}

To further investigate the emission in the dust trap, radial profiles are presented in Fig.~\ref{fig:rad_azi}. These profiles are extracted from an 100$\degree$ wide wedge centred around a position angle of 192$\degree$ east of north, following Paper~I. The NO and methanol emission peak at the same location as the continuum. Furthermore, the emission of NO and methanol decreases slightly or stays relatively constant towards the position of the star due to the large beam of the NO and methanol observations. 
The steep decrease in the \ce{^13CO} $3-2$ emission towards the star is indicative of the 25~AU gas cavity that is also seen in the \ce{^12CO}, \ce{^13CO}, and \ce{C^18O} $6-5$ transitions as these all have a high spatial resolution compared to that of the NO and methanol observations \citep{Bruderer2014, vanderMarel2016}. The azimuthal profile is presented in Fig.~\ref{fig:obs_azi_prof}.

\subsection{NO column density}  \label{sec:NO_obs}
The peak NO column density derived in this work is $7.7\times~10^{14}$~cm$^{-2}$ and the NO column density assuming all NO emission is emitted from the same region as the continuum ($1.4\times 10^{-11}$~sr) $N_{\mathrm{ext}}$ is $\sim 6.7\times 10^{14}$~cm$^{-2}$ for an assumed $T_{\mathrm{ex}} = 40$~K. As the NO and continuum emission are unresolved in the radial direction, the true emitting area may be smaller than this value.
The NO column densities found in this work are a factor of 4-10 lower than the value of $3\times 10^{15}$~cm$^{-2}$ found in Paper~II. The analysis in Paper~II was done using a spectrum in K. This spectrum was computed by applying the Rayleigh Jeans law to the data cube and then \textit{integrating} the emission over an $3.4\times 10^{-11}$~sr region, instead of the regular method of first applying the Rayleigh Jeans law and then \textit{averaging} over these pixels. As the region used to compute the spectrum was 5 times larger than the $0\farcs55\times 0\farcs44$ ($6.4\times 10^{-12}$~sr) ALMA beam, the fluxes and hence the column densities reported in Paper~II are overestimated by a factor of~5. The corrected NO spectrum in K is presented in Fig.~\ref{fig:spec_NO_K}. The abundance ratios reported in Paper~II are not affected by this inconsistency as the difference of a factor of~5 drops out.

\subsection{Upper limits on nitrogen bearing molecules and C/O~ratio}
\ce{N2O}, \ce{NO2}, and \ce{NH2OH} are expected to be closely linked to the chemistry of NO as \ce{NH2OH} is the product of its hydrogenation on icy grains and \ce{N2O} and \ce{NO2} are the by-products \citep{Congiu2012, Fedoseev2012} but none of these are detected. In Table~\ref{tab:N}, the upper limits on their column densities are reported for two different temperatures of 40 and 100~K, following the excitation temperature used to model the NO in Paper~II and the excitation temperature found for \ce{CH3OH} in Paper~I. As the NO lines listed in Table~\ref{tab:obs} all have an upper energy level of 36~K, its excitation temperature is not tightly constrained. Increasing the assumed excitation temperature of NO to 100~K increases the (upper limit) on its column density by a factor $\sim$2. For both temperatures, the upper limits on the \ce{N2O}, \ce{NO2}, and \ce{NH2OH} column densities are comparable to the NO column density at the dust trap.

In contrast to many other disks, CN is not detected in the IRS~48 disk and its peak column density is constrained to be $<~4.1\times 10^{13}$~cm$^{-2}$ assuming a temperature of 40~K. This is a factor of $6.6-47$ lower than the peak CN column densities derived for the five disks that were observed as part of the MAPS large program with masses $8-360$ 
times larger than that of the IRS~48 disk \citep{Bergner2021, Oberg2021MAPS, Zhang2021}. 
Our assumed emitting region is similar to the emitting region of \ce{^13CO} as both CN and \ce{^13CO} are expected to trace large radial extents of the disk \citep[e.g.,][]{Law2021, Nomura2021}, though some radial differences may be expected \citep{Cazzoletti2018, vanTerwisga2019, Paneque-Carreno2022}. A CN/NO ratio of $<0.05$ is found. This points to a low C/O ratio, consistent with that found from the CS/SO ratio \citet{Booth2021irs48}. This conclusion is strengthend by the non-detection of \ce{C2H}, a molecule that becomes abundant when C/O $>1$ \citep{Bergin2016, Bergner2019, Miotello2019, Bosman2021}. We will further explore this qualitatively in Sect.~\ref{sec:extra_H2O_NH3} and \ref{sec:north}, and Appendix~\ref{sec:CO_ratio_CN} using thermo-chemical disk models.

\begin{figure}
   \centering
  \includegraphics[width=1\linewidth]{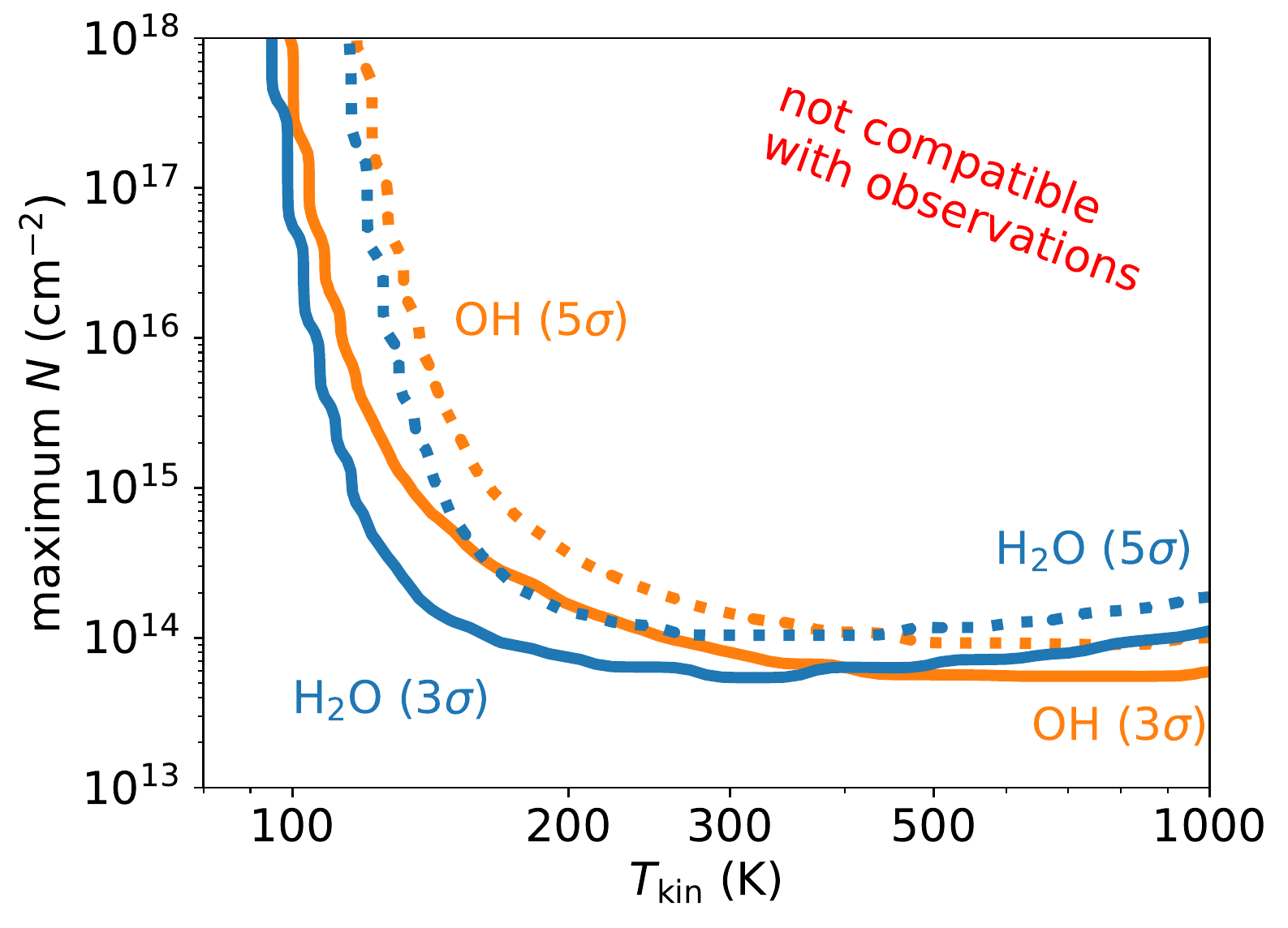}
      \caption{Maximum \ce{H2O} (blue) and OH (orange) column density above the optically thick dust that is compatible with the 3$\sigma$ (solid) and 5$\sigma$ (dotted) upper limits on the \ce{H2O} and OH flux using \textit{Herschel} PACS observations.  } 
         \label{fig:Herschel_H2O_OH}
   \end{figure}

\subsection{Upper limits on \ce{H2O} and OH}
Far-infrared lines of \ce{H2O} and OH are covered in the \textit{Herschel} DIGIT survey using the PACS instrument \citep{Fedele2013}. Similar to the carbon and nitrogen bearing molecules CN, \ce{C2H, N2O, NO2}, and \ce{NH2OH}, the IRS~48 disk shows no detection of \ce{H2O} and OH lines. We revisited these upper limits using the LTE model of \citet{Fedele2013}, assuming that the emission originates from a slab with homogeneous column density and temperature, and including line opacity effects (see \citealt{Bruderer2010} for further details). The emitting area of \ce{H2O} and OH is assumed to be that of the trap (5000 AU$^2$ or $1.4\times 10^{-11}$~sr). This allows us to convert the non-detection into constraints on the column density of \ce{H2O} and OH, and temperature. The joint constraints on the column density above the optically thick dust are presented in Fig.~\ref{fig:Herschel_H2O_OH} using the 3$\sigma$ and 5$\sigma$ upper limits.

In this figure, two regimes can be distinguished, depending on the assumed temperature. First, if the temperature of the \ce{H2O} and OH emitting regions is $< 100$~K, no constraint can be put and the trap can be rich in \ce{H2O} and OH. On the other hand, if the temperature exceeds 200~K, both the \ce{H2O} and OH column densities are constrained to be $\lesssim 10^{14}$~cm$^{-2}$, with \ce{H2O} and OH far-IR lines being optically thin. The excitation temperature of \ce{H2CO} of $173^{+11}_{-9}$~K derived in Paper~I shows that there is a layer of gas above the dust trap where the gas reaches these high temperatures and \ce{H2CO} can exist. Furthermore, \ce{H2O} is expected to be frozen out on the dust grains if the temperature is lower than $\lesssim 100$~K based on its 5770~K binding energy \citep{Fraser2001}. Therefore, the temperature range of $\sim 150-200$~K is likely the most relevant range for our analysis given the complementary observations. Finally, the \textit{Herschel} PACS observations only probe the column of \ce{H2O} and OH above the optically thick dust at $51-220$~$\mu$m, therefore the full \ce{H2O} and OH column density may be higher than the upper limits derived here. 

\begin{figure}
   \centering
\includegraphics[width=\hsize]{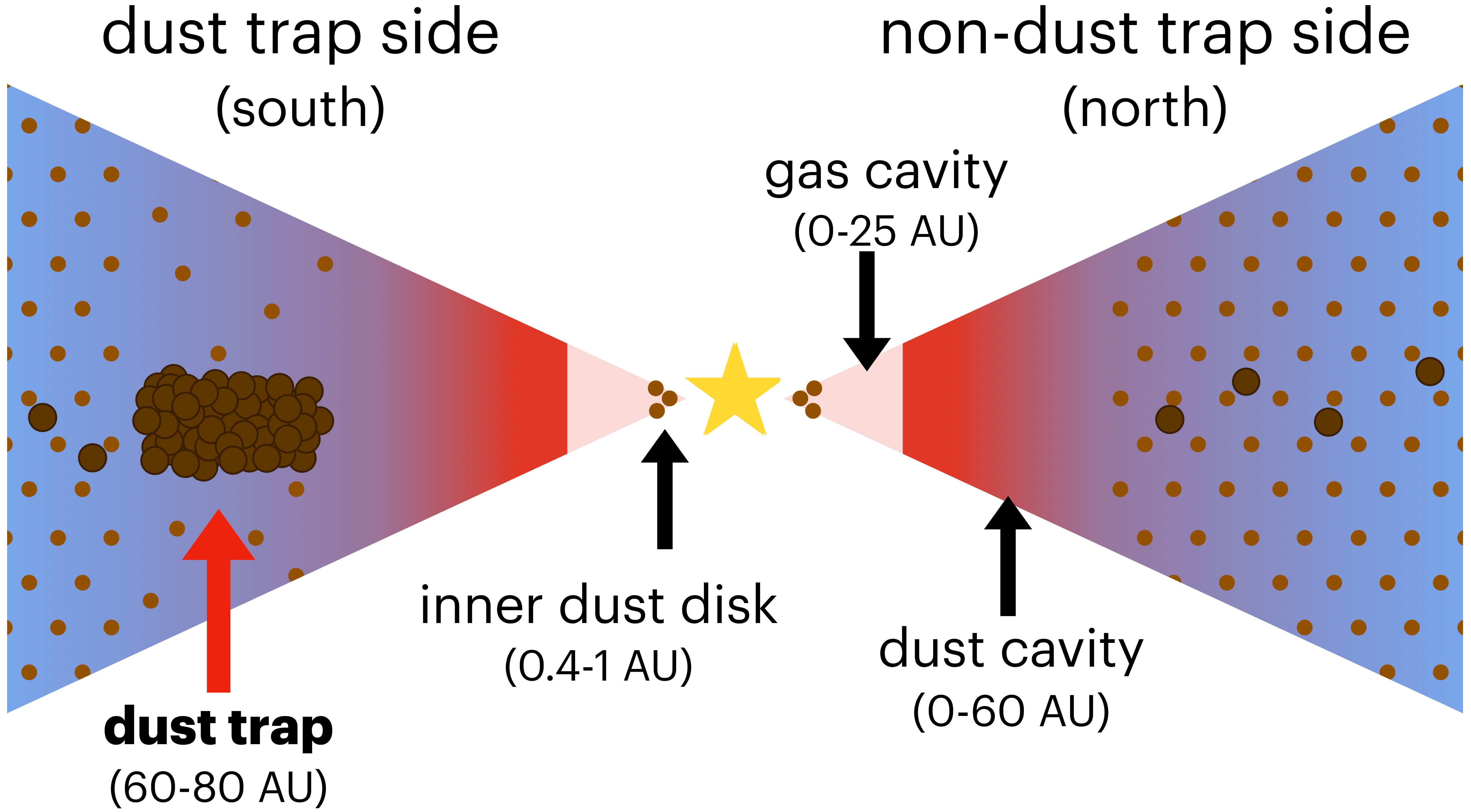}   
      \caption{Cartoon of the dust trap side (south) and non-dust trap side (north) of the disk. The background color indicates the gas and its temperature. The faded red inside 25~AU indicates the deep gas cavity. The small and large brown circles indicate the small and large dust, respectively. Large dust is mainly present between 60 and 80~AU in the dust trap  side (south) of the disk.  }
         \label{fig:cartoon}
\end{figure}

\section{Models} \label{sec:dali}
The observations show a clear north-south asymmetry in the NO emission that is possibly related to the dust trap. To investigate the origin of this asymmetry and the chemical composition of the gas and ice in the IRS~48 disk we run a grid of thermochemical models using DALI \citep{Bruderer2012, Bruderer2013}. We test 3 scenarios: (1) a fiducial model, (2) a model with sublimating \ce{H2O} and/or \ce{NH3} ice, and (3) a model with sublimating NO ice.

The models are based on the previously published, azimuthally symmetric IRS~48 disk model with a fixed gas density and a high dust surface density in the dust trap \citep{Bruderer2014, vanderMarel2016}, see Fig.~3 and Fig.~C.1 in Paper~I. We improve on this model by modelling the chemistry in the dust trap (south side) and the non-dust trap (north side) of the disk separately using the model parameters in Table~\ref{tab:dali} for models 1-3. In particular, we model the effect of the ice trap by increasing the initial abundance of gas-phase species to O/H and N/H ratios above the typical ISM values, mimicking their enhanced abundances due to radial drift and subsequent sublimation. 

\begin{figure}
  \centering
    \includegraphics[width=1\linewidth]{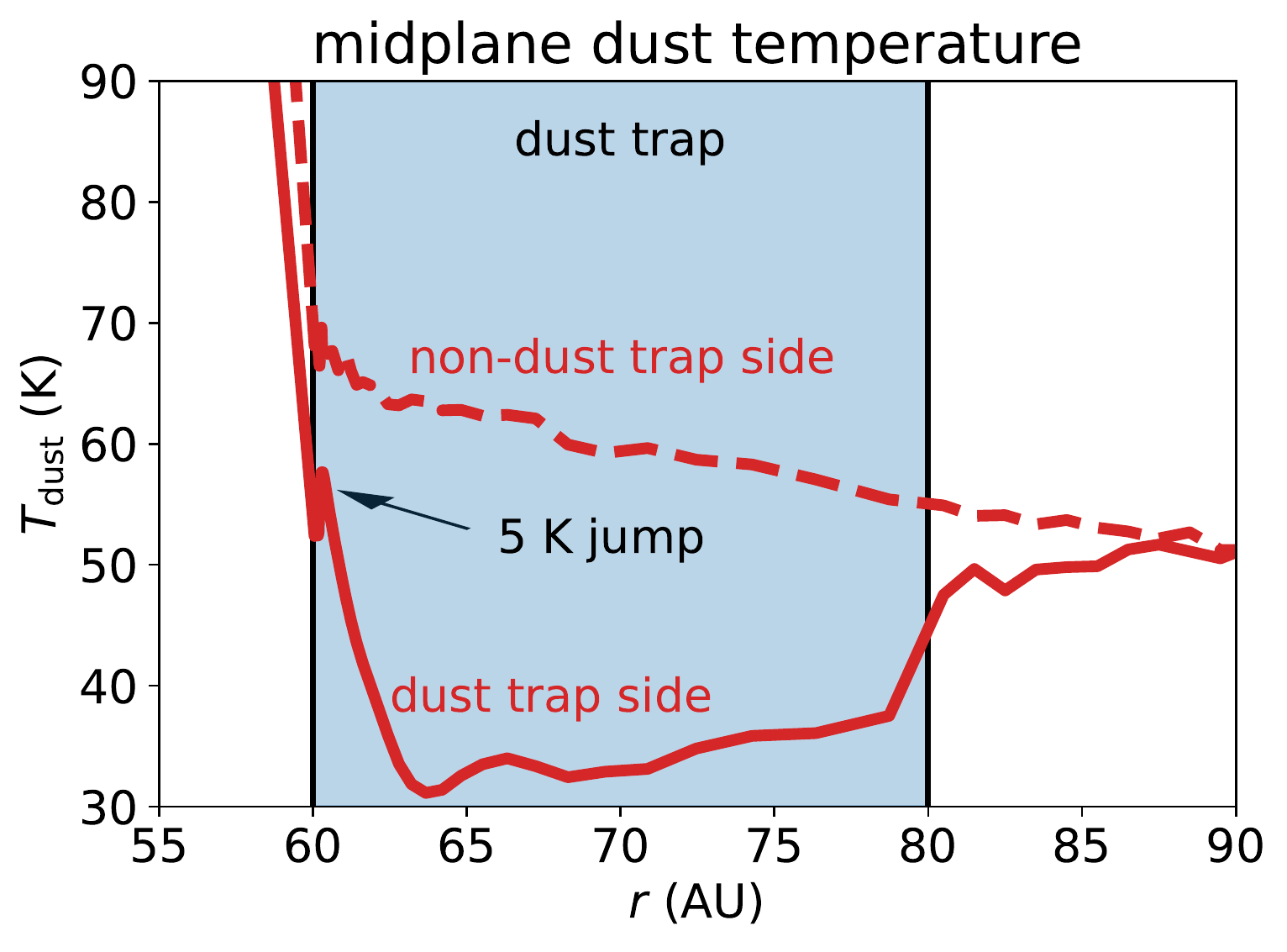}
      \caption{Midplane dust temperature in the DALI model for the dust trap side (solid) and non-dust trap side (dashed). Note the 5~K jump at the edge of the dust trap. }
         \label{fig:dali_Tdust_midplane}
\end{figure}

The typical gas number density at the dust trap edge is $10^8$~cm$^{-3}$ and the corresponding dust number density $n_{\mathrm{dust}}$\footnote{Defined as $n_{\mathrm{gas}}/n_{\mathrm{dust}} = \rho_{\mathrm{gas}}/\rho_{\mathrm{dust}} = 0.04-0.009$, with $\rho$ the mass density in g~cm$^{-3}$ \citep{vanderMarel2016, Ohashi2020}.} in that region increases up to $7.5\times 10^{10}$~cm$^{-3}$. Furthermore, the large dust is settled to the disk midplane with a scale height that is a factor of 0.1 of that of the gas and small dust. The midplane dust temperature in Fig.~\ref{fig:dali_Tdust_midplane} increases from 52~K at the inner edge of the dust trap at 60~AU to 58~K at 60.3~AU and then decreases to 31~K at 64~AU and increases slightly to 38~K at the outer edge at 80~AU. A full description of the model is given in Appendix~\ref{app:dali} together with an overview of the gas and dust density, temperature and the UV field in Fig.~\ref{fig:dali2D_general}. A cartoon of the gas and dust distribution is presented in Fig.~\ref{fig:cartoon}. First, we investigate the effect of the dust trap on the model in Sect.~\ref{sec:dali_dust}. Second, the effects of sublimating \ce{H2O} and \ce{NH3} ices are investigated in Sect.~\ref{sec:extra_H2O_NH3}. Third, the effect of additional NO in the disk is explored in Sect.~\ref{sec:extra_NO}.

\begin{figure*}
   \centering
  \begin{subfigure}{0.99\columnwidth}
  \centering
  \includegraphics[width=1\linewidth]{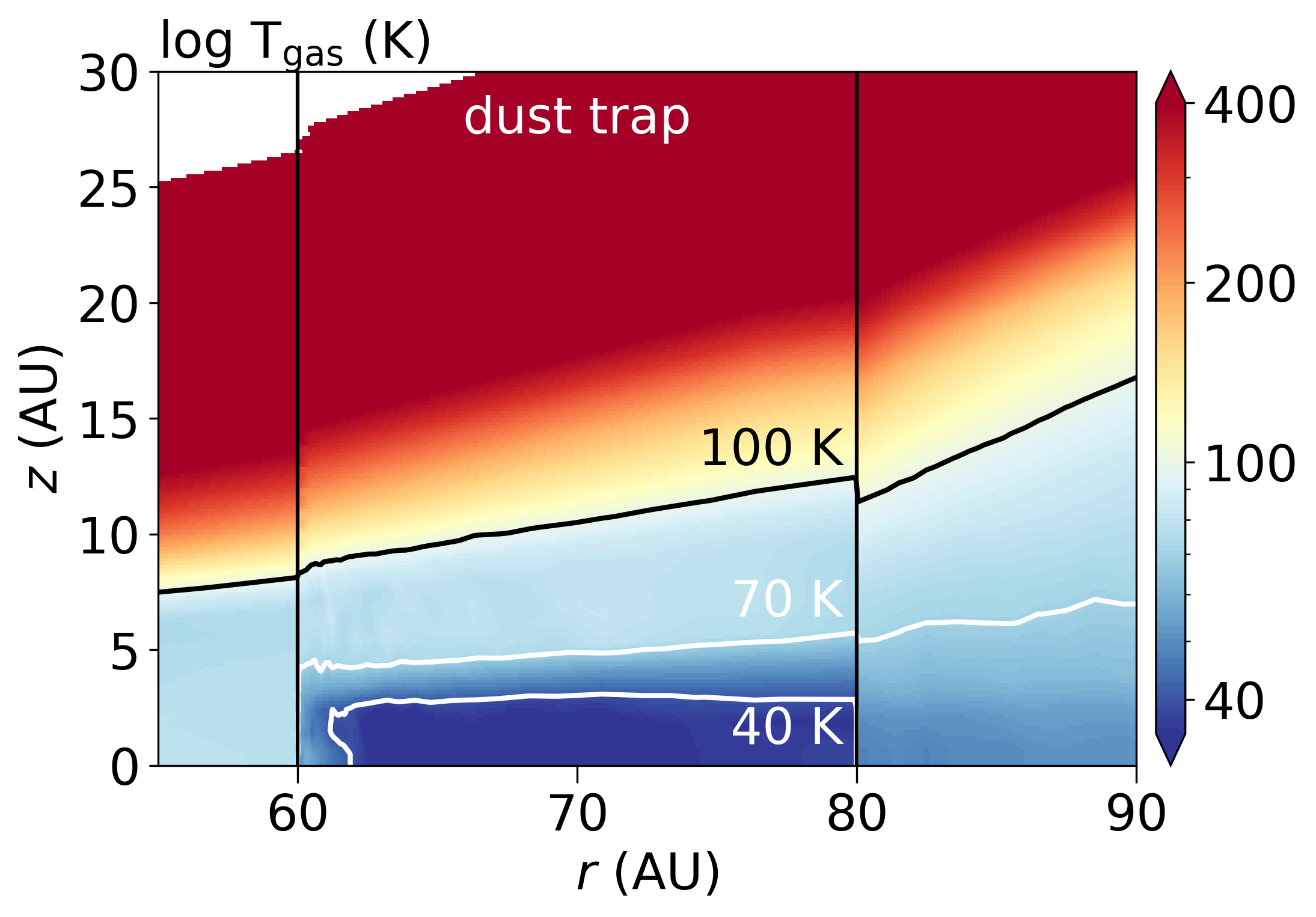}
\end{subfigure}%
\begin{subfigure}{0.99\columnwidth}
  \centering
    \includegraphics[width=1\linewidth]{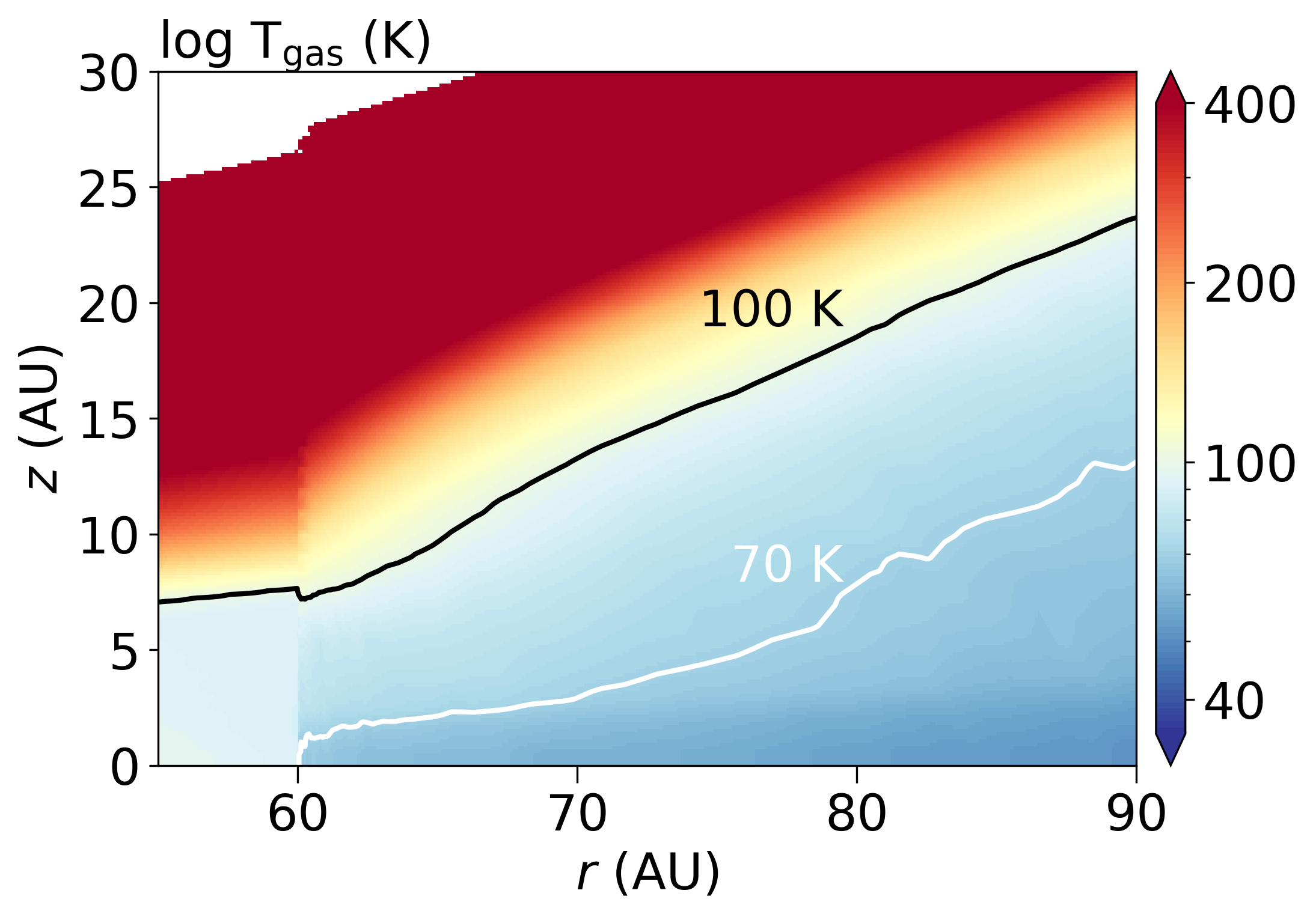}
\end{subfigure}
      \caption{Gas temperature structure in the IRS~48 disk model with (left) and without (right) a dust trap between 60 and 80~AU. The dust trap midplane is up to $\sim$30~K colder than the corresponding region in the non-dust trap side of the disk. At $z>10$~AU the dust trap model is warmer than the non-dust trap side due to the intense radiation field above the dust trap. The gas and dust temperature are similar below the $T_{gas} = 70$~K contour. Note that these figures are zoomed in on the dust trap region. The vertical black lines in the left hand panel indicate the inner and outer radius of the dust trap. }
         \label{fig:daliTgas}
\end{figure*}

The chemical abundances are computed using an updated version of the nitrogen chemistry network first presented in \citet{Visser2018}, \citet{Cazzoletti2018}, and \citet{Long2021}, (for details see Appendix~\ref{app:dali_chem}). This network is suited to evaluate the abundance of nitrogen bearing species such as NO and CN, but also small hydrocarbons such as \ce{C2H}. This network includes freeze-out and desorption but it does not include any ice-phase chemistry except for the formation of \ce{H2, H2O, CH4, NH3}, and HCN ices. More complex nitrogen bearing molecules such as \ce{N2O, NO2}, and \ce{NH2OH} are not included in this network.  
Carbon bearing molecules up to \ce{C2H3} (ethylenyl) are included in this model. The binding energy of \ce{C2H3} is set to $10^4$~K to mimic the effect of \ce{C2H3} being converted to other larger carbon-chain molecules that have a high binding energy. A lower binding energy results in gas-phase \ce{C2H3} that photodissociates into smaller molecules producing unphysically high abundances of other molecules such as \ce{C2H}. \ce{C2H3} thus acts as a sink of hydrocarbons in this model \citep{Wei2019}. For NO we use a binding energy of 1600~K, consistent with its 40-50~K desorption temperature in laboratory experiments \citep{Collings2004, Garrod2006methylformate, Wakelam2017}. \ce{NH2OH} typically desorbs at a much higher temperature of $170-250$~K under laboratory conditions \citep{Zheng2010, Congiu2012, Ioppolo2014, He2015, Fedoseev2016, Jonusas2016, Tsegaw2017}. Using the Redhead equation \citep{Redhead1962}, a binding energy of $\sim 6500$~K can be derived assuming an pre-exponential factor of $10^{13}$~s$^{-1}$ \citep{Congiu2012}. Similarly, for \ce{N2O} the Redhead equation together with an assumed pre-exponential factor of $10^{13}$~s$^{-1}$ and a measured peak desorption temperature of $\sim75$~K under laboratory conditions results in a binding energy of $\sim2500$~K, consistent with the literature value \citep{Congiu2012, Ioppolo2014}. 

The fiducial network starts with molecular initial conditions where all nitrogen starts in gas-phase \ce{N2} with an abundance of $3.1\times 10^{-5}$ w.r.t. hydrogen, and all oxygen is initially distributed over gas-phase CO ($1.3\times 10^{-4}$ w.r.t hydrogen) and \ce{H2O} ($1.9\times 10^{-4}$ w.r.t. hydrogen; see Table~\ref{tab:dali}). The models start with all molecules initially in the gas-phase to mimic the very quick sublimation form the ices. The models are run time dependently up to 100~yr and $10^3$~yr inside the dust trap. These time scales correspond to the typical freeze-out time scale just above the dust trap at $r=70$~AU and $z=5$~AU, and the time scale on which the dust trap could have been formed based on hydrodynamical models \citep{vanderMarel2013}, respectively. The default time scale is 100~yr. Outside the dust trap in the south and in the entire north side of the disk, the disk chemistry is evolved to 1~Myr, the typical disk formation timescale of a protoplanetary disk. As the youngest Keplerian rotating disks are detected in Class~0 objects with an approximate age on the order of $10^5$~yr \citep{Evans2009, Tobin2012, Murillo2013} and the typical disk life time is $\sim6$~Myr \citep[e.g.,][]{Haisch2001}. Evolving the disk for shorter (0.5~Myr) or longer (up to 10~Myr) timescales does not change the NO column density in the south side of the disk outside the dust trap by more than 1.4\%. 

Synthetic images are made using the raytracer in DALI to compare the models to the observations. The excitation temperature in each cell is calculated explicitly, without assuming LTE, using the collisional rate coefficients for NO with He, scaled down with a factor 1.4 to account for collisions with \ce{H2} from the LAMDA database \citep{Varberg1999, Schoier2005, Lique2009}. The images cubes are then convolved to the $0\farcs55 \times 0\farcs44\ (79.8\degree)$ beam of the observations. The DALI raytracer does not take line blending into account. Instead, the three NO lines at $\sim351.05$~GHz were raytraced separately and added afterwards. This is appropriate if these NO lines are optically thin, which is the case for all models except those with a very high initial NO abundance.

\subsection{Dust trap cools the disk midplane} \label{sec:dali_dust}

\begin{figure*}
   \centering
  \begin{subfigure}{0.99\columnwidth}
  \centering
  \includegraphics[width=1\linewidth]{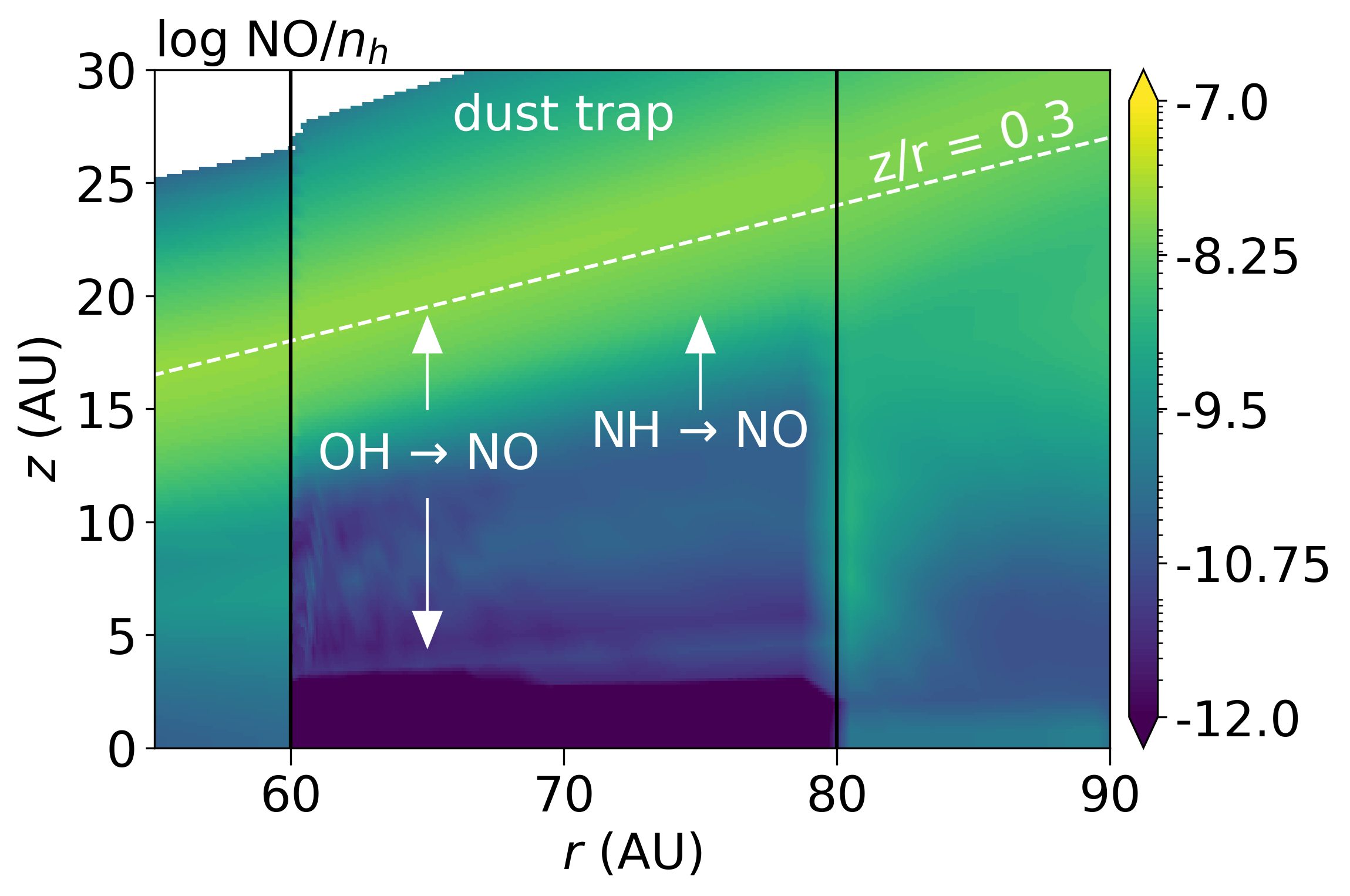}
\end{subfigure}%
\begin{subfigure}{0.99\columnwidth}
  \centering
    \includegraphics[width=1\linewidth]{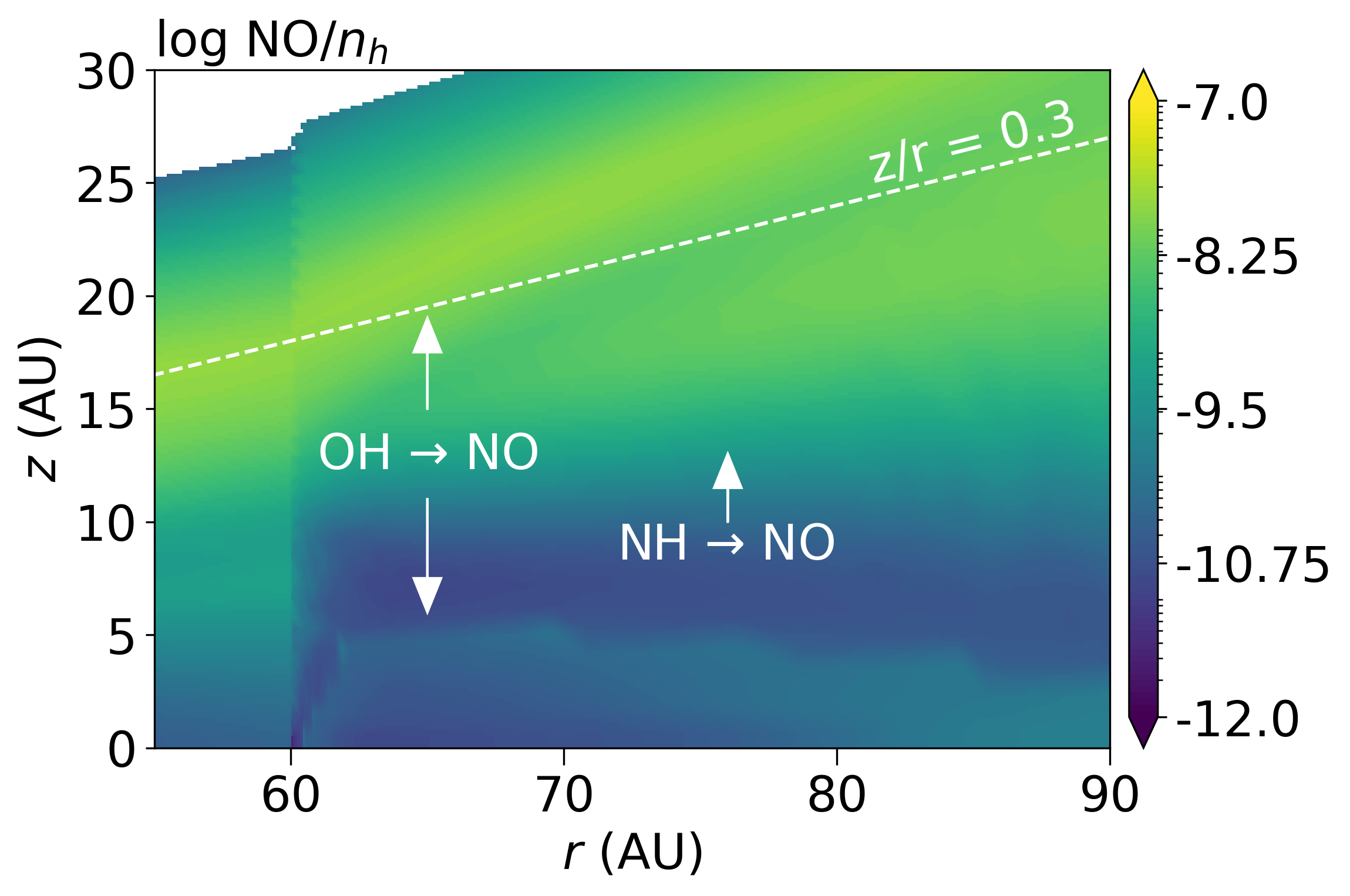}
\end{subfigure}
      \caption{NO abundance in the model with a dust trap (left) and without a dust trap (right). Note that the NO abundance inside the dust trap is lower than in the model without a dust trap except for a thin layer at $z/r \sim$0.3 (white dashed line), opposite of what is observed in the observations. This shows that the fiducial model is not sufficient to reproduce the observations.}
         \label{fig:dali2D_NO}
\end{figure*}

\begin{figure*}
   \centering
  \includegraphics[width=1\linewidth]{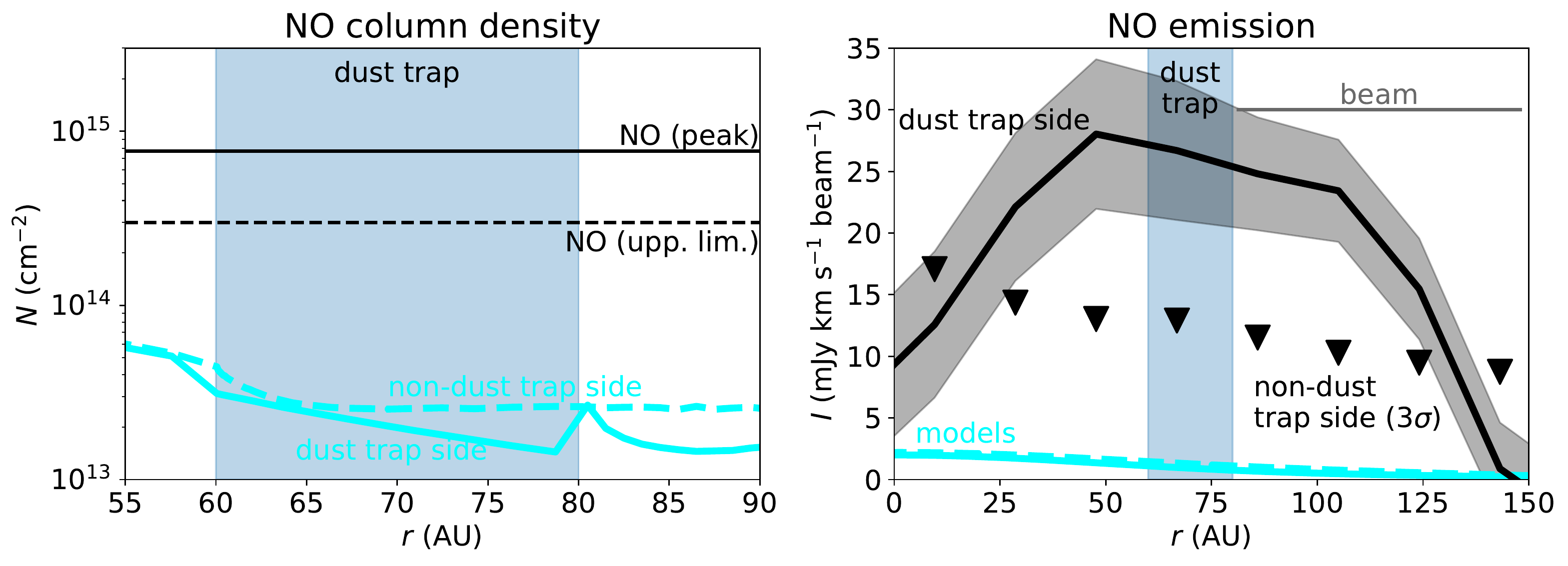}
      \caption{NO column density (left) and emission (right) predicted by the model for the dust trap side (solid blue) and non-dust trap side (dashed blue). The horizontal black lines in the left panel indicate the NO column density or the upper limit derived by the observations. The black line in the right panel indicates the horizontal cut of the NO emission through the dust trap and the triangles indicate the 3$\sigma$ upper limit on the intensity in the north. The horizontal grey line in the right panel indicates the beam size. Note that the two panels have a different horizontal axis. }
         \label{fig:dali_col_dens_em_NO}
\end{figure*}

The resulting gas temperature structure in the dust trap (south) and the non-dust trap side (north) are presented in Fig.~\ref{fig:daliTgas}. The gas temperature is up to 32~K lower inside the bulk of the dust trap than it is in the same disk region without the dust trap. This is due to the very high dust density of the large grains in the dust trap shielding the UV radiation that heats the disk. Therefore, despite their 30-60 times lower opacity to UV radiation, the large grains absorb almost all UV that is incident on the dust trap, lowering the temperature. Notably, a small jump of 5~K in the temperature is seen at 60.3~AU due to the intense radiation field (see Fig.~\ref{fig:dali_Tdust_midplane}). Furthermore, both sides of the disk are too warm for CO freeze-out (at $\sim20$~K). 

The 32~K drop in the dust temperature in the dust trap is not sufficiently cold for NO freeze-out ($T\sim 25$~K in the dust trap). As the non-dust trap side is warmer, NO is not freezing out in the IRS~48 disk model. The dust trap could still contain ices that have formed from NO in earlier phases, for example, \ce{N2O}, \ce{NO2}, and \ce{NH2OH} that are products formed in NO hydrogenation experiments and have larger binding energies \citep{Congiu2012, Yildiz2013, Fedoseev2012, Fedoseev2016} but as \ce{N2O}, \ce{NO2}, and \ce{NH2OH} are not included in the chemical network, their contributions cannot be studied directy with our model. Therefore, the dust trap can contain a reservoir of nitrogen bearing ices build up from the colder earlier phases.

The NO abundance predicted by the dust-trap model is presented in the left panel of Fig.~\ref{fig:dali2D_NO}. The NO has a moderate abundance of $1.4\times 10^{-8}$ in the surface layer at $z/r = 0.3$ (indicated with the dashed white line). In the top half of this layer, NO is mainly formed in the gas through:
\begin{align}
\ce{N + OH &\to NO + H}.
\end{align}
The OH is formed through the photodissociation of water as water photodissociates into OH, H, and O, where the latter reacts with molecular hydrogen to form more OH. In the lower half of the NO surface below $z/r = 0.3$, a second pathway through NH appears:
\begin{align}
\ce{NH + O \to NO + H}
\end{align}
where NH is formed from the reaction of N with vibrationally excited \ce{H2}. Deeper down in the disk including the dust trap itself, the NO abundance decreases due to a lack of OH and NH, the two main ingredients needed to form NO. The main destruction path of NO is through photodissociation. 

The NO abundance in the non-dust trap side of the model is presented in the right panel of Fig.~\ref{fig:dali2D_NO}. The formation and destruction pathways for NO are similar in the north side of the disk. The surface layer of NO splits into two layers: one where NO is mainly formed through OH and one where NO is mainly formed through NH. Furthermore, between the midplane and $z = 5$~AU, an NO layer with an abundance of $\sim 10^{-10}$ is present, indicated by the downward arrow in the right panel of Fig.~\ref{fig:dali2D_NO}. This layer is located just above the water snow surface, where some sublimating water photodissociates into OH and then forms NO. This layer can also be seen just above the dust trap in the dust trap model (left panel of Fig.~\ref{fig:dali2D_NO}).

The column density and emission predicted by the model are compared to the observations in Fig.~\ref{fig:dali_col_dens_em_NO}. The model column density in the non-dust trap side (dashed blue line) is $7-12$ times lower than the upper limit that is derived by the observations. Similarly, the emission predicted by the non-dust trap side is consistent with the non-detection of the NO in the north. On the other hand, the dust-trap side of the model underpredicts the NO column density by a factor of $25-50$ and the emission by a factor of 15. Furthermore, the dust trap model does not reproduce the characteristic shape of the NO emission that is only present at the location of the dust trap. Therefore, the NO abundance in the south needs to be enhanced compared to the fiducial model to explain the observations.

\subsection{Effect of sublimating ices: \ce{H2O} and \ce{NH3}} \label{sec:extra_H2O_NH3}

\begin{figure}
   \centering
  \includegraphics[width=1\linewidth]{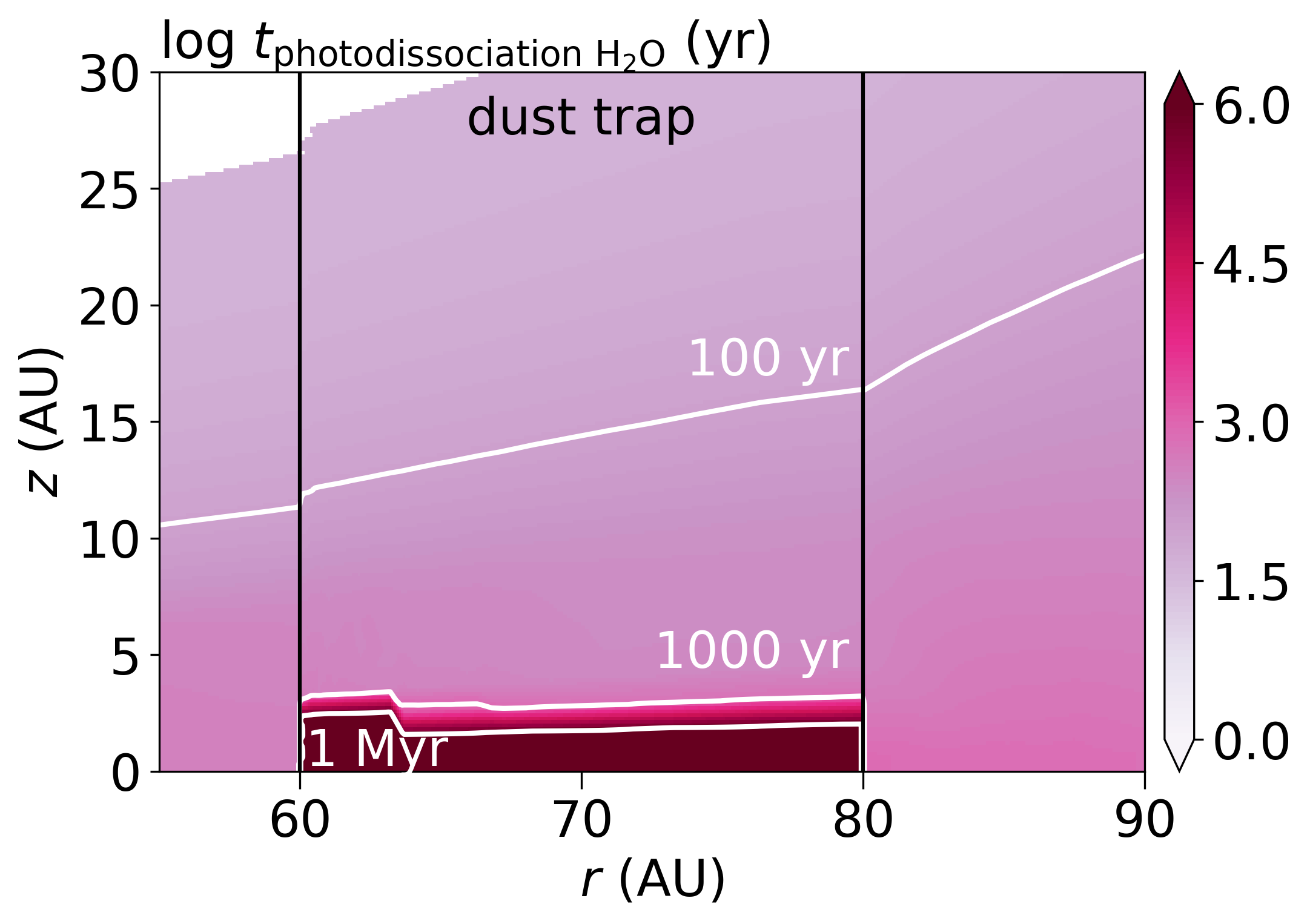}
      \caption{Photodissociation timescale of \ce{H2O}. In the dust trap midplane the photodissociation timescale is long due to the shielding of UV radiation by dust. Above the dust trap, the photodissociation timescale is much shorter, from $10^3$~yr just above to trap to $<100$~yr above $\sim$15~AU.}
         \label{fig:tphotodiss}
\end{figure}

\begin{figure*}
  \centering
  \includegraphics[width=1\linewidth]{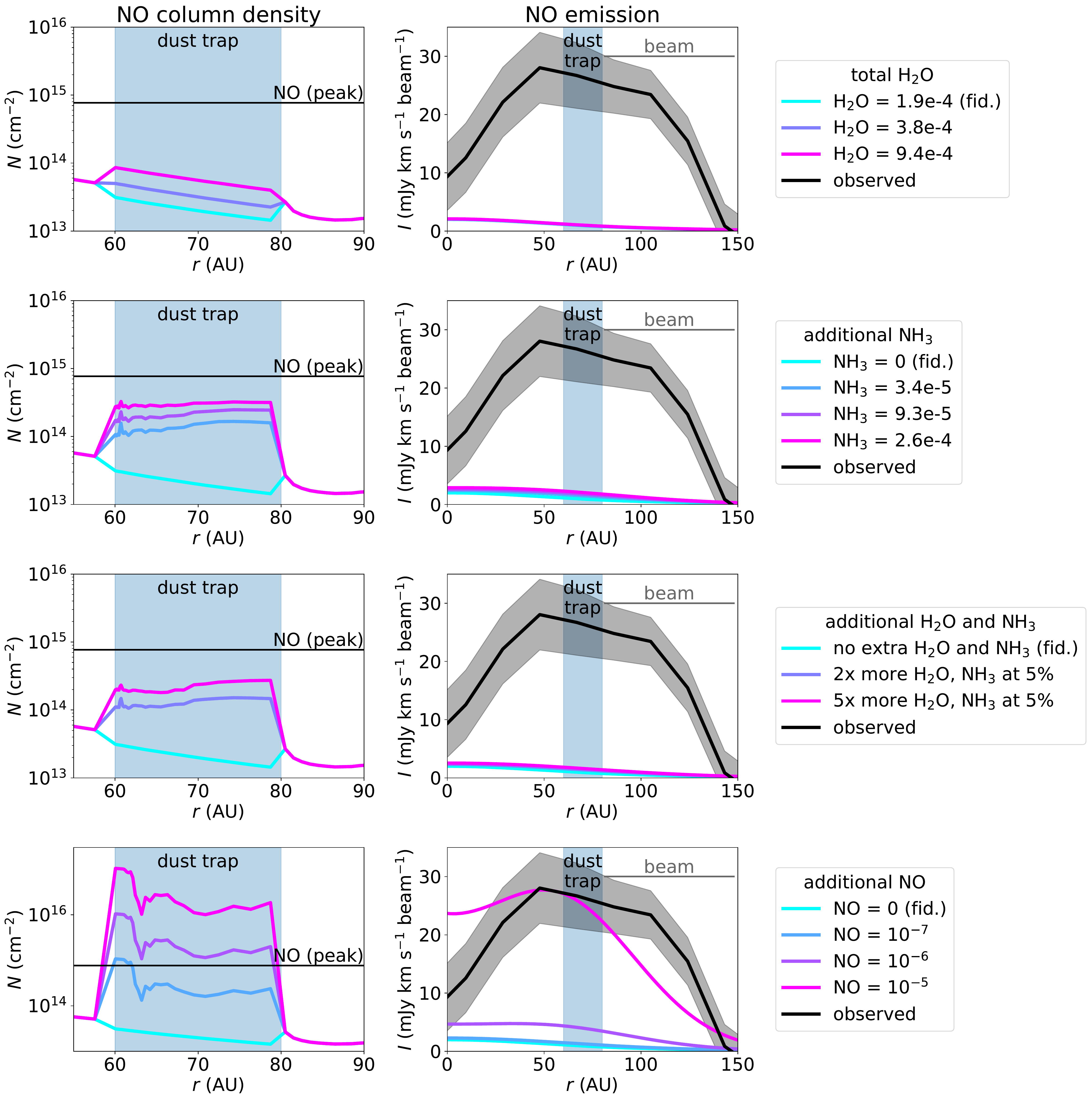}
      \caption{NO column density (left) and emission (right) for the dust trap models with different initial conditions. The different rows indicate the effect of additional \ce{H2O} (first row), \ce{NH3} (second row), \ce{H2O} and \ce{NH3} (third row), and NO (final row). The NO column density in the bottom row is dominated by that hidden underneath the $\tau_{\mathrm{dust}} = 1$ layer. The chemical network inside the dust trap is evolved for 100~yr whereas that outside is evolved for 1~Myr. The horizontal black line in the left column indicates the NO column density derived from the observations and the black line in the right column indicates the observed emission. Note the difference in the vertical axis of the bottom left panel. }
         \label{fig:dali_grid_NO92}
\end{figure*}

\begin{figure}
   \centering
  \includegraphics[width=1\linewidth]{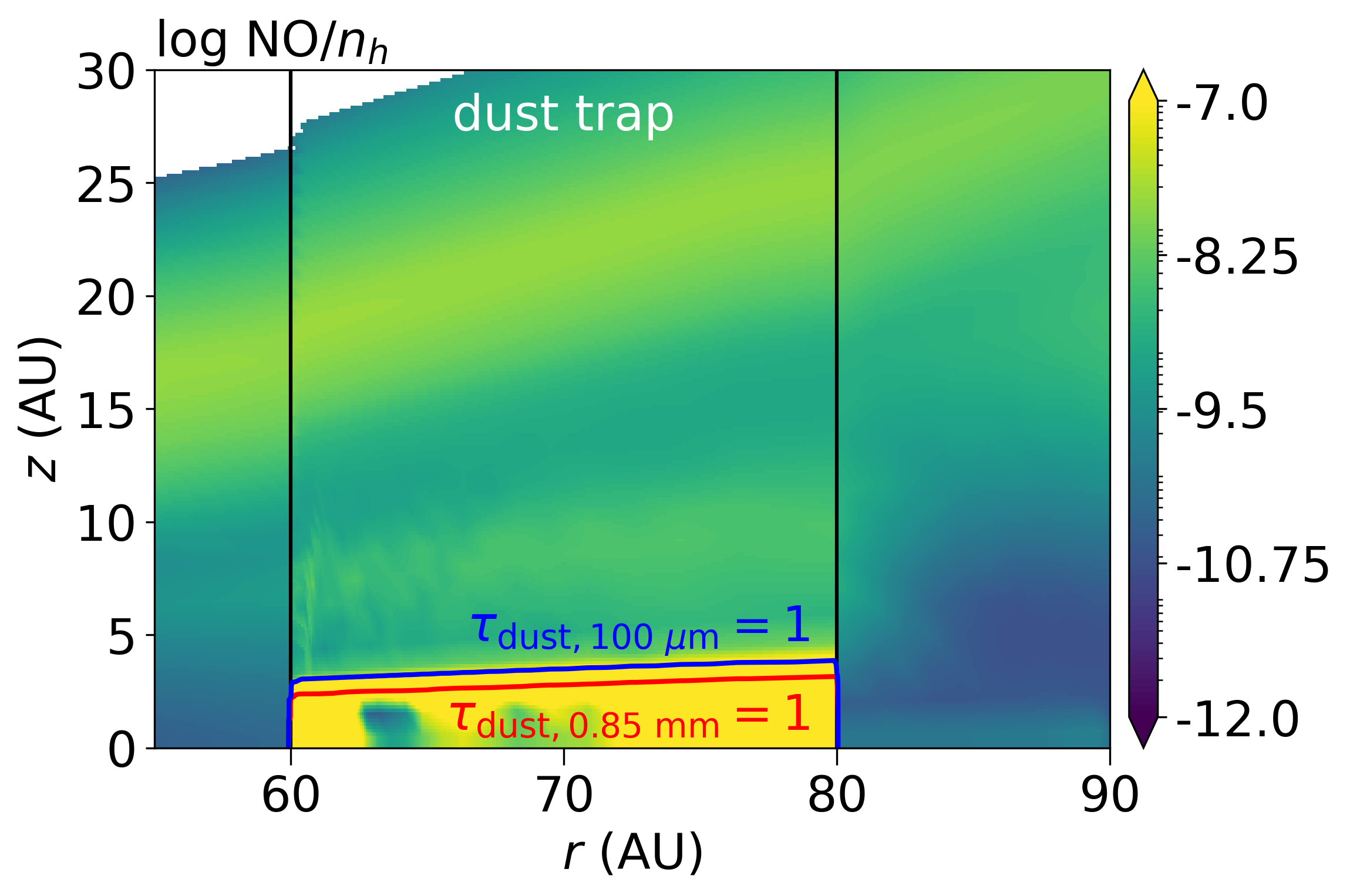}
      \caption{NO abundance after 100~yr in the dust trap model with an initial NO abundance of $10^{-5}$. Most of the NO is hidden below the $\tau_{\mathrm{dust, 0.85\ \mathrm{mm}}} = 1$ surface at 0.85~mm indicated by the red contour. The blue contour indicates $\tau_{\mathrm{dust, 100\ \mathrm{\mu m}}} = 1$ at 100~$\mu$m, representative of the wavelength of the \ce{H2O} and OH upper limits. }
         \label{fig:dali_2D_NO_tau}
\end{figure}

Observations of ices in young stellar objects and in comets show that oxygen is primarily locked up in water ice, whereas nitrogen is mostly in \ce{NH3} ice \citep{Boogert2015, Rubin2019}. The gas-phase detections of \ce{CH3OH}, \ce{H2CO}, SO, \ce{SO2}, and other complex organic molecules in the IRS~48 disk show that the dust trap is an ice trap \citep[Paper~I and II,][]{Booth2021irs48}. Therefore, \ce{H2O} ice is also likely sublimating above the dust trap. The photodissociation timescale of water is presented in Fig.~\ref{fig:tphotodiss} using a photodissociation rate attenuated by grown dust particles appropriate for protoplanetary disks:
\begin{align}
t = \frac{1}{k E_2(\gamma A_{\mathrm{V}})},
\end{align}
with $t$ the photodissociation timescale, $k = 7.7\times 10^{-10}$~s$^{-1}$ the photodissociation rate of \ce{H2O}, $E_2(x)$ the second order exponential integral, $\gamma = 0.41$ the dust shielding factor for \ce{H2O} for large grains, and $A_{\mathrm{V}}$ the visual extinction \citep{Heays2017}. Water dissociates in less than $10^3$ years just above the dust trap and in less than 100 years for the layers above the dust trap at $z > 15$~AU, providing a reservoir of OH that is partially used to form NO. 

Increasing the initial ice and thus gas-phase water abundance by a factor of 2 and 5 only increases the NO column density with a factor of $1.6-2.8$, up to a peak column density of $8.6\times 10^{13}$~cm$^{-2}$ as presented in the top left panel of Fig.~\ref{fig:dali_grid_NO92}. This is still an order of magnitude smaller than the NO column density found from the observations. Thus, much higher initial abundances of water ice, translating into enhanced gas-phase water would be needed to explain the NO column density. However, this is inconsistent with the upper limit on the warm \ce{H2O} and OH column density derived from the \textit{Herschel} PACS observations, suggesting that a different formation pathway of NO is dominant.

The column density of warm ($>150$~K) water predicted by the fiducial model and the models with additional \ce{H2O} is a factor of $\sim 14$ higher than the upper limit derived from \textit{Herschel} at that temperature, see Figs.~\ref{fig:Herschel_H2O_OH}, \ref{fig:dali_grid_big_H2O}, and \ref{fig:dali_grid_big_OH}. Similarly, the warm OH column density is a factor of $\sim30$ higher than its observed upper limit. Additionally, the upper limit on the CN column density constrains the minimum initial water abundance to be $\sim 1.9\times 10^{-5}$ if all nitrogen starts as \ce{N2} and $\gtrsim 9.4\times 10^{-5}$ if this is not the case. This is because lower water abundances overproduce the upper limit on the CN column density (see Fig.~\ref{fig:dali_grid_big_CN}). Hence, the C/O is constrained to be $\lesssim 1$ (if all nitrogen starts as \ce{N2}) or $\lesssim 0.6$ (if some nitrogen starts as N or \ce{NH3}) at the dust trap. Thus, sublimating water alone cannot explain the high NO column density.

The second pathway to NO in the model is through NH. In the disk surface layers, NH is created by the reaction of N with vibrationally excited \ce{H2}, but in the dust trap it could be created through sublimating \ce{NH3} that photodissociates. This is modelled as an additional gas-phase \ce{NH3} abundance of $3.4\times 10^{-5} - 2.6\times 10^{-4}$. At the dust trap edge and in the bulk of the dust trap, the NO column density predicted by these models is a factor of $3-8$ lower than that derived from the observations (see second row in Fig.~\ref{fig:dali_grid_NO92}). Additionally, the highest initial \ce{NH3} abundance of $2.6\times 10^{-4}$ matches the upper limit on the CN column density so the abundance cannot be higher. Finally, the observed NO emission predicted by these models is an order of magnitude too low (second row of Fig.~\ref{fig:dali_grid_NO92}).

All in all, adding gas-phase water or adding gas-phase \ce{NH3} leads to an increase in the NO column density, but neither pathway is sufficient to explain the observed NO emission because the level required is inconsistent with the upper limits on OH, \ce{H2O}, and CN. In astronomical ices the abundance of \ce{NH3} with respect to water is typically 5\% \citep{Boogert2015}. This is modelled by adding a water abundance of $1.9\times 10^{-4}$ and $7.6\times 10^{-4}$ and an \ce{NH3} abundance that is 5\% of that to the fiducial initial water abundance of $1.9\times 10^{-4}$ and initial \ce{NH3} abundance of 0 (i.e., increasing water by a factor of 2 and 5 and increasing \ce{NH3} accordingly to $1.9\times 10^{-5}$ and $4.7\times 10^{-5}$). The results are very similar to just adding \ce{NH3} and can be seen in the third row of Fig.~\ref{fig:dali_grid_NO92}. In summary, sublimating \ce{H2O} and/or \ce{NH3} ices alone cannot explain the NO observations, even if the NO inside the dust trap would be mixed up to higher layers.

\subsection{Effect of sublimating ices: additional NO} \label{sec:extra_NO}
The formation pathways to form NO from simple molecules are not efficient enough to explain the observations. Here, we investigate the opposite scenario: NO is the photodissociation product of a larger molecule with an NO bond that is frozen out in the dust trap. As our chemical network does not include species like this, we model this scenario as an initial NO abundance of $10^{-7}, 10^{-6}$, and $10^{-5}$. The resulting column densities are presented in the bottom left panel of Fig.~\ref{fig:dali_grid_NO92}. An initial NO abundance of $10^{-7}$ results in an NO column density that matches that of the observations between 60 and 62~AU, but it underpredicts the column density by a factor of $\sim$4 in the bulk of the dust trap. Increasing the NO abundance by one and two additional orders of magnitude results in an NO column density that increases that same amount up to a peak NO column density of $10^{16}$~cm$^{-2}$ and $10^{17}$~cm$^{-2}$ respectively.

Only the model with an initial NO abundance of $10^{-5}$ matches the observed NO emission (see bottom right panel of Fig.~\ref{fig:dali_grid_NO92}). The three NO lines in this model are optically thick with optical depth of $1.4-2.7$. As the raytracer in DALI does not do line blending, we raytraced these lines separately and added the intensities of these three lines. If line blending was taken into account, the expected intensity of the three blended NO lines would be at least the intensity of 1 NO line raytraced separately, and at most the sum of the three NO lines raytraced separately. As the three NO lines have similar line strengths, the predicted line intensity is expected to be a factor of $2-3$ lower when line blending is taken into account. In the observations, the 351.044~GHz line can marginally be separated from the other two at 351.052~GHz. Therefore, the true NO intensity expected from this is expected to be a factor $2-3$ lower than that plotted in the bottom right panel of Fig.~\ref{fig:dali_grid_NO92}. The NO intensity in the model with an initial NO abundance of $10^{-6}$ is only marginally optically thick with a maximum optical depth of 0.3. In summary, a high initial NO abundance is needed to match the observations within a factor of $2-3$. 

The discrepancy in the column density needed in the models compared to that derived from the observations is likely due to the different emitting layers. 
Fig.~\ref{fig:dali_2D_NO_tau} shows that most of the NO is distributed inside the dust trap where the dust hides the NO emission. 
The NO column density above the $\tau_{\mathrm{dust}, 0.85~\mathrm{mm}} = 1$ surface is a few $10^{15}$~cm$^{-2}$, which is consistent with the observations within a factor of 5. This factor of 5 could be due to beam dilution indicating that the true emitting region of NO is 5 times smaller than that assumed in Sect.~\ref{sec:NO_obs}. Therefore, the NO column density is higher than the inferred value of $7.7\times 10^{14}$~cm$^{-2}$. Another possibility is that the initial NO abundance in the IRS~48 disk is lower at $10^{-7}-10^{-6}$ and that the NO located below the optically thick dust is mixed up to higher layers, which is not included in our models. Possible parent molecules that could photodissociate into NO are considered in Sect.~\ref{sec:discussion}.

\subsection{Chemical composition of the non-dust trap side} \label{sec:north}
In this section the chemical composition of the north side of the IRS~48 disk is explored. As the large grains with their icy mantles are trapped in the south, the north side of the disk could be depleted in gas-phase water. This is modelled as an initial abundance of gas-phase water of $9.4\times 10^{-5}$, $3.8\times 10^{-5}$, and $1.9\times 10^{-5}$. As these models start without atomic oxygen initially present, the main oxygen reservoir is CO with an abundance of $1.3\times 10^{-4}$ w.r.t. hydrogen. The overall carbon ($1.3\times 10^{-4}$ w.r.t. hydrogen) and nitrogen abundances ($6.2\times 10^{-5}$ w.r.t. hydrogen) are identical to the fiducial model and follow the typical ISM values. The resulting column densities are presented in Fig.~\ref{fig:dali_grid_north_NO}, \ref{fig:dali_grid_north_CN}, and \ref{fig:dali_grid_north_C2H}. Interestingly, the upper limit on the CN column density is more constraining than that of NO in the north and also more constraining than that of \ce{C2H}. Depleting \ce{H2O} by a factor of 2 increases the CN column density by that same factor causing the model to be slightly lower than the derived upper limit. Depleting water with a factor of $5-10$ increases the CN column density to $(6-8)\times 10^{13}$~cm$^{-2}$ in the bulk of the dust trap. This is consistent with the CN upper limit if all nitrogen initially starts as \ce{N2}. On the other hand, if nitrogen starts as solely N, solely \ce{NH3} or a mix of \ce{N2}, N, and \ce{NH3}, the CN constrains the water abundance to be at least $9.4\times 10^{-5}$. Therefore the C/O ratio in this disk region is $\leq 1$ if all nitrogen is initially in \ce{N2} and $\leq 0.6$ if not all nitrogen is initially in \ce{N2}.

\section{Discussion} \label{sec:discussion}
\subsection{Possible parent molecules of NO}

The NO emission is only seen at the location of the dust trap in the observations, indicating that the NO is directly related to the dust trap. Additionally, modelling of the IRS~48 dust trap has shown that an additional source of NO relative to the fiducial model is needed to explain the observations. As the photodissociation timescale in these surface layers of the disk is short, this NO is likely the photodissociation product of a larger molecule that carries an NO bond and is frozen out in the dust trap. 

\begin{figure}
   \centering
  \includegraphics[width=1\linewidth]{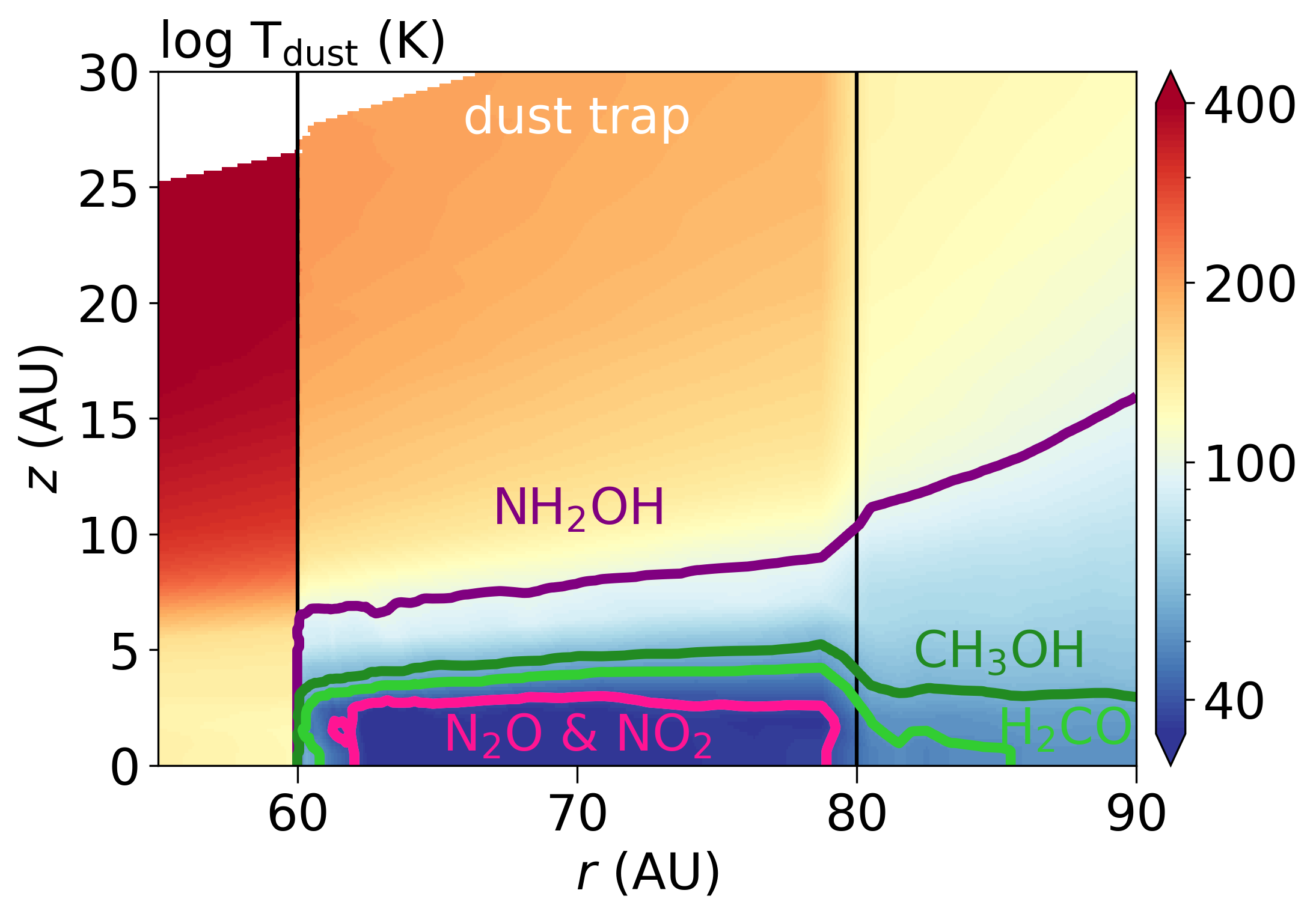}
      \caption{Snowlines of possible parent molecules of NO: \ce{N2O} and \ce{NO2} (pink) and \ce{NH2OH} (purple). For comparison also the \ce{CH3OH} (dark green) and \ce{H2CO} (lighter green) are indicated. The colored background indicates the dust temperature. }
         \label{fig:snowlines}
\end{figure}

One candidate for this is \ce{NH2OH} that can be formed by hydrogenation of NO ice \citep{Fedoseev2012}. However, \ce{NH2OH} gas photodissociates primarily into \ce{NH2 + OH} \citep{Betts1965}. The two-step photodissociation pathway to form NO is less efficient \citep{Gericke1994, Fedoseev2016}. Although \ce{NH2OH} has a higher binding energy of $\sim 6500$~K \citep{Penteado2017} than \ce{CH3OH} (3800~K), its snowline is only 5~AU higher, just above the region where \ce{CH3OH} is expected to be abundant in the models (see Fig.~\ref{fig:snowlines} and Paper~I). The fact that \ce{NH2OH} is not detected in the IRS~48 disk with an upper limit on the column density that is 4 times lower than that of NO suggests that it is not present in the gas-phase. Furthermore, the only detection of \ce{NH2OH} in the ISM yields a low abundance of only a few $10^{-10}$ \citep{Rivilla2020}. Therefore, \ce{NH2OH}, if present, likely remains frozen onto the dust grains. This together with only a minor route that produces NO after photodissociation, makes \ce{NH2OH} an unlikely parent molecule of the detected NO.

Three other molecules that photodissociate into NO are \ce{N2O}, \ce{NO2}, and HNO. No HNO lines are covered in our data so no constraint can be made on its column density. Its photodissociation cross section is $1.7\times 10^{-10}$~s$^{-1}$, which is 2 times smaller than that of NO. Therefore, HNO photodissociates slower than NO by that same factor. If HNO would be the parent molecule of NO, it would have to be at least twice as abundant to compensate for its faster photodissociation. 

The upper limit on the \ce{N2O} column density is comparable to that of the detected NO emission in the south. Furthermore, the \ce{N2O} binding energy is $\sim 2500$~K \citep{Congiu2012, Ioppolo2014}, which is lower than that of methanol. Therefore, if \ce{N2O} ice is present on the grains, it will sublimate and photodissociate in the gas into either NO + N or \ce{N2} + O \citep{vanDishoeck2006, Heays2017}. As the photodissociation rate of \ce{N2O} is 5 times faster than that of NO, the formation of NO outpaces that of its destruction if \ce{N2O} photodissociation is the main pathway for NO. The additional atomic nitrogen slightly enhances the CN, but more importantly, the atomic nitrogen destroys part of the NO, lowering its column density compared to the NO = $10^{-5}$ model by two orders of magnitude down to a peak value of $6.5\times 10^{14}$~cm$^{-2}$. Therefore, \ce{N2O} could be the parent molecule of NO, but the high N abundance produced with it makes this unlikely. 

\ce{NO2} is not detected with an upper limit that is an order of magnitude larger than that of NO. As it does photodissociate into NO, this could be the parent molecule of NO \citep{vanDishoeck2006, Heays2017}, but it cannot be excluded that other molecules not considered here are parents as well. 

Another origin of the NO could be in \ce{OCN-} ice that has a typical abundance of 0.4\% in low-mass young stellar objects and $0.6-1.53$\% w.r.t. \ce{H2O} in massive young stellar objects \citep{vanBroekhuizen2005, Oberg2011, Boogert2022}. Additionally, \ce{OCN-} ice has recently been detected with an abundance of 0.3\% w.r.t. \ce{H2O} in a cloud prior to star formation \citep{McClure2023}. \ce{OCN-} likely quickly reacts with hydrogen to form HNCO or possibly another  molecule upon desorption. HNCO, as well as HCNO and HOCN have been detected in the gas-phase in molecular clouds \citep{Marcelino2010}. Photodissociation of HNCO could also lead to the formation of NO, but quantifying its contribution requires knowledge of its photodissociation cross section which has not been measured to date. In summary, multiple species that are frozen out on the grains in the IRS~48 dust trap could lead to the formation of NO in the gas-phase through photodissociation.

\subsection{Comparison to other sources}
In the IRS~48 disk we find NO/\ce{N2O} $> 1$ assuming that \ce{N2O} would emit at 100~K, similar to \ce{CH3OH}. This ratio is consistent with that of $\sim 8.8-10$ found in the giant molecular cloud Sgr~B2(M) and (N) \citep{Ziurys1994, Halfen2001} but it is inconsistent with the value of $\leq 0.5$ in the young protosolar analog IRAS 16293–2422B \citep{Ligterink2018}. This is interesting because both in IRAS 16293–2422B and in the IRS~48 disk the emission of many molecules is explained by sublimating ices. The difference in the NO/\ce{N2O} ratio could be due to chemical differences between the two sources. 

In the comet 67P, most of the volatile nitrogen is locked up in \ce{NH3} \citep{Rubin2019} though its abundance is very low at only 0.67\% compared to \ce{H2O}. Other nitrogen carriers in this comet include \ce{N2} (13\% all w.r.t. \ce{NH3}), HCN (21\%), HNCO (4\%), some minor contributions ($<1$\%) of \ce{NH2CHO}, \ce{CH3CN}, and \ce{HC3N}. Furthermore, refractory ammonia salts (\ce{NH_4^+X^-}) or CHON particles in the dust grains on 67P may carry a large part of the total nitrogen budget \citep{Rubin2019, Altwegg2020, Altwegg2022}. Therefore, the NO emission in the IRS~48 disk could be enhanced by the sublimation of \ce{NH3} but as we have shown with the models, this pathway alone is not sufficient to match the NO detection and the upper limit on the CN column density.

Another chemical difference between the IRS~48 disk and other sources arises from the \ce{H2O} to \ce{CH3OH} ratio, which is typically found to be $10-100$ in the ice on comets and on the dust around young stellar objects \citep{Bottinelli2010, Boogert2015}. In contrast, this ratio is only $\lesssim 0.2$ in the gas in the IRS~48 disk. This suggests that this ratio is very low in the ices in the IRS~48 dust trap or that the gas seen in the IRS~48 disk does not directly trace the ice abundances. The gas-phase \ce{H2O} to \ce{CH3OH} ratio could be modified compared to that in the ices due to photodissociation and (a lack of) subsequent formation. \ce{H2O} photodissociates 2 times slower than \ce{CH3OH} \citep{Heays2017}, increasing the expected \ce{H2O} to \ce{CH3OH} ratio in the gas w.r.t. that in the ice. Furthermore, \ce{H2O} is formed in the gas-phase through OH if the gas temperature exceeds 300~K, whereas \ce{CH3OH} does not have an efficient gas-phase formation route \citep{Garrod2006methanol, Geppert2006}. This further increases the expected \ce{H2O} to \ce{CH3OH} ratio in the gas compared to that in the ice.

Additionally, this ratio is found to be 6 in the gas disk around the young outbursting source V883~Ori, using an \ce{H_2^18O} column density of $(5.52 \pm 1.2)\times 10^{15}$~cm$^{-2}$ and an \ce{^13CH3OH} column density of $6.9^{+0.70}_{-0.72}\times 10^{15}$~cm$^{-2}$ both scaled to their main isotopologues using \ce{^12C}/\ce{^13C} = 70 and \ce{^16O}/\ce{^18O} = 560 but uncorrected for the dust that may hide part of the line emission \citep{Wilson1994, Milam2005, Lee2019, Tobin_accepted}. The difference between the V883~Ori and IRS~48 disks could be due to a difference in the continuum optical depth. \ce{H2O} and \ce{CH3OH} are both observed at sub-mm wavelengths in the V883~Ori disk but in the IRS~48 disk \ce{H2O} observed at $51-220~\mu$m and the \ce{CH3OH} observed at sub-mm wavelengths. Yet, the $\tau = 1$ surface of the continuum at 100~$\mu$m is located only $\sim$0.5~AU higher than that of the sub-mm. Therefore, the \ce{H2O} and \ce{CH3OH} observations probe up to very similar depths in the disk, suggesting that the optical depth of the 100~$\mu$m continuum is underestimated or that there is a chemical difference between the IRS~48 dust trap and V883~Ori. 

In the IRS~48 disk, our upper limit on the warm gaseous \ce{H2O} column density of $\lesssim~10^{14}$~cm$^{-2}$ inside its snowline at 60~AU results in an upper limit on the abundance of $\lesssim~6\times~10^{-9}$ w.r.t. hydrogen using the hydrogen column density of $1.6\times 10^{22}$~cm$^{-2}$ at 60~AU in the DALI model based on the $J=6-5$ transition of \ce{C^17O} \citep{Bruderer2014, vanderMarel2016}. This upper limit is consistent with the gas-phase \ce{H2O} abundance in the cold, photodesorbed reservoir located outside the water snowline in the HD~100546 and HD~163293 disks that is low due to radial drift. It is not consistent with the abundance of warm gas-phase water in the inner disk (inside the water snowline) of HD~163296 \citep{Pirovano2022}.

The lack of CN emission in this disk constrains its C/O ratio to be $\lesssim 0.6$ both inside and outside the dust trap. 
In the south, this is in line with the C/O ratio that is expected for gas dominated by sublimating ices \citep{Oberg2011}. Yet, in the north side of the disk where no direct evidence of sublimating ices is seen, a low C/O of $\lesssim 0.6$ is more surprising as gas depleted in volatiles has a high C/O ratio of $\sim 1$. Therefore, the low C/O ratio suggests that also the gas in the north side of the disk could be affected by ices that sublimated and then photodissociated. If this is the case, the dust trap must be relatively old such that the gas in the disk has either mixed or the dust trap has at least made one more full orbit than the sub-Keplerian gas.

\section{Conclusions} \label{sec:conclusions}
In this work, we investigated the chemical origin of the first detection of NO in a disk. The special case of the IRS~48 dust trap allows us to investigate its origin, as this disk is a unique laboratory for sublimating ices. The NO column density and emission were modelled using the thermochemical code DALI with a nitrogen chemistry network. The effects of sublimating \ce{H2O} and \ce{NH3} ices were investigated as well as the effect of sublimation of larger molecules that carry an NO bond. In summary, we conclude:

\begin{itemize}
\item[$\bullet$] NO in the observations is co-located with the IRS~48 dust trap in the south, and it is not detected in the north.
\item[$\bullet$] Models indicate that the dust temperature in the dust trap midplane is up to $\sim$30~K colder than the non-dust trap side of the disk. 
\item[$\bullet$] The fiducial model for the north side of the disk is consistent with the observations, but the fiducial model for the south side (dust trap location) underpredicts the NO column density by an order of magnitude. 
\item[$\bullet$] Sublimating water and/or \ce{NH3} enhance the NO column density up to a factor of $10$, but the observed NO emission is still not reproduced. These models overproduce the upper limits on \ce{H2O}, OH column densities. The upper limit on the CN emission constrains the \ce{NH3} abundance to be $\lesssim 2.6\times 10^{-4}$. Therefore, the high NO column density cannot be explained by sublimation of these molecules.
\item[$\bullet$] An additional source of NO is needed to explain the observations. This is likely a molecule that carries an NO bond and freezes out in the dust trap midplane but sublimates in the higher layers of the disk. Possible candidates could be \ce{N2O} and HNCO. 
\item[$\bullet$] The C/O ratio in this disk is $\lesssim 0.6$ both in the dust trap side and in the non-dust trap side due to the lack of CN emission, if not all nitrogen is initially in \ce{N2}. If all nitrogen is initially in \ce{N2}, the C/O ratio is $\leq 1$. 
\end{itemize}

The observed line emission from NO and COMs are co-spatial with the asymmetric dust trap in the IRS~48 disk. This coincidence strongly suggests a direct relation between the detected gas-phase molecules and mm dust grains which are expected host an ice reservoir. The first detection of NO in a protoplanetary disk opens a new opportunity for studying the C/O ratio in disks using NO and CN. Future observations targeting other nitrogen bearing molecules are needed to further constrain the parent molecule that is forming the NO and characterise the partitioning of volatile nitrogen in this disk.

\begin{acknowledgements}
We thank the referee for the constructive comments. Furthermore, we would like to thank A. D. Bosman for the useful discussions. We acknowledge assistance from Allegro, the European ALMA Regional Centre node in the Netherlands.
Astrochemistry in Leiden is supported by the Netherlands Research School for Astronomy (NOVA), by funding from the European Research Council (ERC) under the European Union’s Horizon 2020 research and innovation programme (grant agreement No. 101019751 MOLDISK), and by the Dutch Research Council (NWO) grant 618.000.001.
BT acknowledges support from the Dutch Astrochemistry Network II, grant 648.000.022 from the Dutch Research Council (NWO). N.F.W.L. acknowledges funding by the Swiss National Science Foundation Ambizione grant 193453. 
This paper makes use of the following ALMA data: ADS/JAO.ALMA \#2013.1.00100.S and \#2017.1.00834.S. ALMA is a partnership of ESO (representing its member states), NSF (USA) and NINS (Japan), together with NRC (Canada), MOST and ASIAA (Taiwan), and KASI (Republic of Korea), in cooperation with the Republic of Chile. The Joint ALMA Observatory is operated by ESO, AUI/NRAO and NAOJ.
\end{acknowledgements}

\bibliographystyle{aa}
\bibliography{refs.bib}

\begin{appendix}

\section{Observations}

\begin{figure*}
  \centering
    \includegraphics[width=1\linewidth]{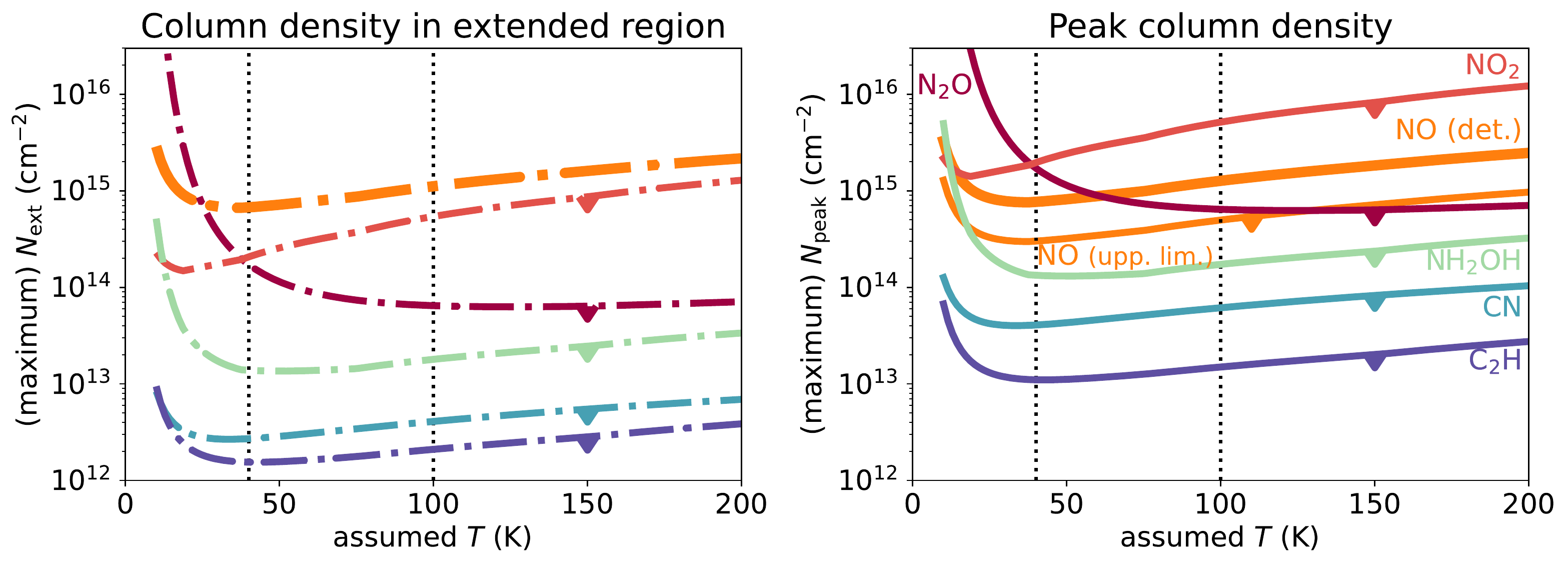}
      \caption{Derived (upper limits on the) column densities in the IRS~48 disk as a function of assumed temperature. The left panel presents column densities computed in an extended region ($N_{\mathrm{ext}}$; dash-dotted lines) and the right panel those in a single beam ($N_{\mathrm{peak}}$; solid lines). The NO column density is indicated with the thick orange line. The upper limit on the \ce{N2O}, \ce{NO2}, \ce{NH2OH}, \ce{CN}, and \ce{C2H} are indicated with the dark red, red, green, blue, and purple lines respectively. The upper limit on the NO column density in the north is indicated with the thin orange line in the right panel. The dotted vertical black lines indicate the two assumed temperatures for which the (upper limits on the) column densities are listed in Table~\ref{tab:N}.} 
         \label{fig:T_vs_N}
\end{figure*}

\begin{figure}
  \centering
    \includegraphics[width=1\linewidth]{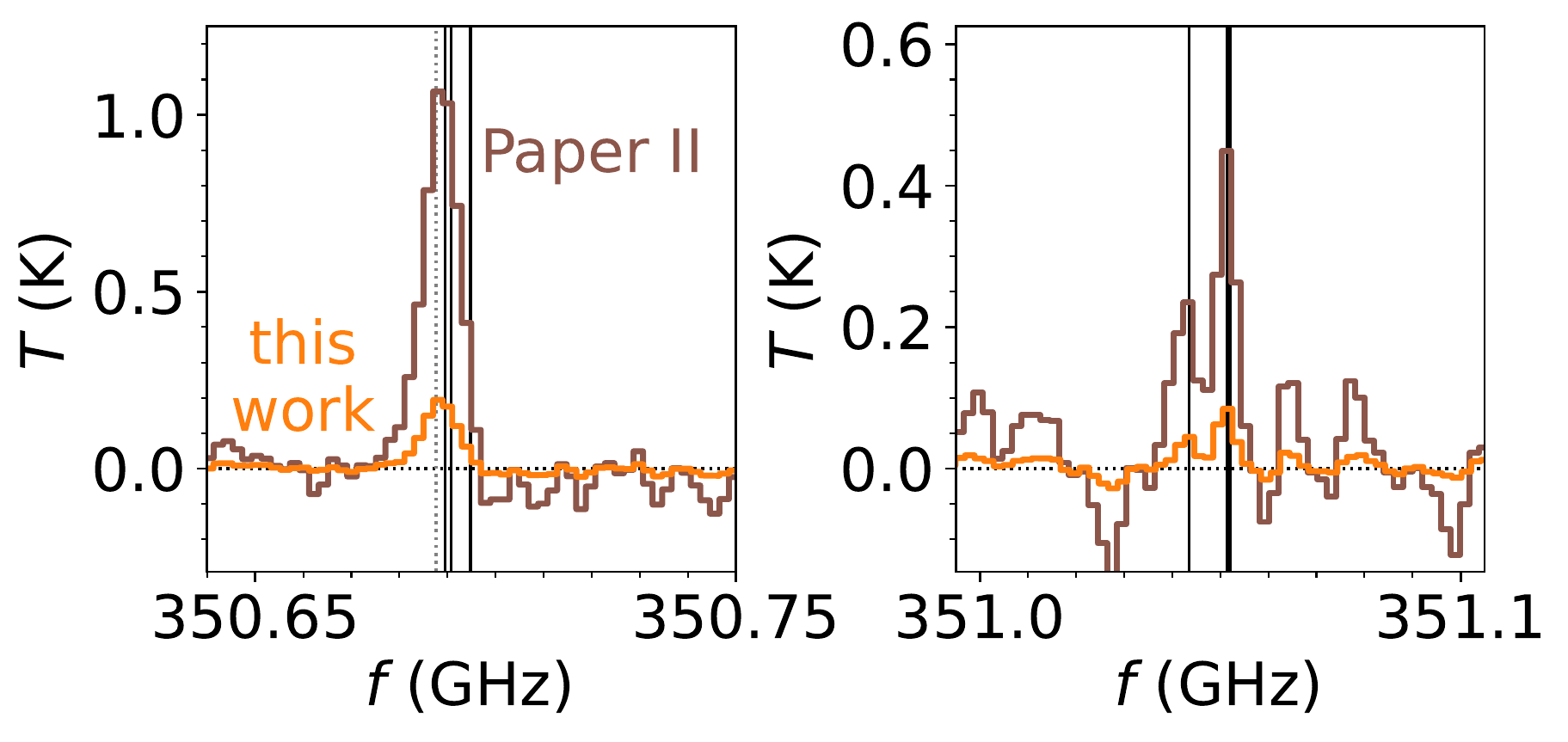}
      \caption{Observed and stacked spectrum of the NO lines detected inside 115~AU in units of K. The brown line indicates the spectrum presented in Paper~II and the orange line presents the corrected spectrum with a 5 times lower flux presented in this work. The detected NO lines are indicated with the vertical black lines. The thick line at 351.052~GHz indicates that there are two NO lines at the same frequency. The methanol line is indicated with the dotted grey line. For details see Sect.~\ref{sec:NO_obs}. Note the difference in the vertical scale between the panels.} 
         \label{fig:spec_NO_K}
\end{figure}

\begin{figure*}
  \centering
    \includegraphics[width=1\linewidth]{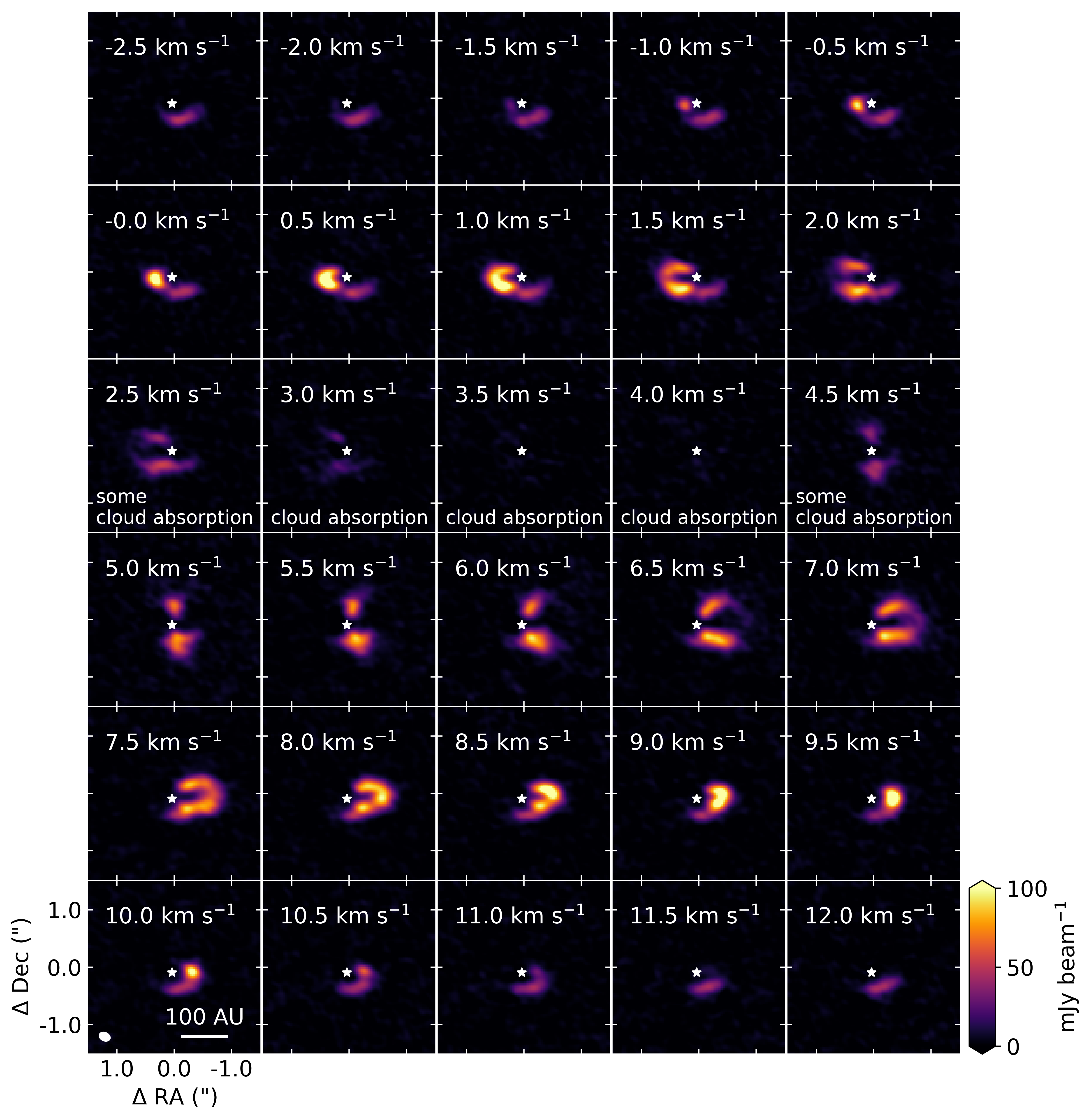}
      \caption{Channel maps of the \ce{^13CO} $J=3-2$ transition before continuum subtraction. The channels between 2.5 and 4.5~km~s$^{-1}$ are affected by (some) cloud absorption. The beam and a 100~AU scale bar are indicated in the bottom left panel. Additionally, the position of the star is indicated with the white star in each panel. }
         \label{fig:obs_13CO32_chans_w_cont}
\end{figure*}

\begin{figure*}
  \centering
    \includegraphics[width=1\linewidth]{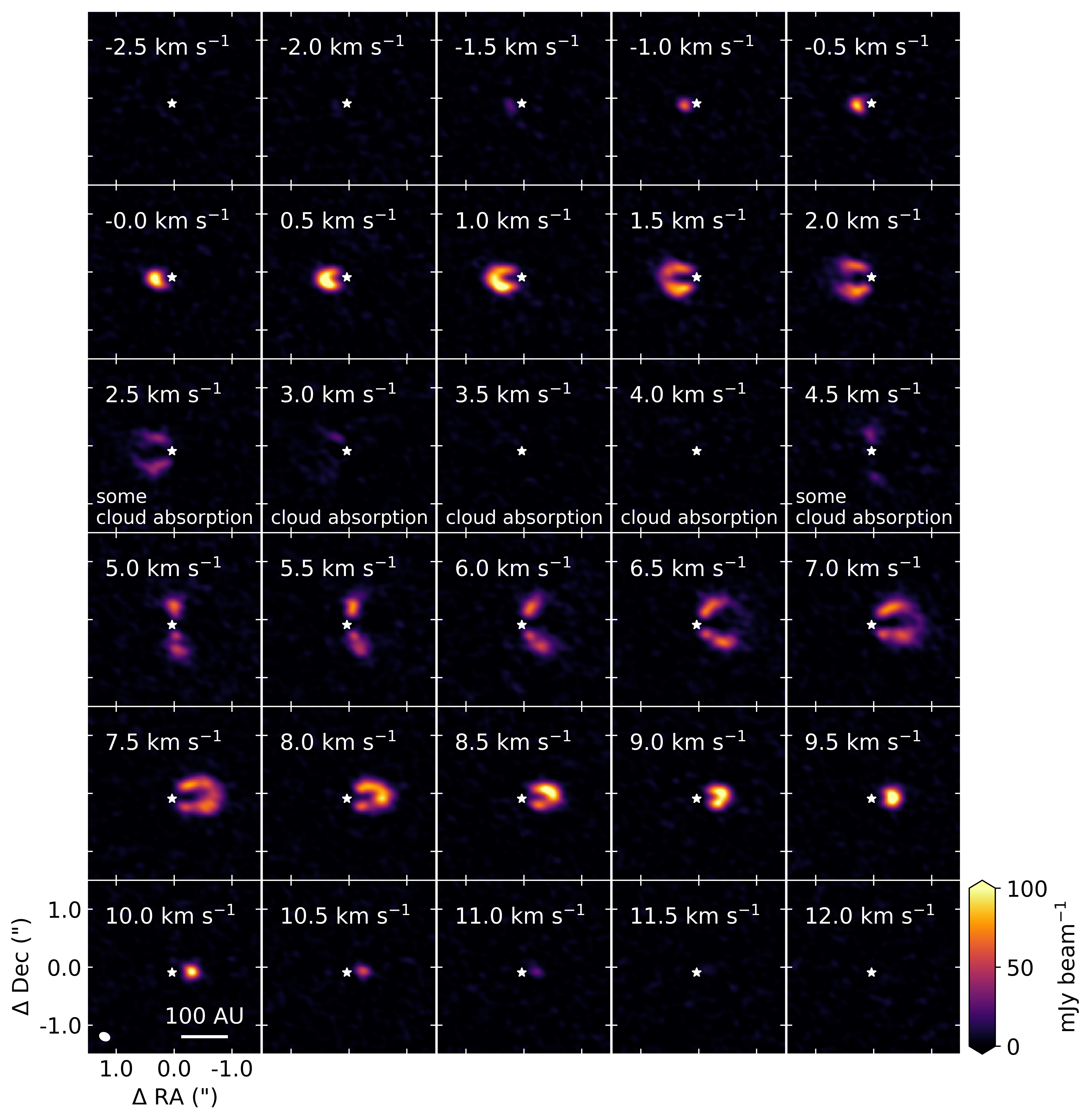}
      \caption{Channel maps of the \ce{^13CO} $J=3-2$ transition after continuum subtraction. The channels between 2.5 and 4.5~km~s$^{-1}$ are affected by (some) cloud absorption. The beam and a 100~AU scale bar are indicated in the bottom left panel. Additionally, the position of the star is indicated with the white star in each panel.}
         \label{fig:obs_13CO32_chans}
\end{figure*}

The NO column density together with the upper limits on the column densities of NO in the north, \ce{N2O}, \ce{NO2}, \ce{NH2OH}, CN, and \ce{C2H} for a range of temperatures is presented in Fig.~\ref{fig:T_vs_N}. The details of this calculation are discussed in Sect.~\ref{sec:N}. The NO column density in this Figure is a factor of 5 larger than that found in Paper~II due to an incorrect conversion to K in Paper II as explained in Sect.~\ref{sec:NO_obs} (see Fig.~\ref{fig:spec_NO_K} for the corrected spectrum). Finally, we present the channel maps of the \ce{^13CO} $J=3-2$ transition in Fig.~\ref{fig:obs_13CO32_chans_w_cont} (before continuum subtraction) and Fig.~\ref{fig:obs_13CO32_chans} (after continuum subtraction).

\begin{figure}
  \centering
    \includegraphics[width=1\linewidth]{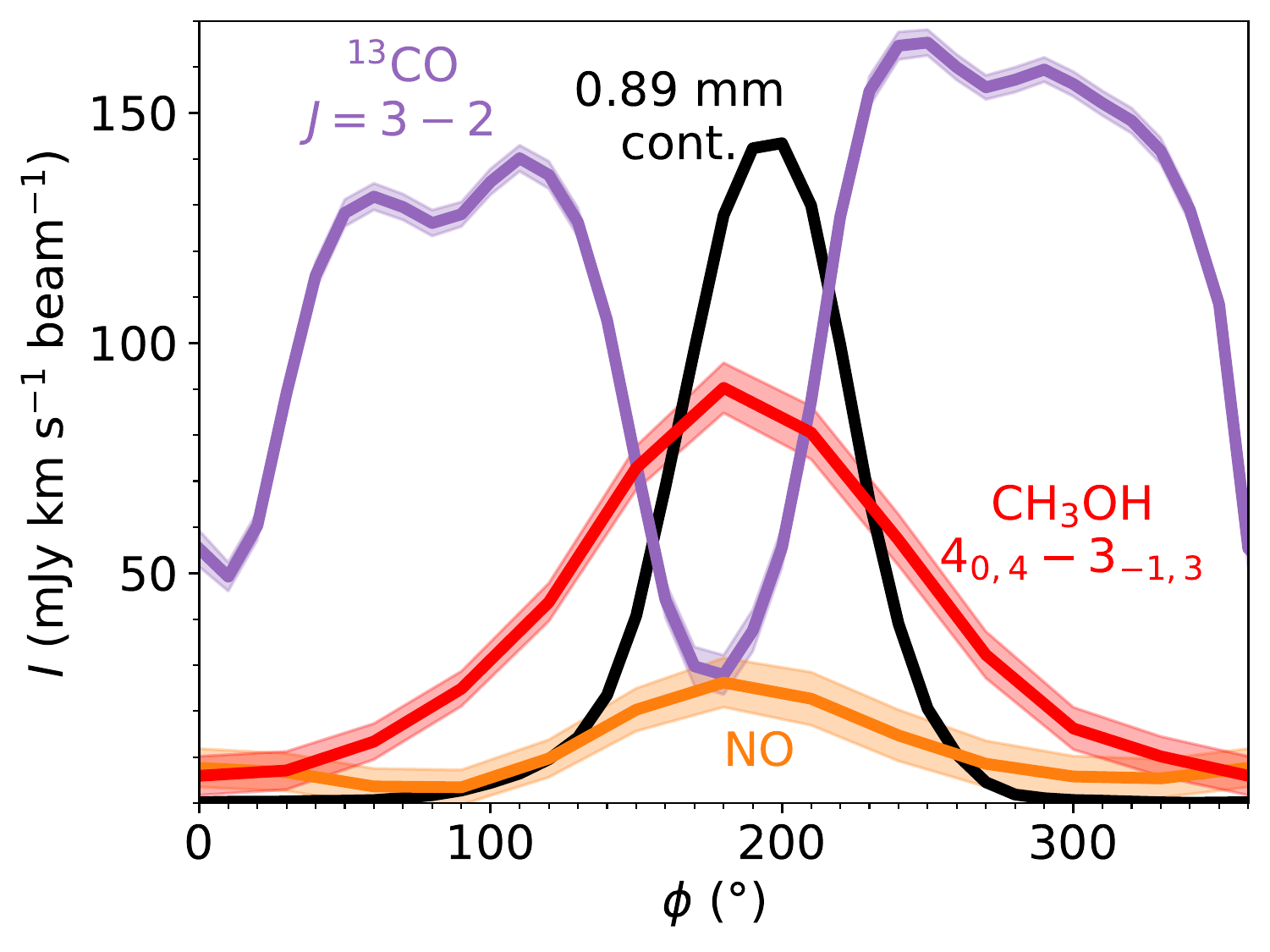}
      \caption{Azimuthal profile of the 0.89~mm continuum (black), NO (orange), \ce{^13CO} $J=3-2$ (purple), and \ce{CH3OH} $4_{0,4}-3_{-1,3}$ (red) emission. For each bin in $\phi$, the emission at that position angle is averaged inside an 60~AU wide ring at a radius of 62~AU, consistent with the radial location of the dust trap.}
         \label{fig:obs_azi_prof}
\end{figure}

The azimuthal profile of NO, \ce{CH3OH}, \ce{^13CO} and the 0.89~mm continuum emission are presented in Fig.~\ref{fig:obs_azi_prof}. These profiles are extracted by averaging the intensity from a 60~AU wide ring located at a radius of 62~AU (i.e. averaging between 32~AU and 92~AU) to compare the emission profiles at the radial location of the dust trap. Similar to the radial profiles presented in Fig.~\ref{fig:rad_azi} the NO and methanol peak at the location of the dust trap. The \ce{^13CO} emission drops at the continuum peak due to optical depth effects.

\section{DALI} \label{app:dali}
The setup of the DALI models used in this work is described in detail in the following subsection. Additionally, the model results for different the CN column density for different C/O ratios are presented in Appendix~\ref{sec:CO_ratio_CN}, the effect of the initial distribution of nitrogen are presented in Appendix~\ref{sec:init_nitrogen}, and the time evolution is presented in Appendix~\ref{sec:timeevolution}. 

\subsection{Model setup}

\subsubsection{Gas and dust density structure}

The IRS~48 DALI disk model is based on the previously published gas and dust surface density model, that was used to reproduce the spatially resolved CO isotopologue emission. The model for the south side reproduces the \ce{^13CO} $J=3-2$, \ce{^13CO} $J=6-5$, \ce{C^18O} $J=6-5$, and \ce{C^17O} $J=6-5$ emission and the total dust continuum flux at 0.44~mm and 0.89~mm within a factor of $<2$ \citep[][Paper~I]{Bruderer2014, vanderMarel2016}. The \ce{CO} $J=6-5$ intensity is reproduced within a factor of $\sim3$ by this model. The model for the north side dominates the 18.7~$\mu$m continuum flux and reproduces the observations within a factor of $<2$ \citep{vanderMarel2013}.

\begin{figure}
  \centering
    \includegraphics[width=1\linewidth]{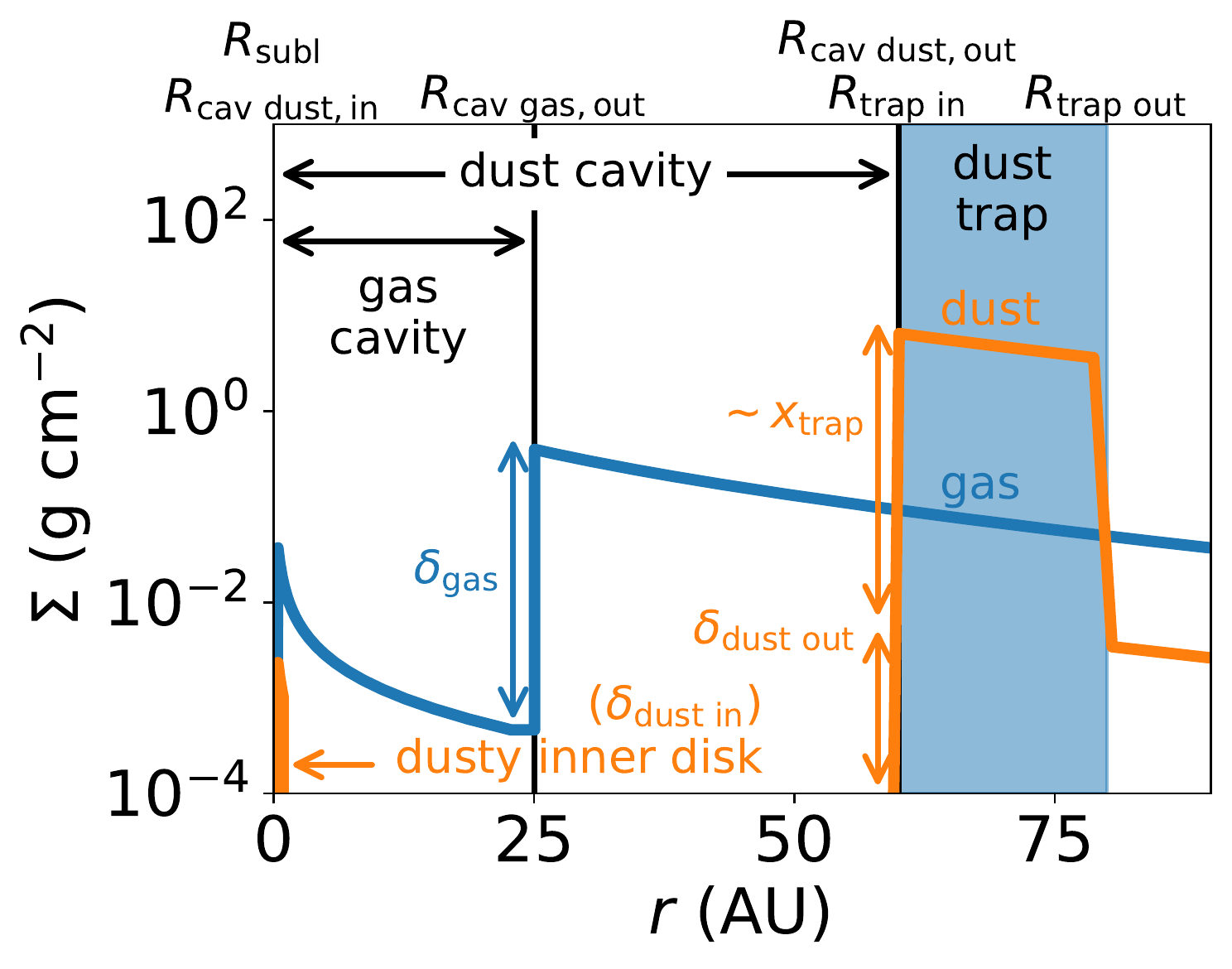}
      \caption{Gas (blue) and dust (orange) surface density in the DALI models of the dust trap (south) side of the IRS~48 disk. The dust trap is indicated with the shaded blue region. }
         \label{fig:dali_surf_dens}
\end{figure}

In the radial direction, the gas surface density profile is that of the self-similar solution of a viscously evolving disk \citet{LyndenBell1974, Hartmann1998}:
\begin{align}
\Sigma_{\mathrm{gas}} (R) &= \Sigma_c \left (\frac{R}{R_c}\right )^{-\gamma} e^{-\left (R/R_c\right )^{2-\gamma}},
\end{align}
with $\Sigma_{\mathrm{gas}} (R)$ the gas-surface density as a function of radius $R$, $\Sigma_c$ the surface density at the characteristic radius $R_c$, and $\gamma$ the power-law exponent. This relation is used outside the dust sublimation radius that is located at 0.4~AU. Inside the dust sublimation radius, the gas number density is set to the ISM value of $n_{\mathrm{gas}} = 7.1\times 10^2$~cm$^{-3}$. The deep gas cavity that is seen in the IRS~48 disk is modelled by reducing the gas surface density between the sublimation radius and the cavity radius $R_{\mathrm{cav\ out,gas}} = 25$~AU by a factor of $10^{-3}$, see Fig.~\ref{fig:dali_surf_dens}. In the vertical direction, the gas follows a Gaussian distribution with a scale height defined as:
\begin{align}
h &= h_c \left (\frac{R}{R_c} \right )^{\psi},
\end{align}
with $h_c$ the scale height at $R_c$ and $\psi$ the flaring index. 

The deep 60~AU dust cavity that is seen in the observations is modelled by reducing the total (small $+$ large) dust density by a factor of $10^{-20}$ between 1 and 60~AU. A small dusty inner disk is created by reducing the total dust density by a factor of $9\times 10^{-4}$ with respect to that of a full disk between 0.4 and 1~AU. 

The dust consists of two dust populations \citep{DAlessio2006} following an MRN distribution \citep{Mathis1977}. The small dust ($0.005-1~\mu$m) follows the number density of the gas scaled with the gas-to-dust mass ratio $\Delta_{\mathrm{gas/dust}} = 20$. On the other hand, the large dust ($0.005~\mu$m$-1~$mm) is settled to the midplane by reducing its scale height with a factor of $\chi$ compared to the gas and small dust. This results in the following mass densities outside the dust trap \citep[following][]{Bruderer2013}:
\begin{align}
\rho_{\mathrm{small}} &= \frac{(1-f_{\mathrm{ls}})\Sigma_{\mathrm{gas}}}{\Delta_{g/d}\sqrt{2\pi}rh}\exp{\left [-0.5 \left (\frac{0.5\pi-\theta}{h} \right )^2\right ]}~\text{and}\\
\rho_{\mathrm{large}} &= \frac{f_{\mathrm{ls}}\Sigma_{\mathrm{gas}}}{\Delta_{g/d}\sqrt{2\pi}rh\chi}\exp{\left [-0.5 \left (\frac{0.5\pi-\theta}{h\chi} \right )^2\right ]},
\end{align}
where $f_{\mathrm{ls}} = 0.85$ sets the mass fraction of the large grains and $\theta$ indicates the latitude coordinate (pole: $\theta =0$; midplane: $\theta= \pi/2$). 

The dust trap itself is modelled by modifying the dust density between 60 and 80~AU, as observations indicate a high surface density of large dust grains \citep{vanderMarel2013, Ohashi2020}. As DALI is 2D and thus unable to produce azimuthal asymmetries, the IRS48 dust trap is simulated by producing models with and without a high-density dust ring. The density of large grains is increased by a factor of $x_{\mathrm{trap}}$, whereas the small grains are decreased by that same factor following Paper~I:
\begin{align}
\rho_{\mathrm{small}} &= \frac{\Sigma_{\mathrm{gas}}}{x_{\mathrm{trap}}\Delta_{g/d}\sqrt{2\pi}rh}\exp{\left [-0.5 \left (\frac{0.5\pi-\theta}{h} \right )^2\right ]}~\text{and}\\
\rho_{\mathrm{large}} &= \frac{x_{\mathrm{trap}}\Sigma_{\mathrm{gas}}}{\Delta_{g/d}\sqrt{2\pi}rh\chi}\exp{\left [-0.5 \left (\frac{0.5\pi-\theta}{h\chi} \right )^2\right ]}.
\end{align}
The fraction of large grains, $f_{\mathrm{ls}}$, that is used outside the dust trap is not included here as the small grains likely grew to larger sizes which is taken into account by $x_{\mathrm{trap}}$. A value of 1000 results in a dust surface density of 6~g~cm$^{-2}$ at 60~AU, consistent with the dust surface density of 2-8~g~cm$^{-2}$ derived from polarization observations \citep{Ohashi2020}. In summary, two models are used: one for the non-dust trap (north) side of the disk and one for the dust trap (south) side, see also the right and left side of Fig.~\ref{fig:cartoon} respectively.

\subsubsection{Stellar spectrum}
The stellar spectrum is modelled as a 14.3~L$_{\odot}$ star with an effective temperature of $9\times 10^3$~K following Paper~I. The stellar accretion rate of $4\times 10^{-9}$~M$_{\odot}$~yr$^{-1}$ \citep{Salyk2013} is modelled by adding a $10^4$~K black body. We have updated the X-ray luminosity of IRS~48 as no X-rays have been detected with Chandra, resulting in an upper limit on the X-ray luminosity (not corrected for extinction) of $<10^{28}$~erg~s$^{-1}$ \citep{Imanishi2001}.

\begin{figure*}
   \centering
  \begin{subfigure}{0.8\columnwidth}
  \centering
  \includegraphics[width=1\linewidth]{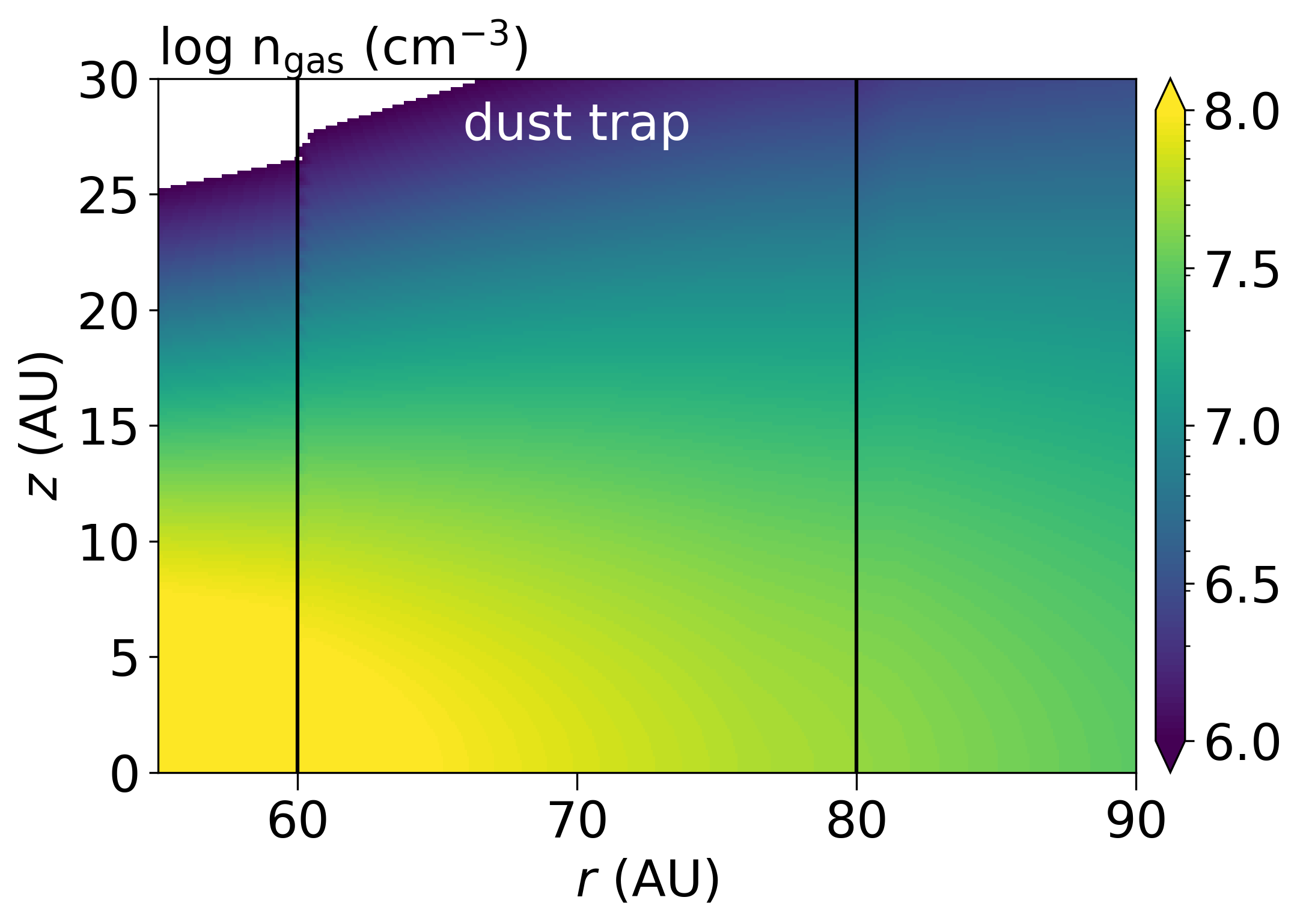}
\end{subfigure}%
\begin{subfigure}{0.8\columnwidth}
  \centering
    \includegraphics[width=1\linewidth]{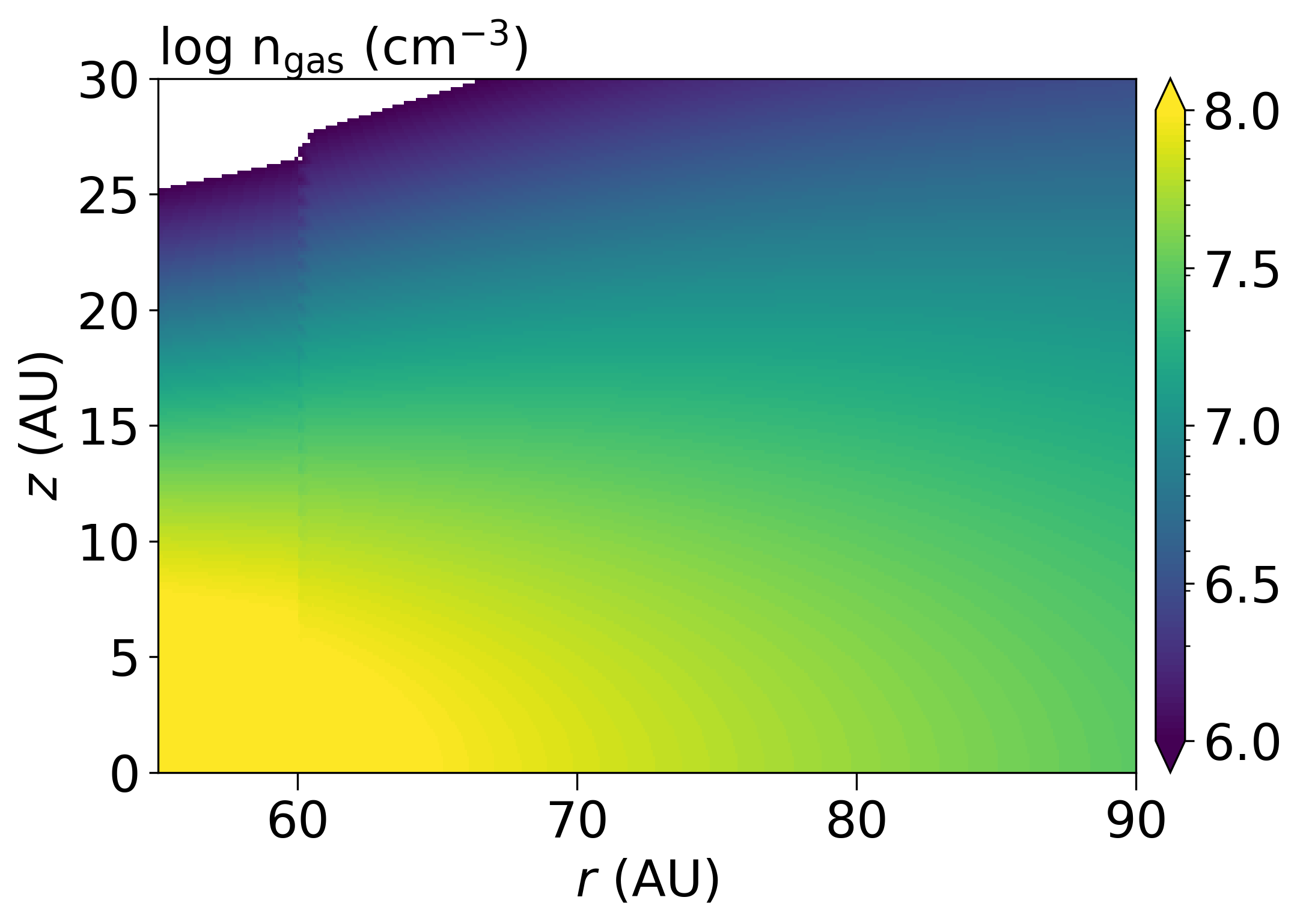}
\end{subfigure}

  \begin{subfigure}{0.8\columnwidth}
  \centering
  \includegraphics[width=1\linewidth]{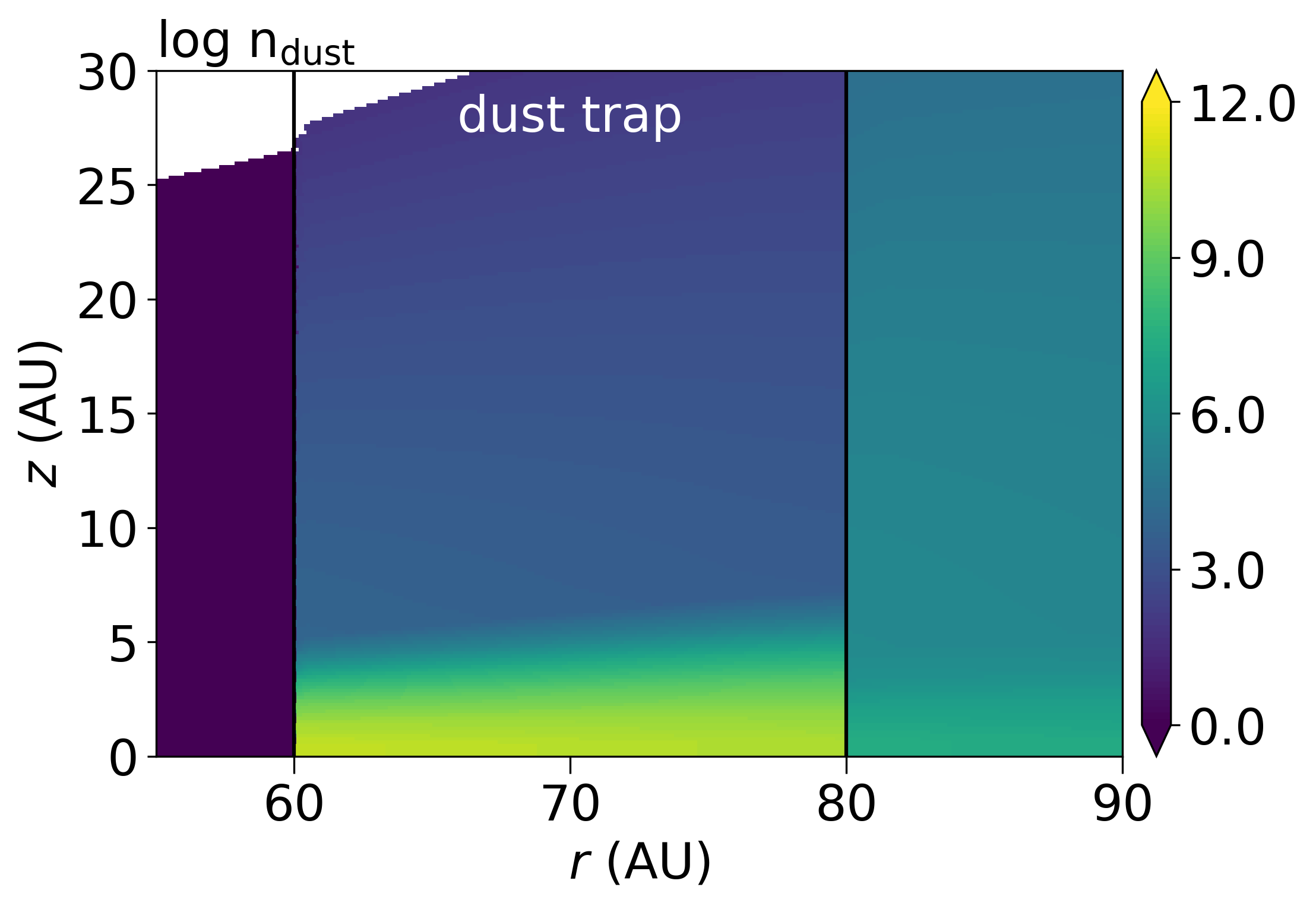}
\end{subfigure}%
\begin{subfigure}{0.8\columnwidth}
  \centering
    \includegraphics[width=1\linewidth]{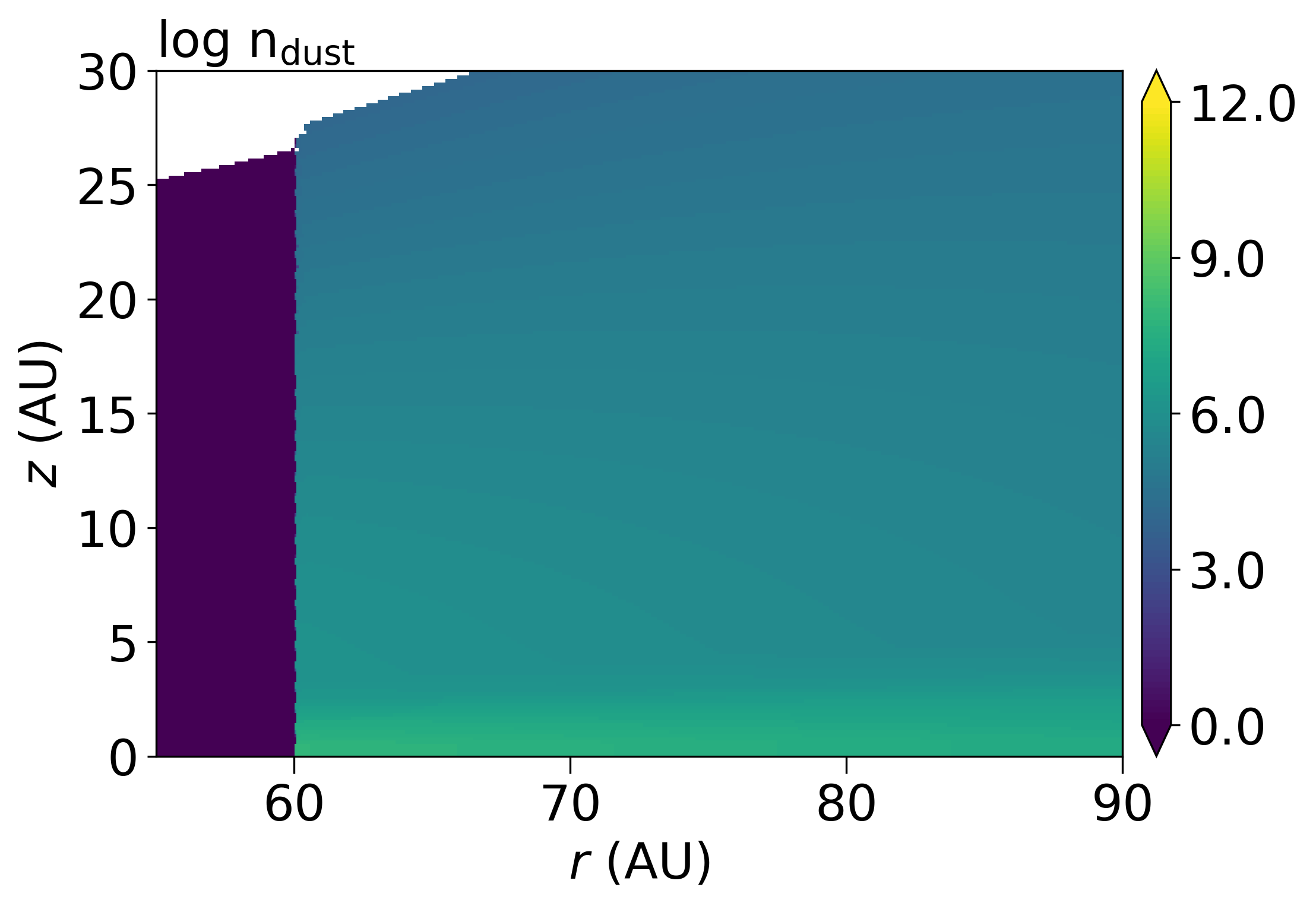}
\end{subfigure}

  \begin{subfigure}{0.8\columnwidth}
  \centering
  \includegraphics[width=1\linewidth]{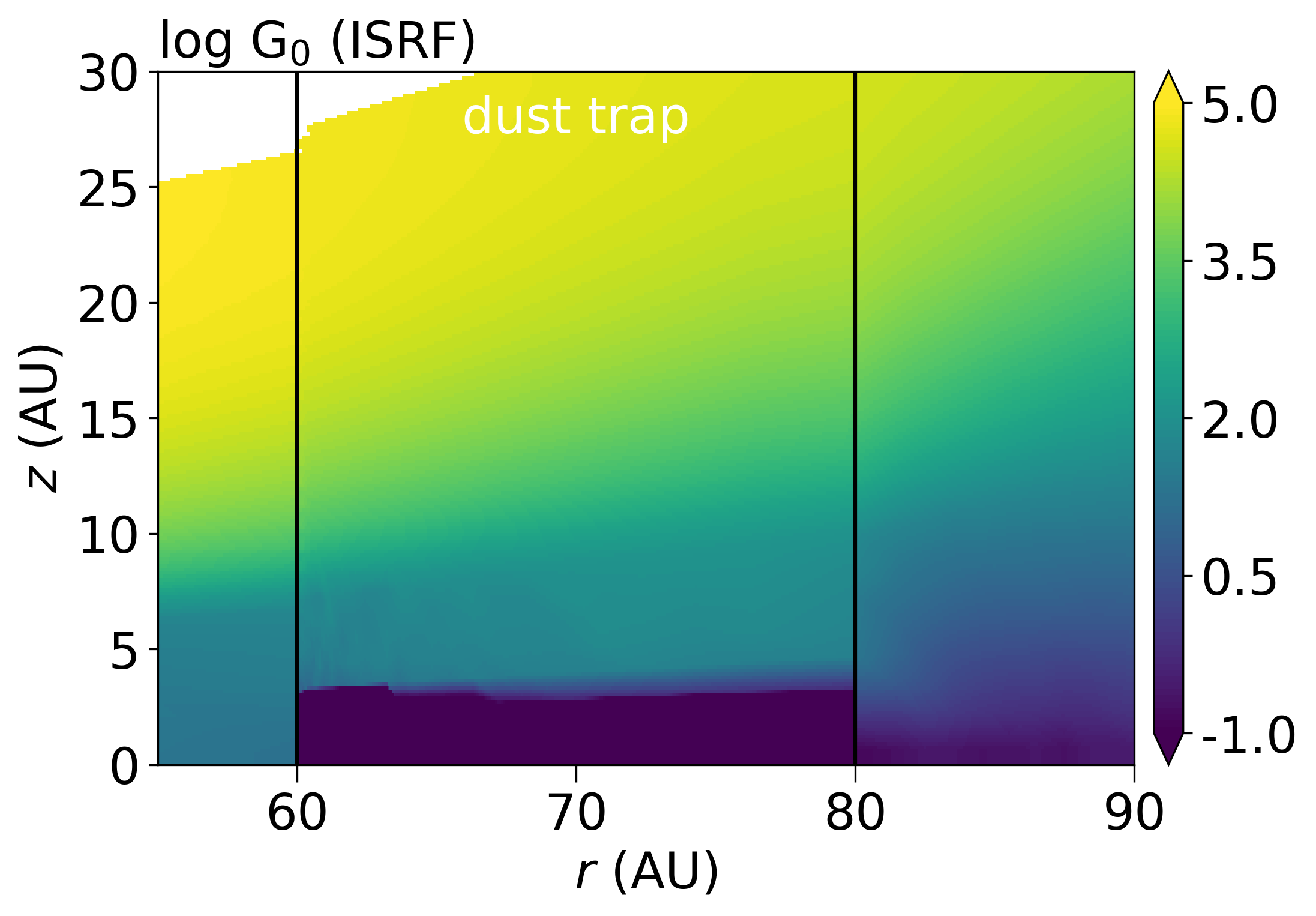}
\end{subfigure}%
\begin{subfigure}{0.8\columnwidth}
  \centering
    \includegraphics[width=1\linewidth]{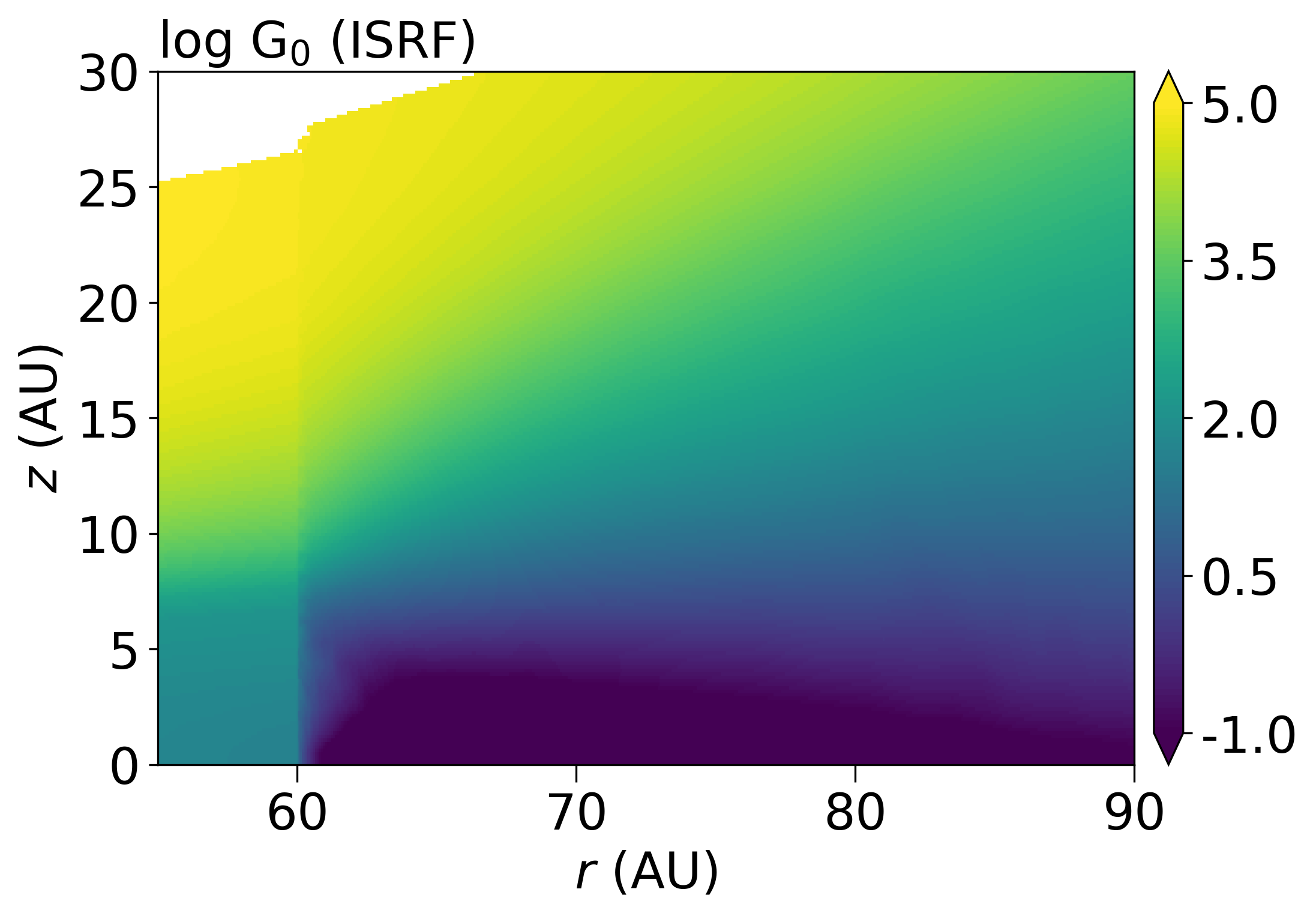}
\end{subfigure}
  \begin{subfigure}{0.8\columnwidth}
  \centering
  \includegraphics[width=1\linewidth]{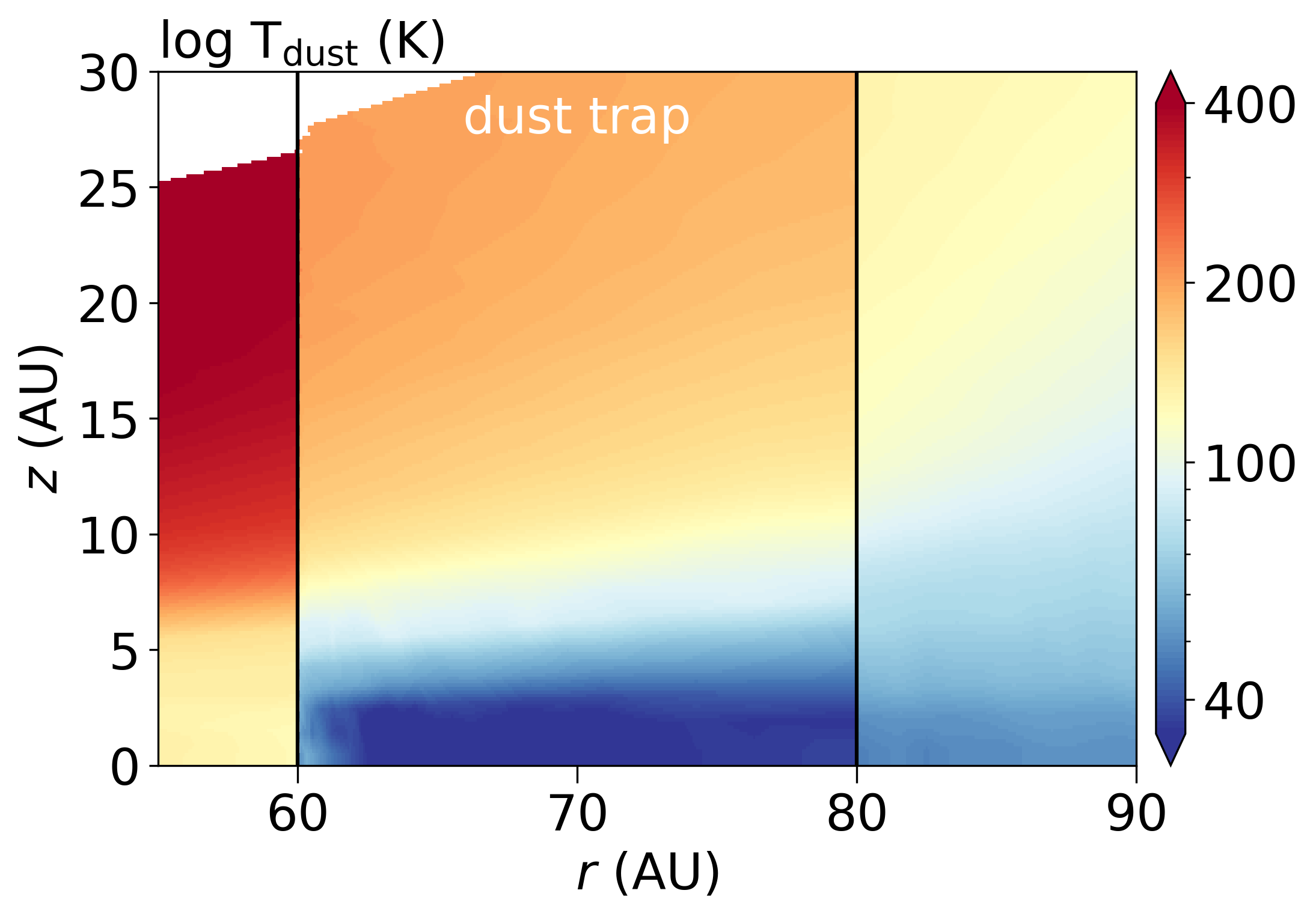}
\end{subfigure}%
\begin{subfigure}{0.8\columnwidth}
  \centering
    \includegraphics[width=1\linewidth]{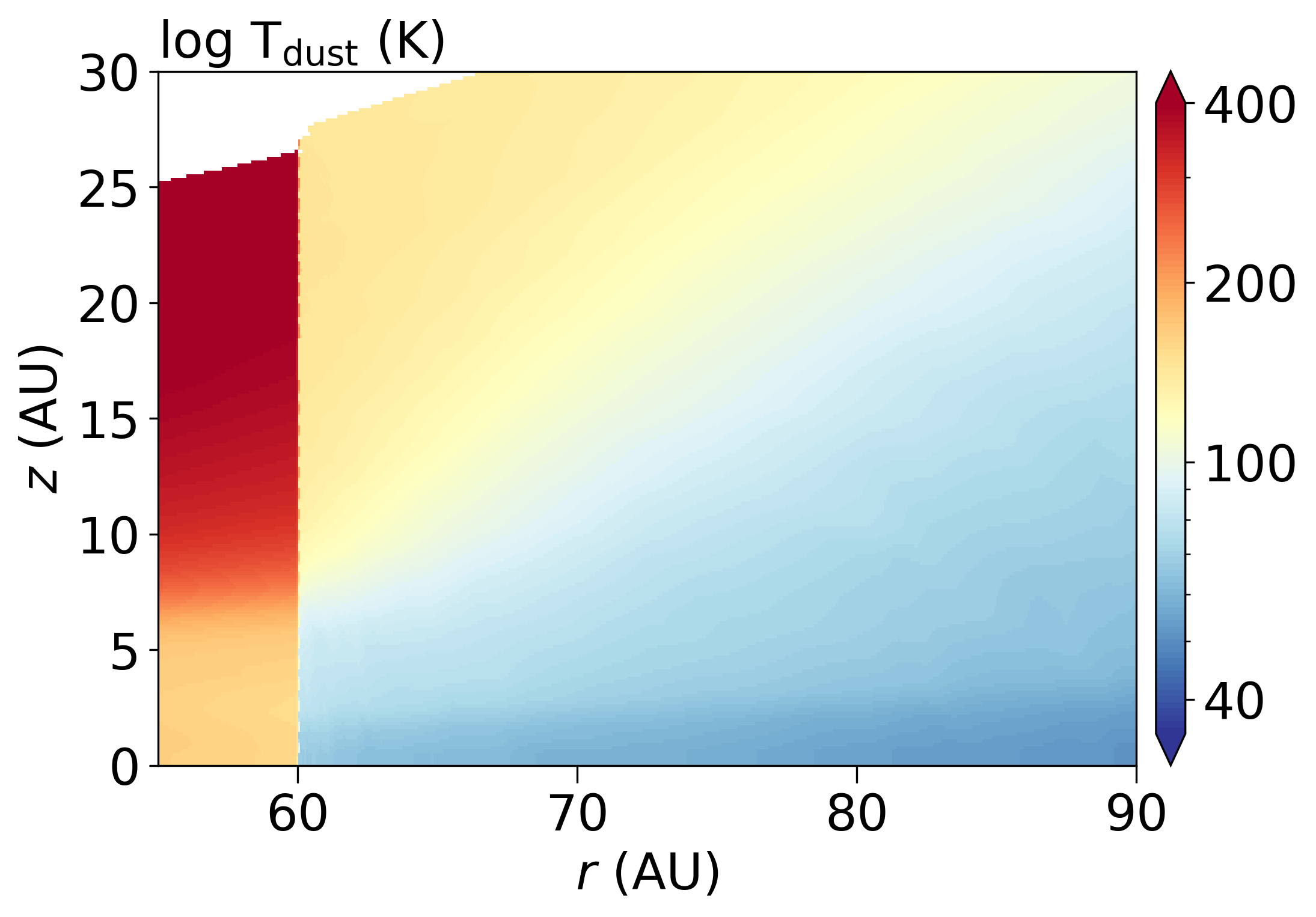}
\end{subfigure}
      \caption{Density, UV field and dust temperature structure of the DALI model for the dust trap side (south, left) and non-dust trap side (north, right). The top row presents the gas density structure (identical in both models), the second row the dust density structure, the third row the resulting UV field, the fourth row the dust temperature.  }
         \label{fig:dali2D_general}
\end{figure*}

\begin{figure*}
   \centering
  \begin{subfigure}{0.78\columnwidth}
  \centering
  \includegraphics[width=1\linewidth]{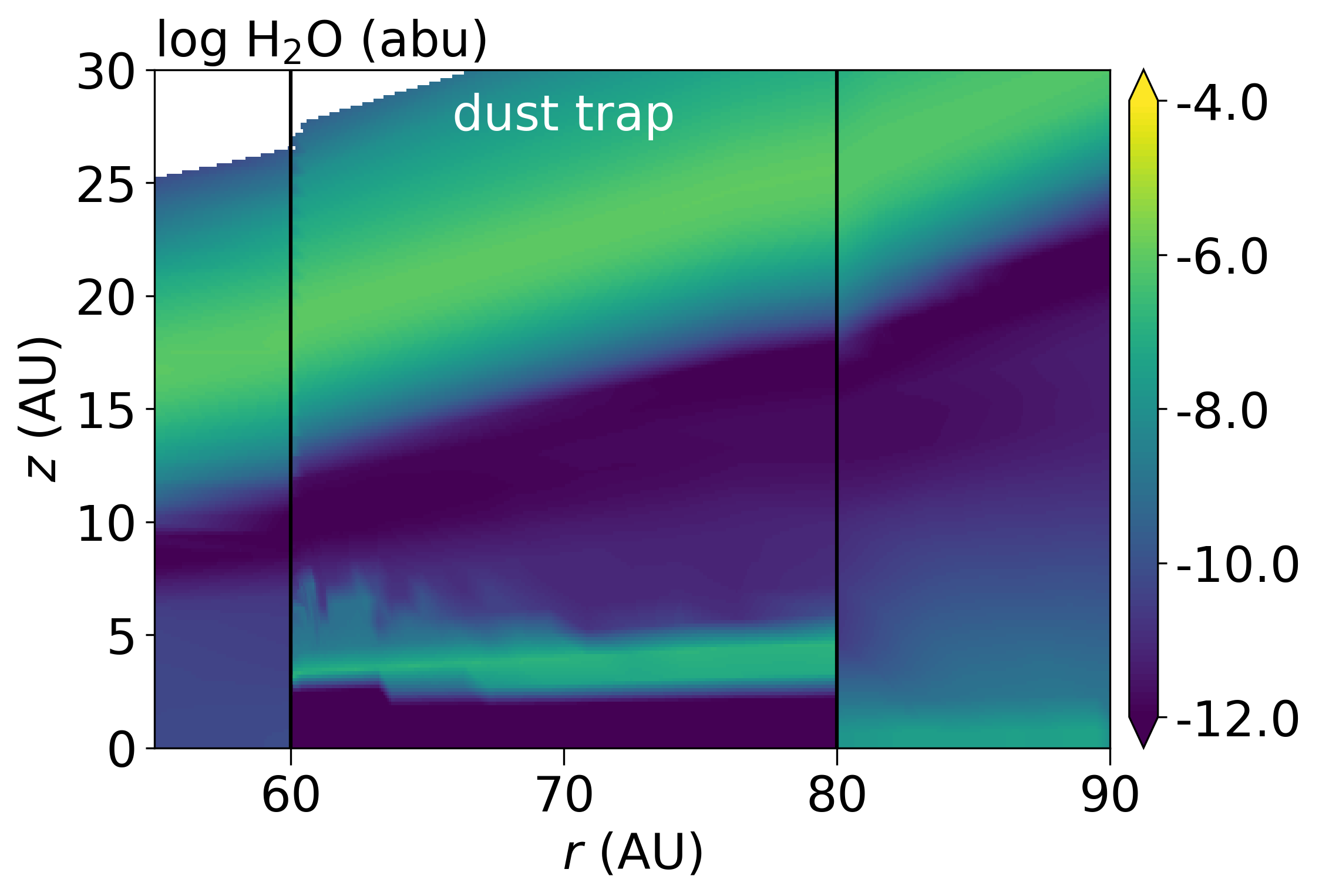}
\end{subfigure}%
\begin{subfigure}{0.78\columnwidth}
  \centering
    \includegraphics[width=1\linewidth]{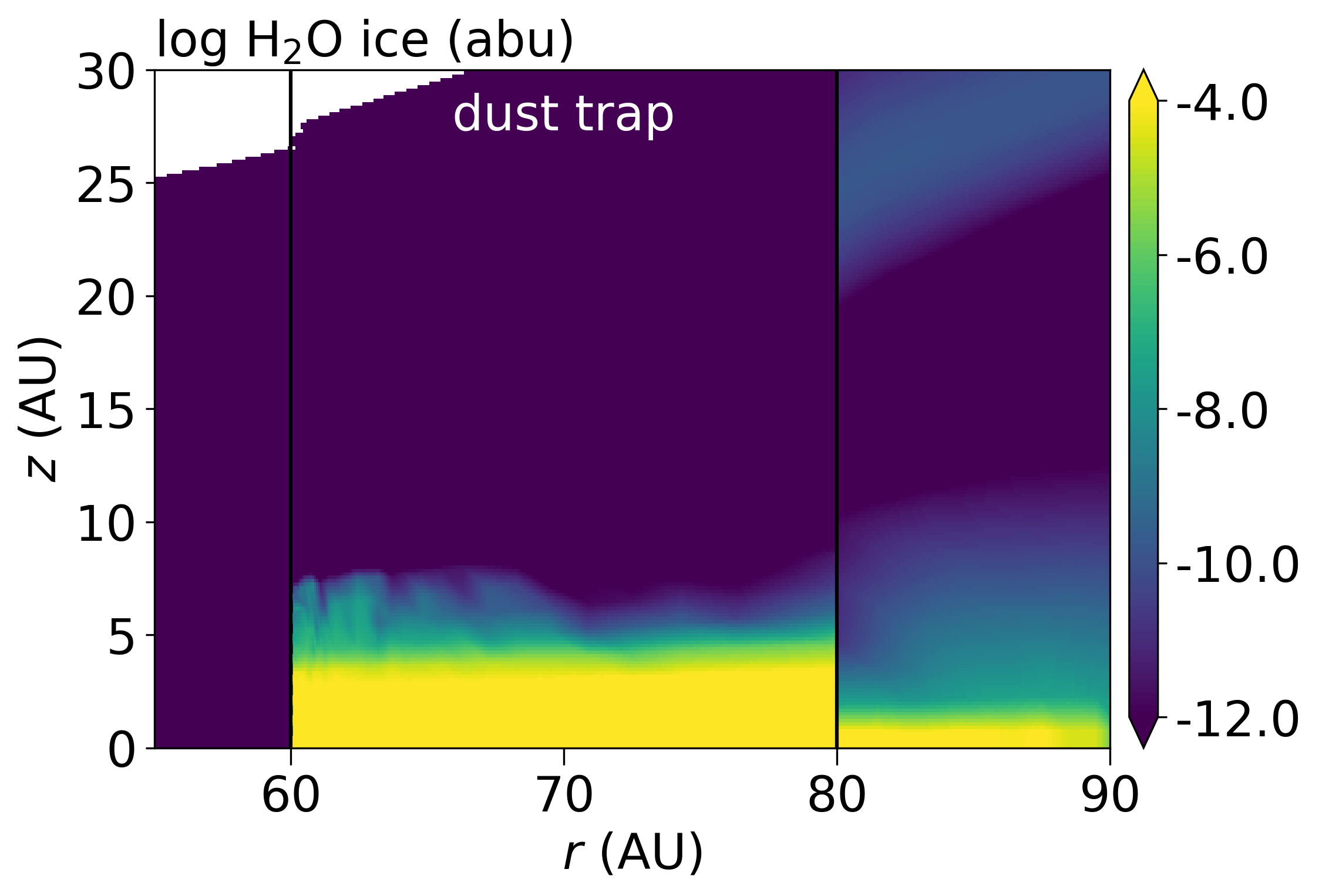}
\end{subfigure}

  \begin{subfigure}{0.78\columnwidth}
  \centering
  \includegraphics[width=1\linewidth]{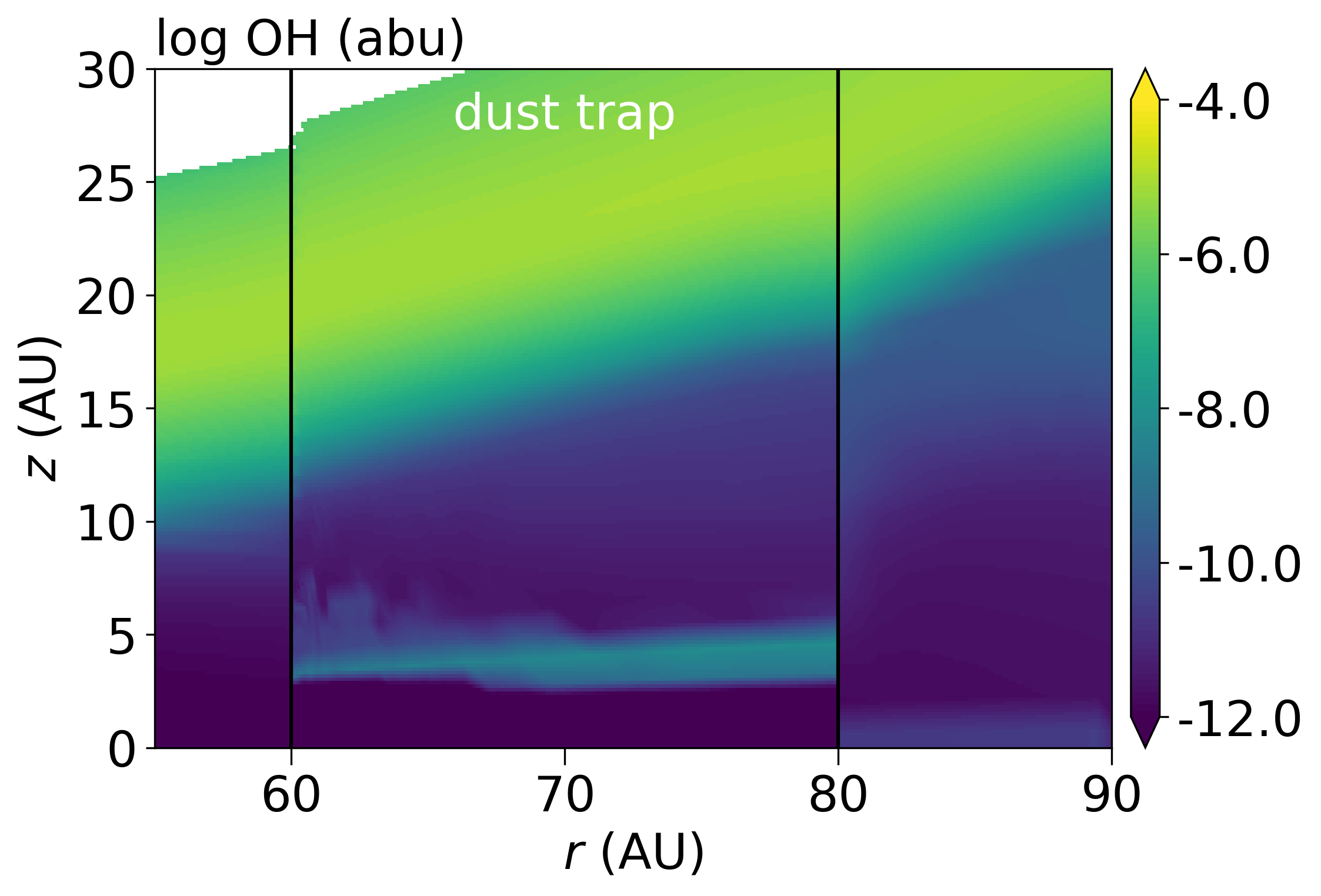}
\end{subfigure}%
\begin{subfigure}{0.78\columnwidth}
  \centering
    \includegraphics[width=1\linewidth]{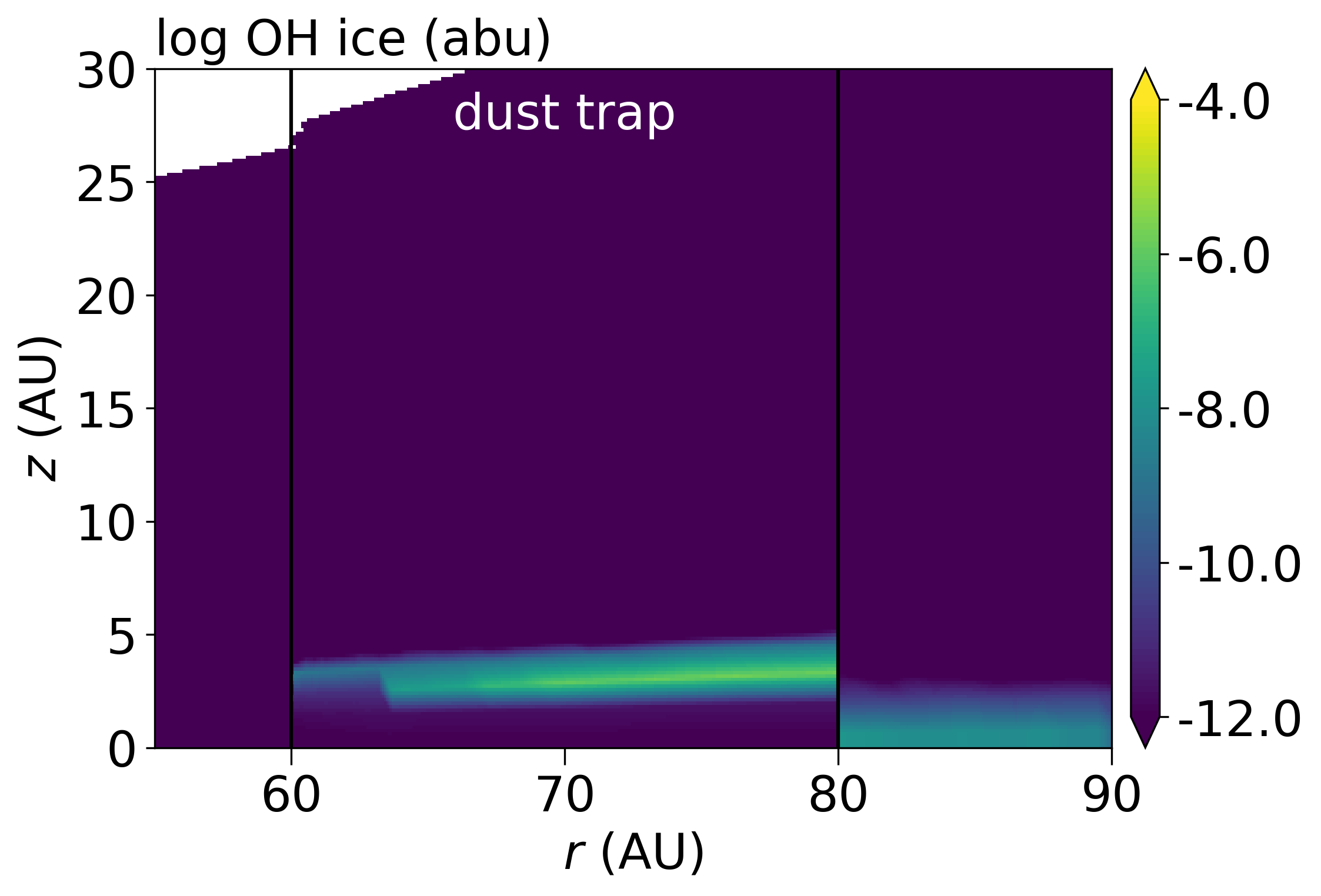}
\end{subfigure}
\\
  \begin{subfigure}{0.78\columnwidth}
  \centering
  \includegraphics[width=1\linewidth]{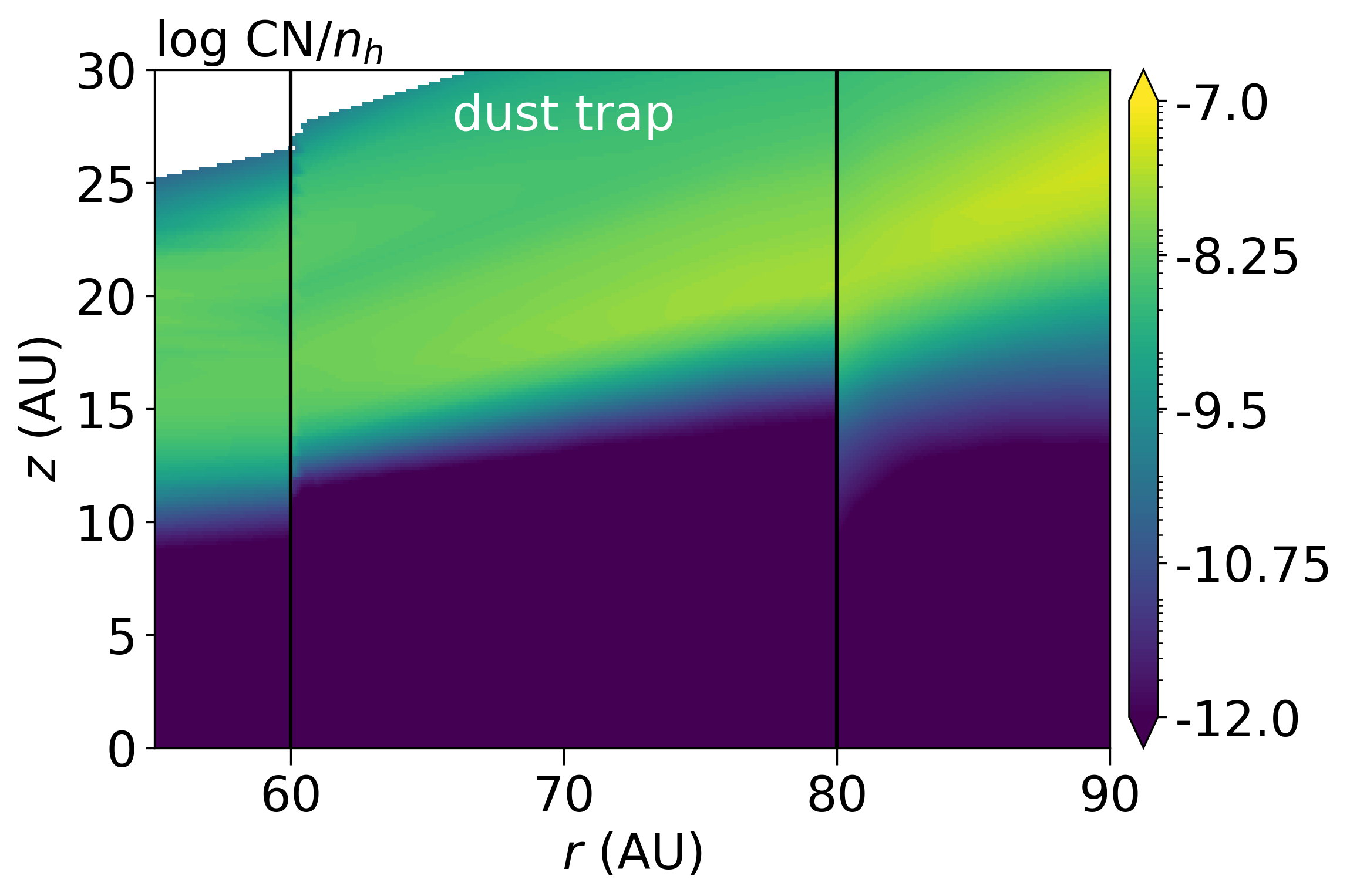}
\end{subfigure}%
\begin{subfigure}{0.78\columnwidth}
  \centering
    \includegraphics[width=1\linewidth]{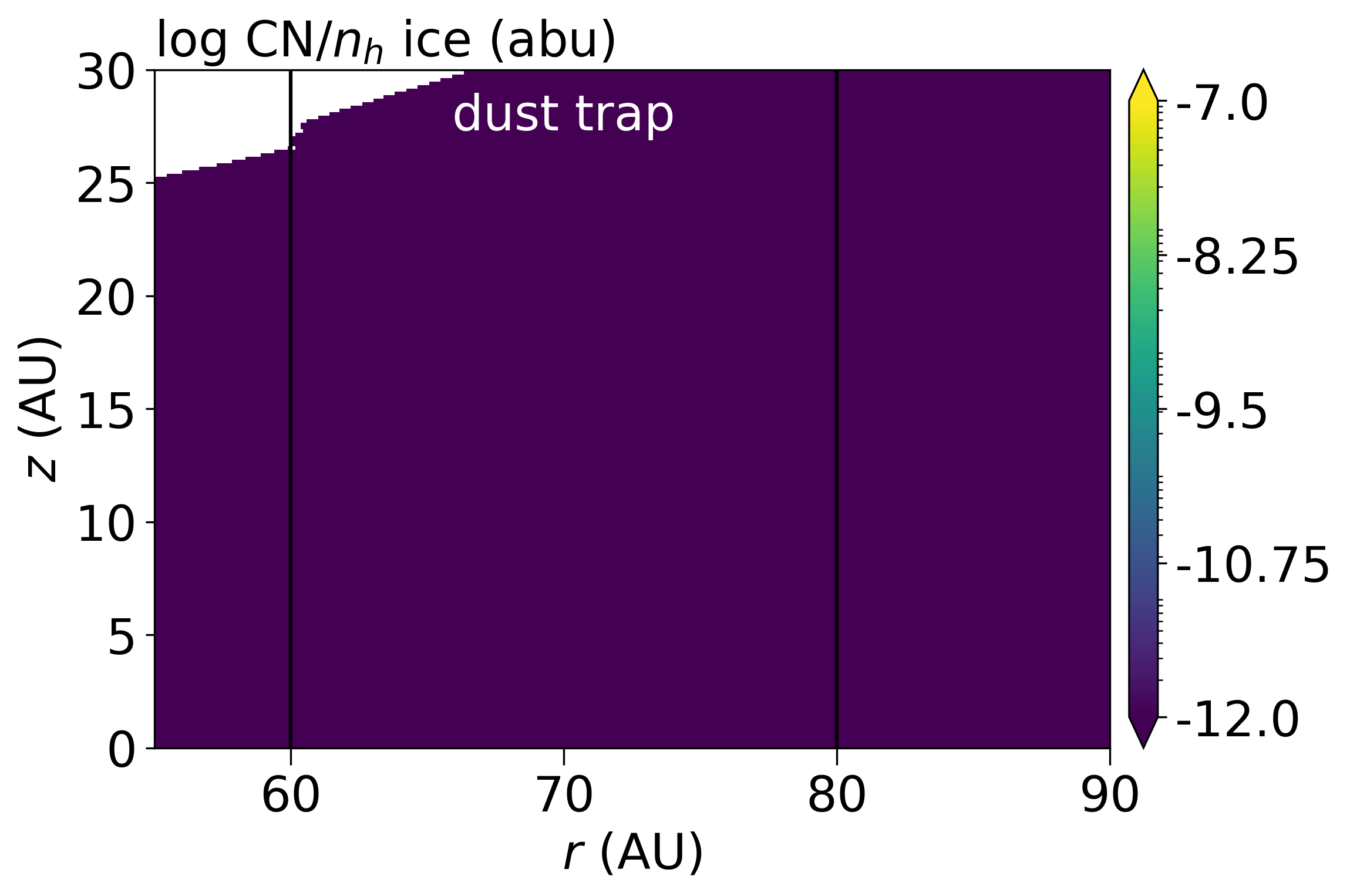}
\end{subfigure}

  \begin{subfigure}{0.78\columnwidth}
  \centering
  \includegraphics[width=1\linewidth]{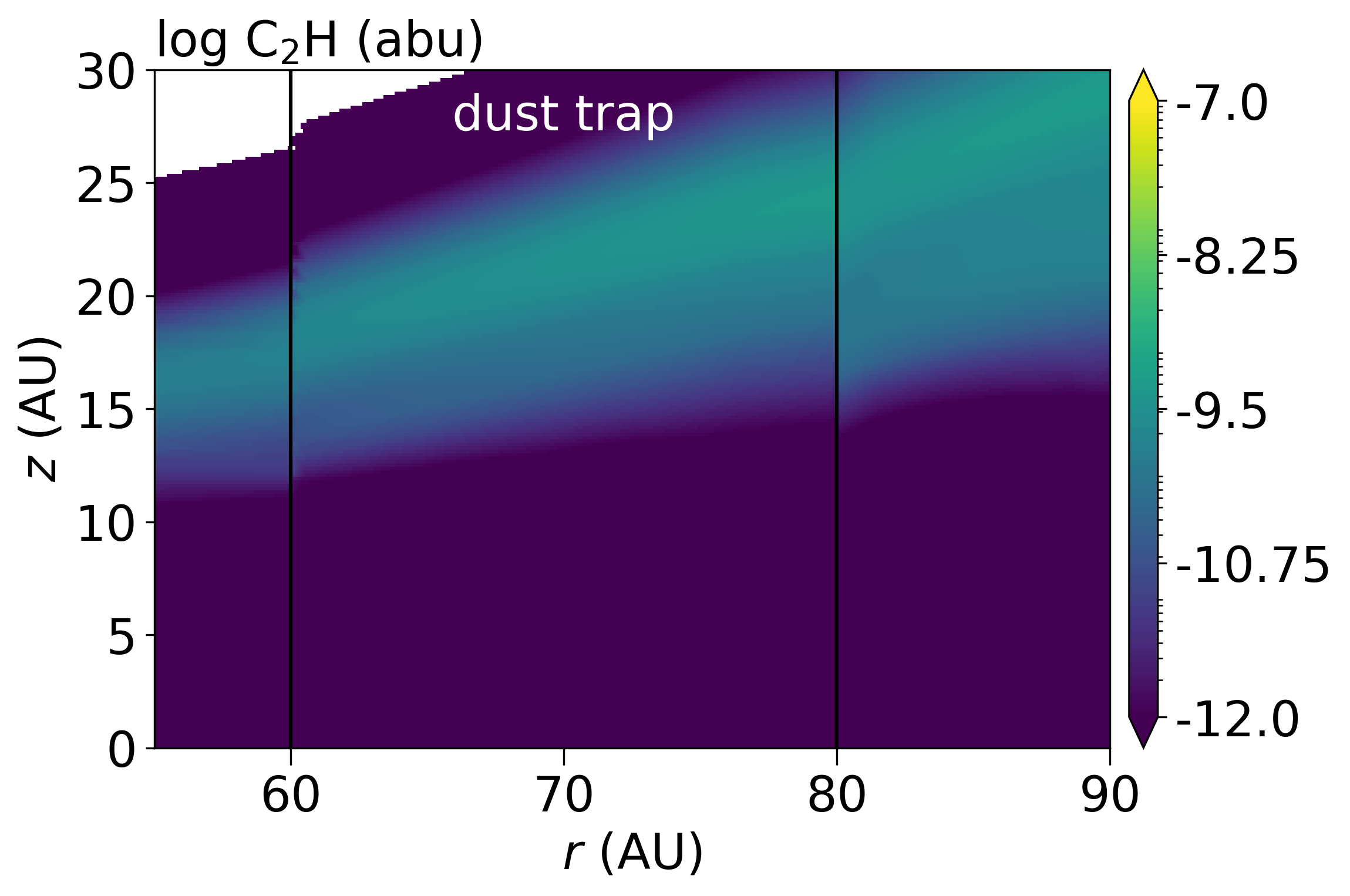}
\end{subfigure}%
\begin{subfigure}{0.78\columnwidth}
  \centering
    \includegraphics[width=1\linewidth]{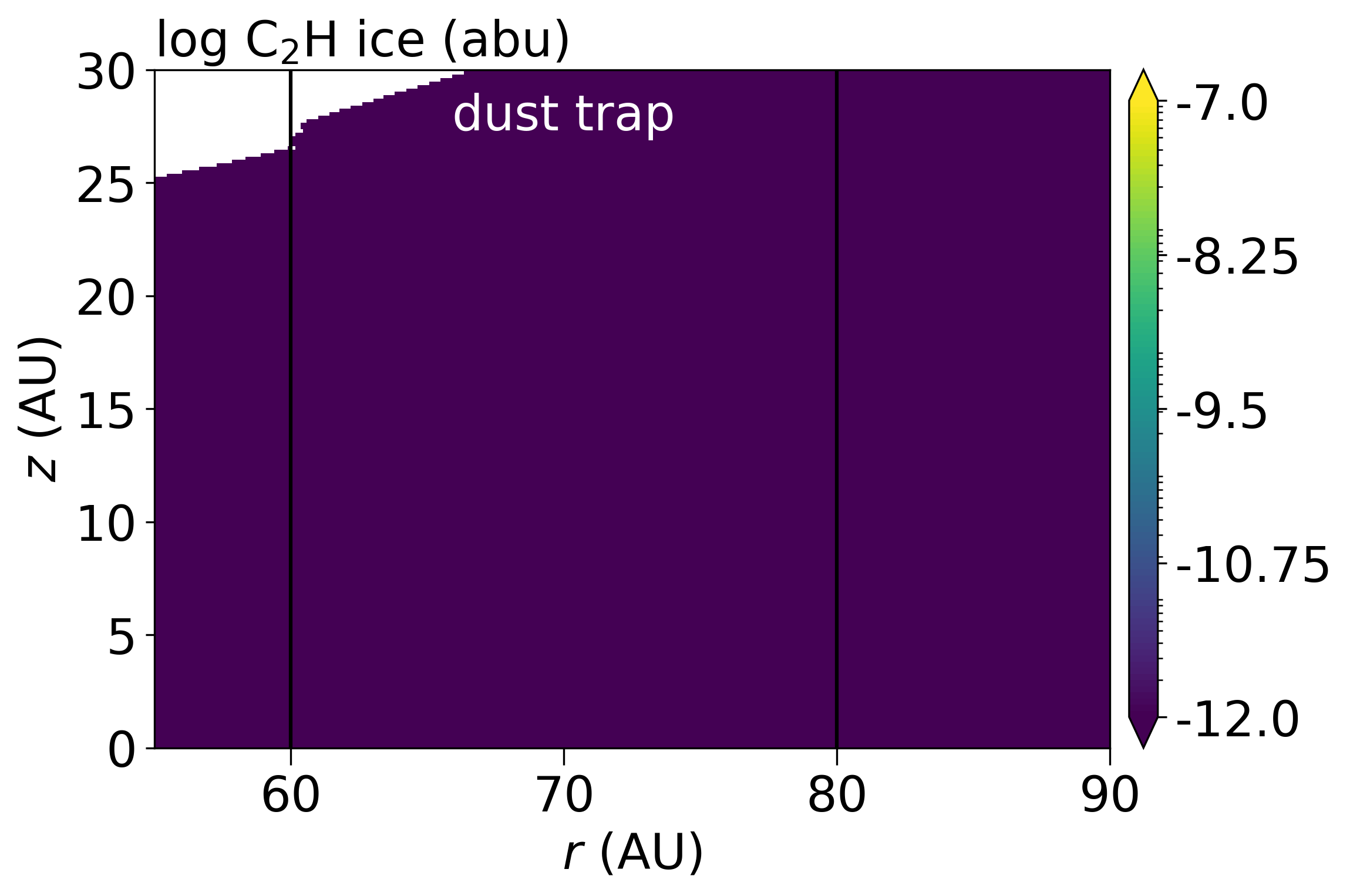}
\end{subfigure}

  \begin{subfigure}{0.78\columnwidth}
  \centering
  \includegraphics[width=1\linewidth]{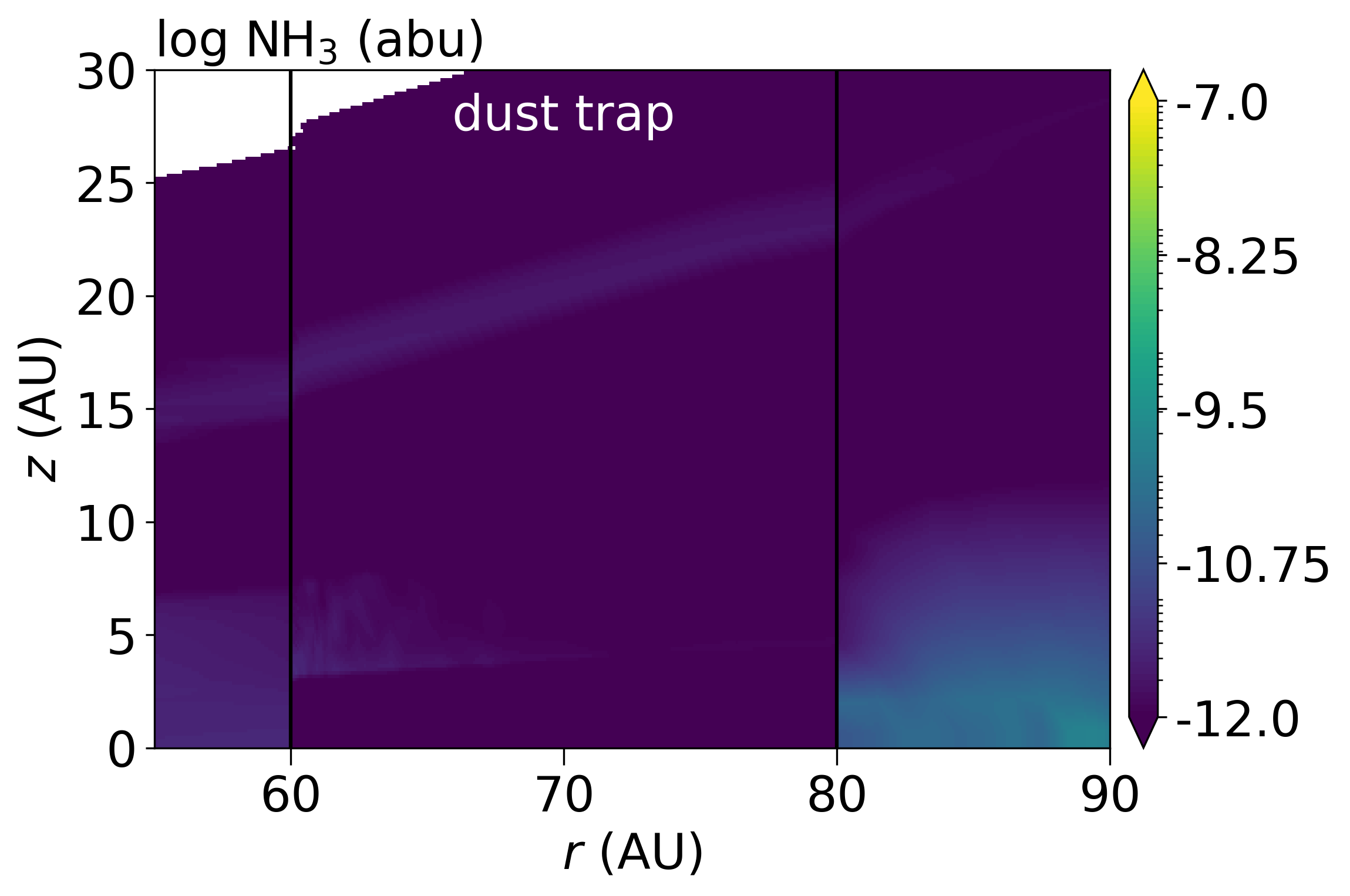}
\end{subfigure}
\begin{subfigure}{0.78\columnwidth}
  \centering
    \includegraphics[width=1\linewidth]{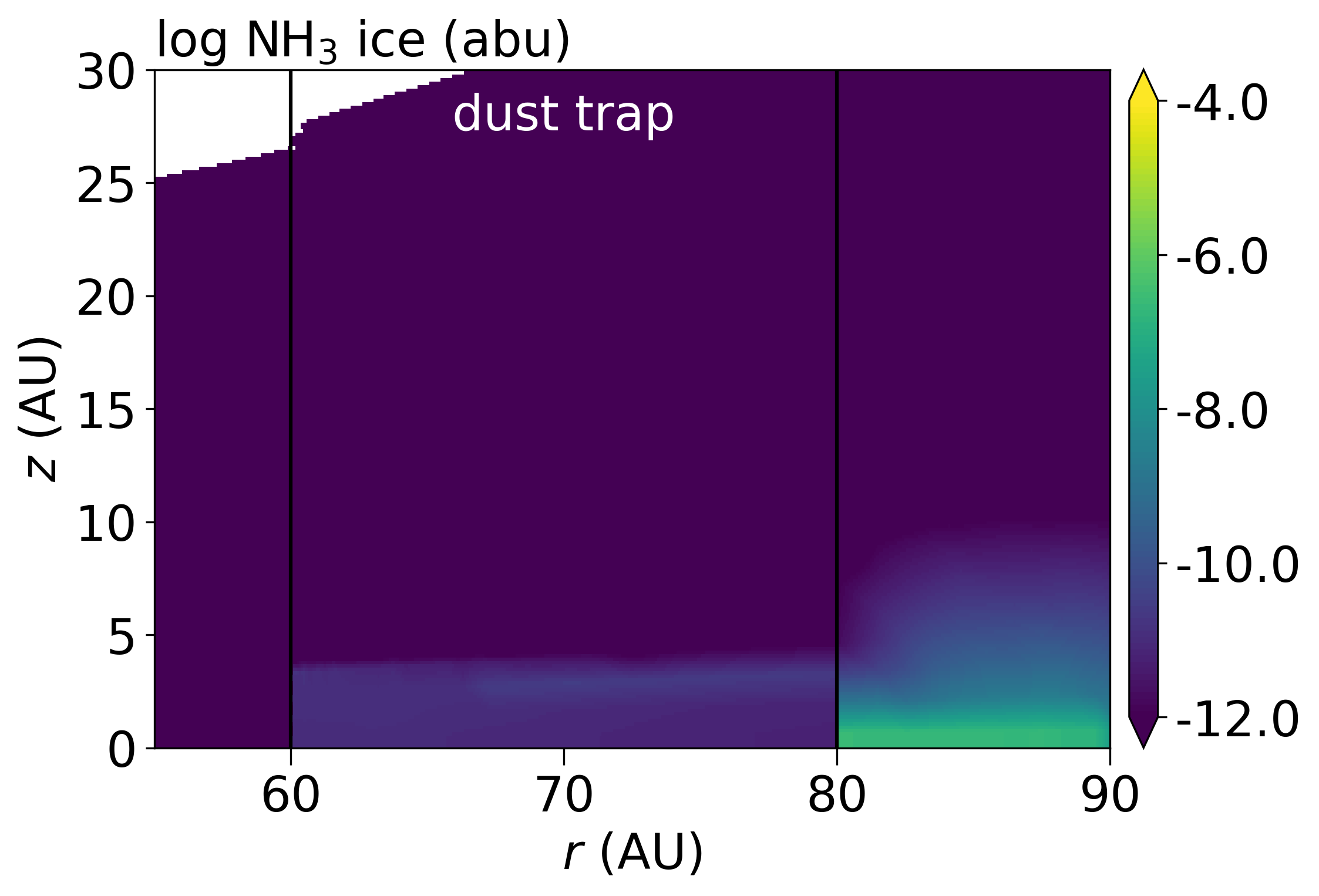}
\end{subfigure}
      \caption{Abundance of gas (left) and ice (right) phase species in the dust trap side model of the IRS~48 disk. The top row presents the \ce{H2O} abundance, the second row that of OH, the third row that of CN, the fourth row shows \ce{C2H} and the final row shows \ce{NH3}, the largest carbon chain in our network that acts as a sink for larger hydrocarbons.  }
         \label{fig:dali2D_abu_dusttrap}
\end{figure*}

\subsubsection{Chemical network} \label{app:dali_chem}
The gas and dust temperature are calculated using the default chemical network in DALI that is based on the UMIST06 chemical network \citep{Woodall2007, Bruderer2013} and starts with atomic initial conditions. This network is relatively small and does not include the necessary reactions to model nitrogen bearing molecules such as NO and CN, and small hydrocarbons such as \ce{C2H}. Therefore, an updated version of the nitrogen chemistry network is used to calculate the molecular abundances using the temperature and density structure obtained in the previous step. This network was first presented in \citet{Visser2018, Cazzoletti2018, Long2021}. For this updated version, we have added the $\ce{HNC + H \to HCN + H}$ and $\ce{HNC + O \to NH + CO}$ reactions with the energy barriers derived in \citet{Hacar2020}. Furthermore, we have set the binding energy of \ce{C2H3} ice to $10^4$~K as this is the largest carbon chain in the network that acts as a sink to model the effect of the formation of larger hydrocarbons that remain frozen out on the grains. Finally, this network is started with molecular initial conditions (see Table~\ref{tab:dali}). For computational reasons, the first network is run in steady state and the second network is run to 100~yr, 1~kyr (inside the dust trap), or 1~Myrs (outside the dust trap), appropriate for the dust trap around IRS~48. The effect of the initial distribution of nitrogen is further discussed in Appendix~\ref{sec:init_nitrogen}.

\begin{table*}
\caption{DALI model parameters. }             
\label{tab:dali}      
\centering          
\begin{tabular}{p{0.3\columnwidth}>{\centering\arraybackslash}p{0.3\columnwidth}>{\centering\arraybackslash}p{0.3\columnwidth}p{1.\columnwidth}}    
    \hline\hline
        \centering\arraybackslash Model parameter & \centering\arraybackslash dust trap side (south) & \centering\arraybackslash non-dust trap side (north) & \centering\arraybackslash Description \\ \hline
        \\[-0.7em]
        \textit{Physical structure} & & & \\
        $R_{\mathrm{subl}}$ & \multicolumn{2}{c}{0.4~AU} & Sublimation radius.\\        
        $R_{\mathrm{cav\ in,\ dust}}$ &  \multicolumn{2}{c}{1~AU} & Inner radius of the dust cavity.  \\        
        $R_{\mathrm{cav\ out,\ gas}}$ & \multicolumn{2}{c}{25~AU}& Outer radius of the gas cavity.  \\        
        $R_{\mathrm{cav\ out,\ dust}}$ & \multicolumn{2}{c}{60~AU} & Outer radius of the dust cavity. \\ 
        $R_{\mathrm{trap\ in}}$  & 60~AU & - & Inner radius of the dust trap. \\  
        $R_{\mathrm{trap\ out}}$ & 80~AU & - & Outer radius of the dust trap. \\                                                    
        $R_{\mathrm{c}}$ & \multicolumn{2}{c}{60~AU}& Characteristic radius of the surface density profile.  \\        
        $R_{\mathrm{out}}$ & \multicolumn{2}{c}{120~AU}& Outer radius of the disk.  \\
        $\Sigma_{\mathrm{c}}$ & \multicolumn{2}{c}{0.25 g$\ \mathrm{cm^{-2}}$}& Sets the gas surface density at the characteristic radius $R_{\mathrm{c}}$. \\    
        $M_{\mathrm{disk}}$ & \multicolumn{2}{c}{$3.3\times 10^{-4}~\mathrm{M_{\odot}}$}& Mass of the disk. \\   
        $\gamma$ & \multicolumn{2}{c}{1} & Power law index of the surface density profile.  \\
        $h_{\mathrm{c}}$ & \multicolumn{2}{c}{0.14} & Scale height angle at the characteristic radius $R_{\mathrm{c}}$. \\
        $\psi$ & \multicolumn{2}{c}{0.22}& Flaring index of the disk surface density. \\
        PAH abundance & \multicolumn{2}{c}{0.1}& Gas-phase abundance of PAHs w.r.t. to ISM value.  \\
        $\delta_{\mathrm{gas}}$ & \multicolumn{2}{c}{$10^{-3}$} & Relative drop in gas density inside the gas cavity ($R_{\mathrm{subl}} < R < R_{\mathrm{cav\ out,\ gas}}$). \\
        $\delta_{\mathrm{dust,\ in}}$ & \multicolumn{2}{c}{$9\times 10^{-4}$}& Relative drop in dust density in the dusty inner disk w.r.t. the outer disk ($R_{\mathrm{subl}} < R < R_{\mathrm{cav\ in,\ dust}}$). \\
        $\delta_{\mathrm{dust,\ out}}$ & \multicolumn{2}{c}{$10^{-20}$}&  Relative drop in dust density inside the dust cavity ($R_{\mathrm{cav\ in,\ dust}} < R < R_{\mathrm{cav\ out,\ dust}}$). \\
        \\[-0.3em]
        \textit{Dust properties} & & &  \\
        $\chi$ & \multicolumn{2}{c}{0.1} & Settling of large grains. \\
        $f_{\mathrm{ls}}$ & \multicolumn{2}{c}{0.85} & Mass-fraction of grains that is large.  \\
        $x_{\mathrm{trap}}$ & 1000 & - & Increase of large dust grains inside the dust trap.  \\
        $\Delta_{\mathrm{gas/ dust}}$ & \multicolumn{2}{c}{20} & Gas-to-dust mass ratio.   \\
        \\[-0.3em]
        \textit{Stellar properties} & & & \\          
        $M_{\star}$ & \multicolumn{2}{c}{2~$\mathrm{M_{\odot}}$} & Mass of the central star.  \\
        $\dot{M}_{\star}^{(2)}$ & \multicolumn{2}{c}{$4\times 10^{-9}~\mathrm{M_{\odot}}$~yr$^{-1}$} & Mass accretion rate of the central star.   \\        
        \\[-0.3em]
        \textit{Stellar spectrum} & & &  \\     
        $L_{\star}$ & \multicolumn{2}{c}{14.3~$\mathrm{L_{\odot}}$} & Luminosity of the central star.   \\
        $L_{\mathrm{X}}$ & \multicolumn{2}{c}{$1\times 10^{27}$ erg s$^{-1}$} & X-ray luminosity of the central star.  \\
        $T_{\mathrm{eff}}$ & \multicolumn{2}{c}{$9\times 10^3$~K} & Effective temperature of the central star. \\
        $T_{\mathrm{X}}$ &  \multicolumn{2}{c}{$7\times 10^7$~K}  & Effective temperature of the X-ray radiation.   \\  
        $\zeta_{\mathrm{c.r.}}$ & \multicolumn{2}{c}{$1\times 10^{-18}\mathrm{s^{-1}}$} & Cosmic ray ionization rate. \\ 
        \\[-0.3em]
        \multicolumn{3}{l}{\textit{Observational geometry}}&  \\
        $i$ & \multicolumn{2}{c}{$50\degree$ } & Disk inclination ($0\degree$ is face-on).\\
        $d$ & \multicolumn{2}{c}{135~pc} & Distance to the star. \\ 
        \\[-0.3em]
        \textit{Chemistry}$^{(3)}$ & &&  \\
        H        &  \multicolumn{2}{c}{$5.2\times 10^{-5}$}  & \\
        He       &  \multicolumn{2}{c}{$1.4\times 10^{-1}$}  & \\
        \ce{H2}  &  \multicolumn{2}{c}{$5.0\times 10^{-1}$}  & \\   
        \ce{H2O} &  \multicolumn{2}{c}{$1.9\times 10^{-4}$}  & \\             
        \ce{CO}  &  \multicolumn{2}{c}{$1.3\times 10^{-4}$}  & \\             
        \ce{N2}  &  \multicolumn{2}{c}{$3.1\times 10^{-5}$}  & \\             
        \ce{Mg+} &  \multicolumn{2}{c}{$1.0\times 10^{-11}$} & \\             
        \ce{Si+} &  \multicolumn{2}{c}{$1.0\times 10^{-11}$} & \\             
        \ce{S+}  &  \multicolumn{2}{c}{$1.0\times 10^{-11}$} & \\             
        \ce{Fe+} &  \multicolumn{2}{c}{$1.0\times 10^{-11}$} & \\ \hline               
\end{tabular}
\tablefoot{$^{(2)}$~The stellar accretion rate is converted to a $10^4$~K black body and then added to the stellar spectrum. An accretion rate of $4\times 10^{-9}~\mathrm{M_{\odot}}\ \mathrm{yr}^{-1}$ corresponds to a UV luminosity of 0.2~L$_{\odot}$. $^{(3)}$~Abundance w.r.t. the total number of hydrogen atoms. All molecules start as gas-phase species.}
    \begin{tablenotes}
      \small
      \item 
    \end{tablenotes}
\end{table*}

\subsection{Model results}

\subsubsection{Initial distribution of nitrogen} \label{sec:init_nitrogen}
The fiducial model discussed in the main text is initialized with all nitrogen in \ce{N2}. However, the initial distribution of nitrogen over the main nitrogen carries such as \ce{N, N2, NH3} is unknown. Therefore, we run all models with initially all nitrogen in \ce{N2} (fiducial models), all nitrogen initially in N following \citet{Fogel2011}, all nitrogen in \ce{NH3} following \citet{Maret2006} and \citet{Daranlot2012}, and a mixture of these three. The mixture is based on the dark cloud model in \citet{Walsh2015} and consists of an initial atomic nitrogen abundance of $3.3\times 10^{-5}$, \ce{N2} initially $2\times 10^{-5}$, and \ce{NH3} initially $8.8\times 10^{-6}$ with respect to hydrogen. All molecules start in the gas phase as the model is evolved for a short time. 

\begin{figure*}
   \centering
  \includegraphics[width=1\linewidth]{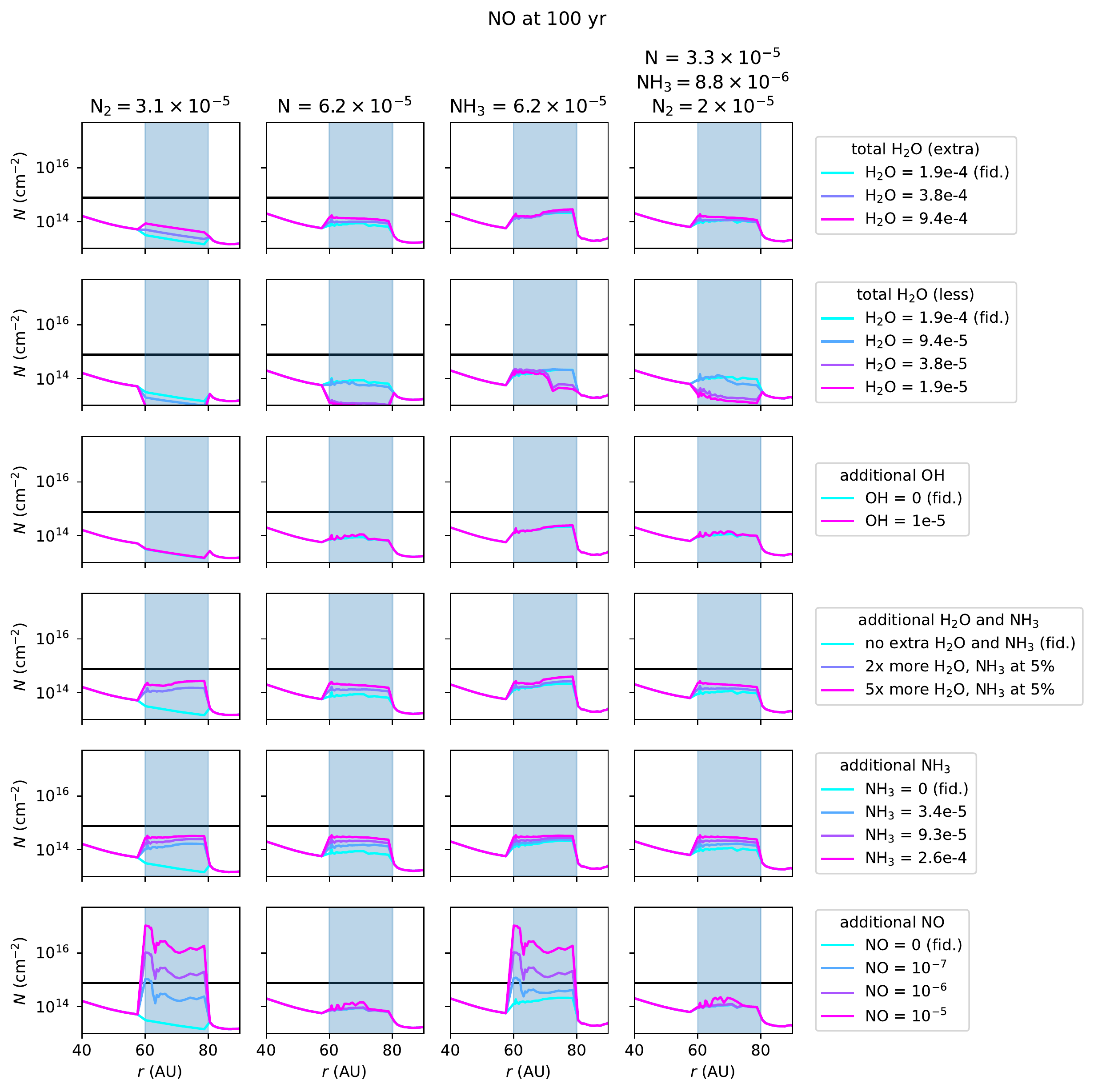}
      \caption{Predicted NO column density for different distributions of nitrogen and sublimating ices. Nitrogen initially starts all in \ce{N2} (left column), N (second column), \ce{NH3} (third column), and a mixture (fourth column). The effect of sublimating ices is presented in the different rows: sublimating water (top row), water depletion (second row), extra OH (third row), sublimating water and \ce{NH3} simultaneously (fourth row), sublimating \ce{NH3} (fifth row), and sublimating NO (final row). The horizontal black line indicates the NO column density derived from the observations. Only the models with a high initial NO abundance and all nitrogen initially in \ce{N2} or \ce{NH3} match the observed NO column density.}
         \label{fig:dali_grid_big}
\end{figure*}

An overview of the model results for different initial distributions of nitrogen and the effect of sublimating ices is presented in Fig.~\ref{fig:dali_grid_big}. The initial distribution of nitrogen has the most profound effect when NO gas is added to the initial conditions. Only when atomic nitrogen is initially absent, NO can survive. The reason for this is that NO is destroyed by atomic nitrogen to form \ce{N2}. The models that start with atomic nitrogen, \ce{NH3} or a mixture of \ce{N, NH3}, and \ce{N2} predict a higher NO column density when no sublimating ices are added to the initial conditions. Still these models underpredict the NO column density. Furthermore, this difference is diminished when the effect of sublimating \ce{NH3}, \ce{NH3} \& \ce{H2O}, and NO is modelled. Only in the case of sublimating water, the NO column density in the \ce{N2} model is slightly lower than that in the other models. 
In summary, varying the initial distribution of nitrogen shows that the initial abundance of atomic nitrogen must be low to explain the observed NO emission.

\subsubsection{Effect of evaporating ices on CN and \ce{C2H}} \label{sec:CO_ratio_CN}
\begin{figure*}
   \centering
  \includegraphics[width=1\linewidth]{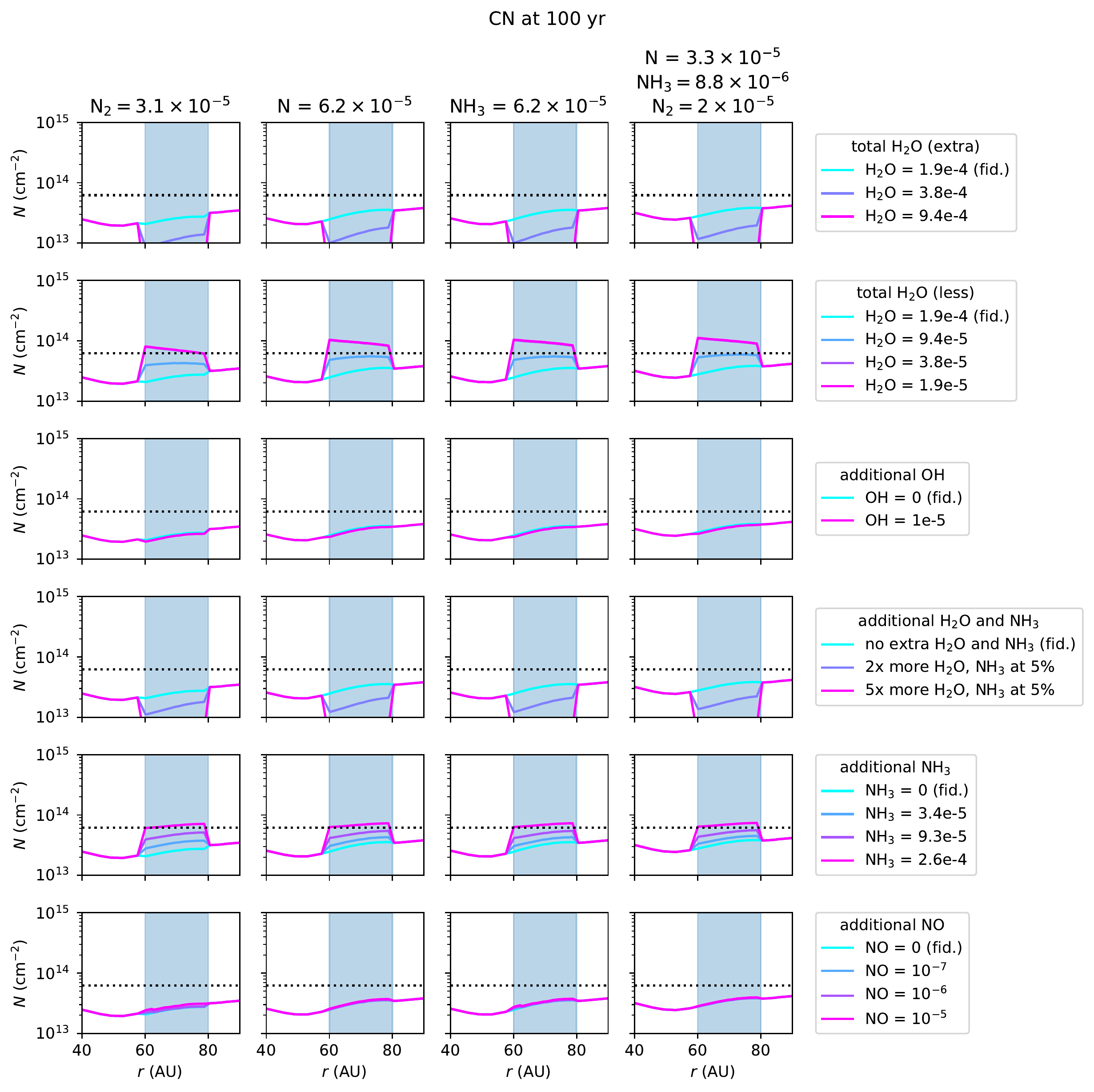}
      \caption{Same as Fig.~\ref{fig:dali_grid_big} but then for CN. The horizontal black dotted line indicates the upper limit on the CN column density derived from the observations. The initial \ce{H2O} abundance needs to be at least  $1.9\times 10^{-5}$ (if all nitrogen starts as \ce{N2}) or $3.8\times 10^{-5}$ (if nitrogen starts as \ce{NH3}, \ce{N}, or a mix of \ce{NH3}, N and \ce{N2}. Similarly, the initial (additional) abundance of \ce{NH3} is $\lesssim 2.6\times 10^{-4}$ to match the CN upper limit.}
         \label{fig:dali_grid_big_CN}
\end{figure*}

The dust trap in the IRS~48 disk is an ice trap. Sublimating \ce{H2O} and \ce{NH3} cannot boost the NO column density to match the observations. In this section, their effect on the upper limits on CN and \ce{C2H} is investigated. The CN column density presented in Fig.~\ref{fig:dali_grid_big_CN} and is consistent with the upper limit derived from the observations for all models except those with an initial water abundance of $\gtrsim 3.8\times 10^{-5}$ or an initial \ce{NH3} abundance $\gtrsim 2.6\times 10^{-4}$. Furthermore, the CN column density is slightly lower for the models that start with all nitrogen initially in \ce{N2}. The reason for this is that the \ce{N2} bond needs to be broken before CN can be formed in these models. 

\begin{figure*}
   \centering
  \includegraphics[width=1\linewidth]{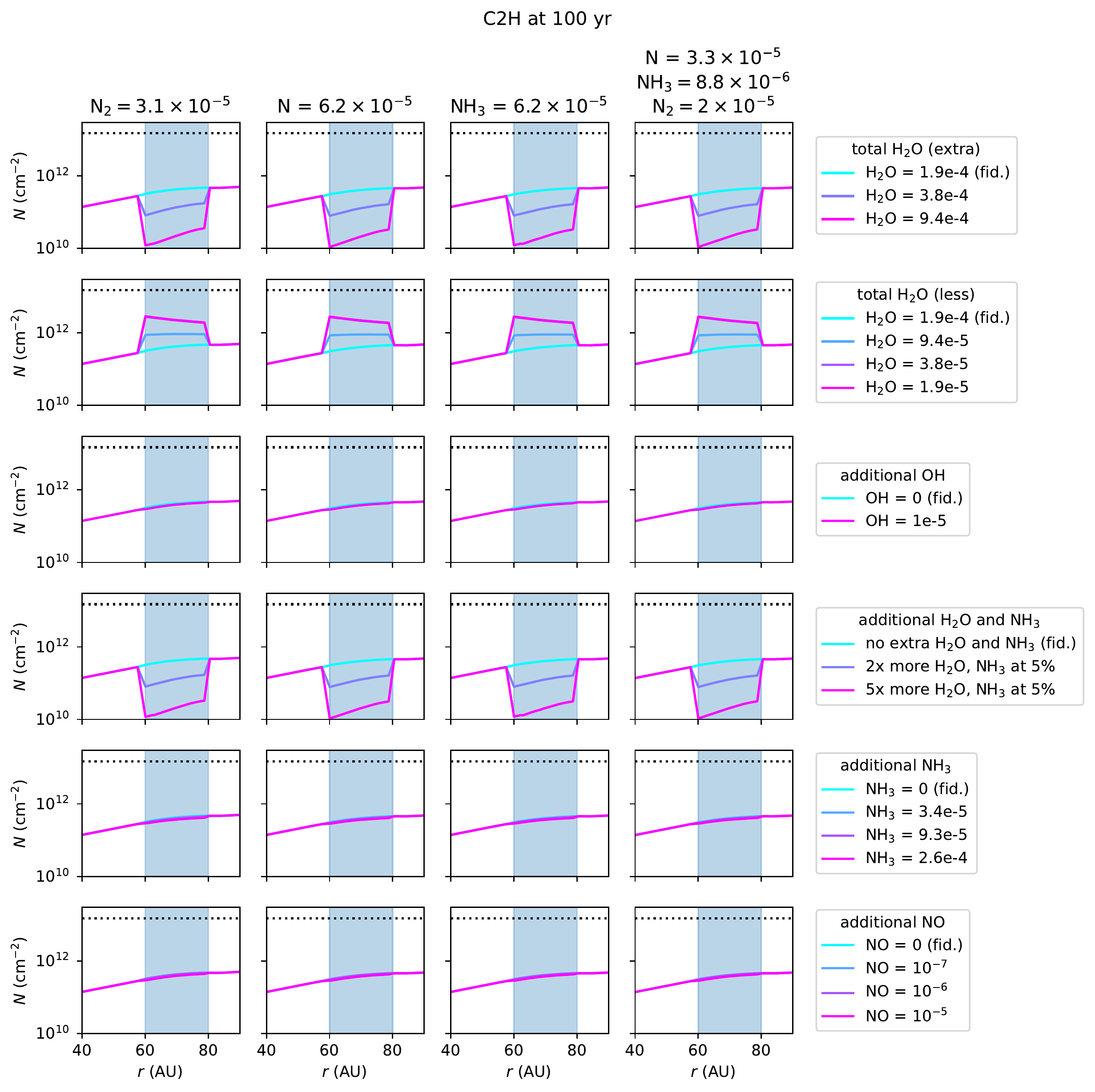}
      \caption{Same as Fig.~\ref{fig:dali_grid_big} but then for \ce{C2H}. The horizontal black dotted line indicates the upper limit on the \ce{C2H} column density derived from the observations. All models are compatible with the upper limit on the \ce{C2H} from the observations. }
         \label{fig:dali_grid_big_C2H}
\end{figure*}

The results for the \ce{C2H} column density are presented in Fig.~\ref{fig:dali_grid_big_C2H}. The \ce{C2H} column density is sensitive to the C/O ratio in this disk as it decreases with an increasing abundance of water (first two rows of Fig.~\ref{fig:dali_grid_big_C2H}). The highest C/O ratio in the models explored here is 0.99, so just below a C/O ratio of 1 when \ce{C2H} is expected to become very bright \citep{Bergin2016, Bergner2019, Miotello2019, Bosman2021}. In summary, the CN emission places the strongest constraint on the C/O ratio in this disk. Deeper observations of both CN and \ce{C2H} are needed to pinpoint the C/O ratio to an accuracy better than $\lesssim 0.6$. 

\begin{figure*}
   \centering
  \includegraphics[width=1\linewidth]{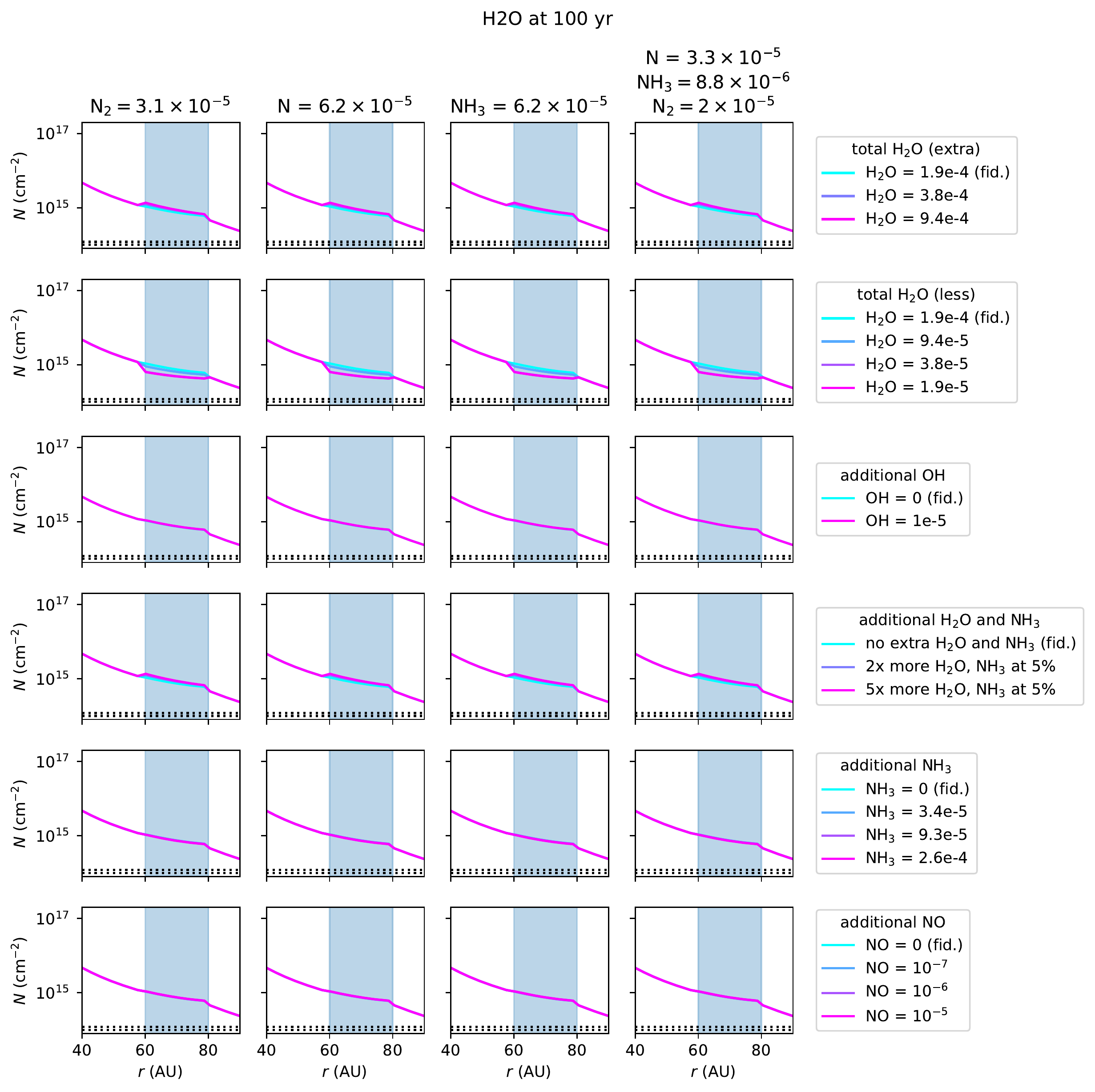}
      \caption{Same as Fig.~\ref{fig:dali_grid_big} but then for the \ce{H2O} column density in the region where $T_{\mathrm{gas}} \geq 150$~K, the temperature range probed by the \textit{Herschel} PACS observations. The horizontal black dotted lines indicate the upper limit on the \ce{H2O} column density derived from the observations at 150 and 200~K. All models overpredict the upper limit on the \ce{H2O} column density. This could be due to optically thick dust at $51-220~\mu$m hiding part of the \ce{H2O} column in the IRS~48 disk.}         
      \label{fig:dali_grid_big_H2O}
\end{figure*}

\begin{figure*}
   \centering
  \includegraphics[width=1\linewidth]{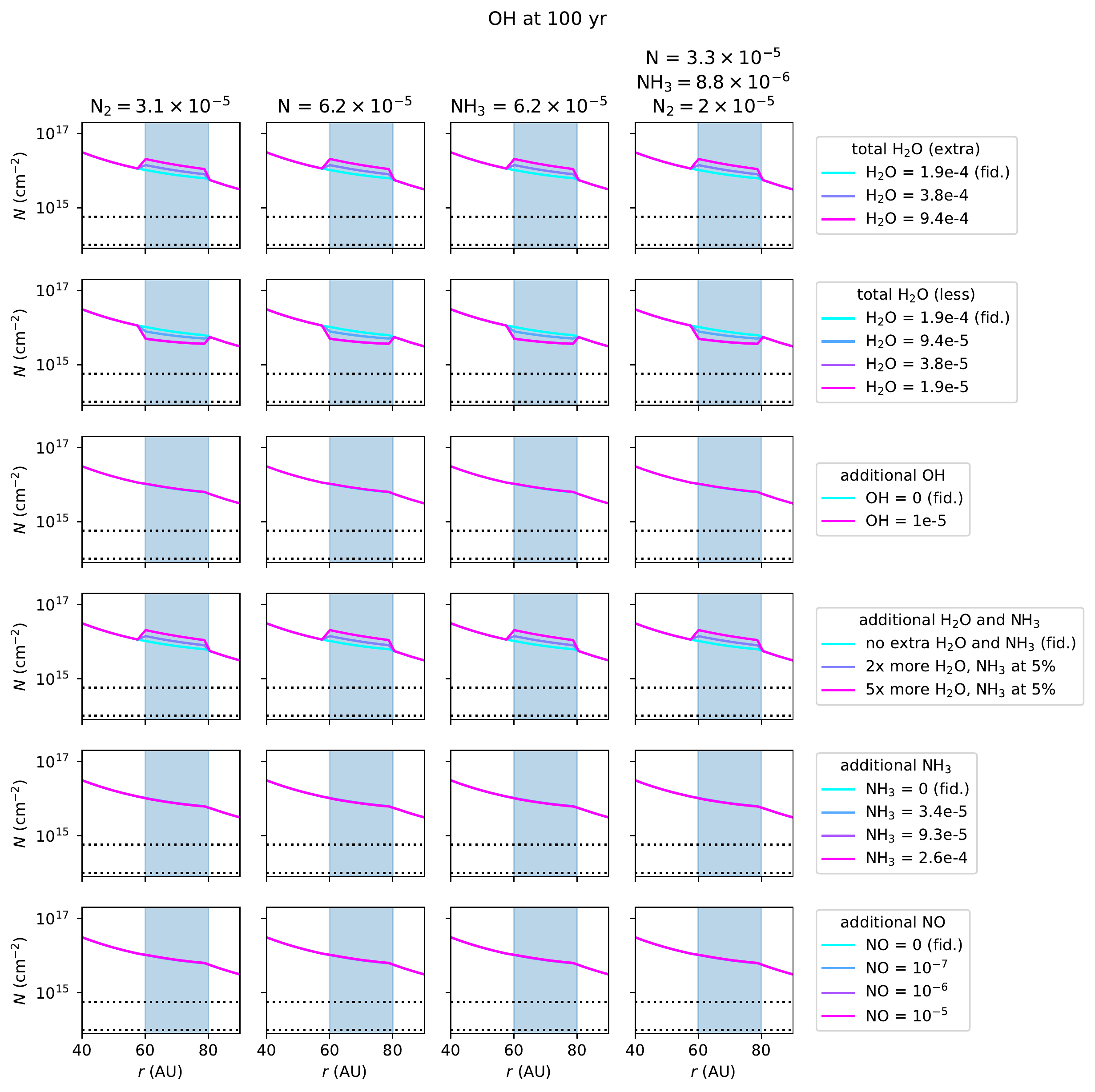}
      \caption{Same as Fig.~\ref{fig:dali_grid_big} but then for the OH column density in the region where $T_{\mathrm{gas}} \geq 150$~K, the temperature range probed by the \textit{Herschel} PACS observations. The horizontal black dotted lines indicate the upper limit on the \ce{OH} column density derived from the observations 150 and 200~K. Similar to \ce{H2O}, all models overpredict the upper limit on the OH column density. This could be due to optically thick dust at $51-220~\mu$m hiding part of the OH column in the IRS~48 disk.}
         \label{fig:dali_grid_big_OH}
\end{figure*}

Finally, the warm \ce{H2O} and OH column densities are presented in Fig.~\ref{fig:dali_grid_big_H2O} and \ref{fig:dali_grid_big_OH}. Only the \ce{H2O} and OH in the regions with $T_{\mathrm{gas}} \geq 150$~K are taken into account as this is the temperature range probed by the \textit{Herschel} PACS observations. Even at these high OH columns, formation of NO through OH are not efficient enough to explain the observations further strengthening the hypothesis that the NO is the photodissociation product of a larger molecule.

\subsubsection{Chemical composition of the non-dust trap side}
\begin{figure*}
   \centering
  \includegraphics[width=1\linewidth]{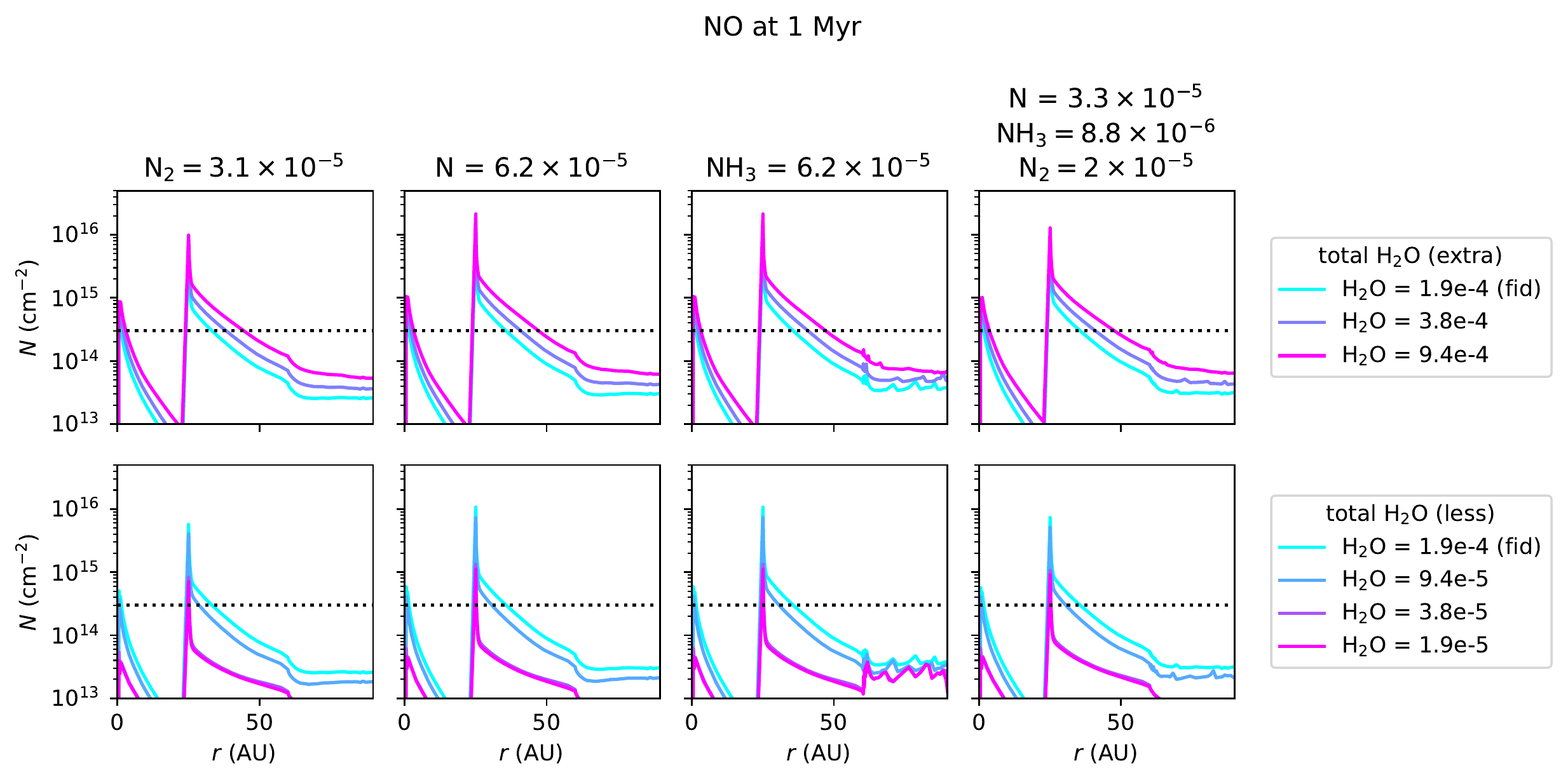}
      \caption{NO column density in the non-dust trap (north) side of the disk after 1~Myr. The black horizontal dotted line indicates the upper limit on the NO column density derived from the observations. Despite the narrow peak at 25~AU, the models are consistent with the derived upper limit on the NO column density in the north side of the IRS~48 disk. }
         \label{fig:dali_grid_north_NO}
\end{figure*}

\begin{figure*}
   \centering
  \includegraphics[width=1\linewidth]{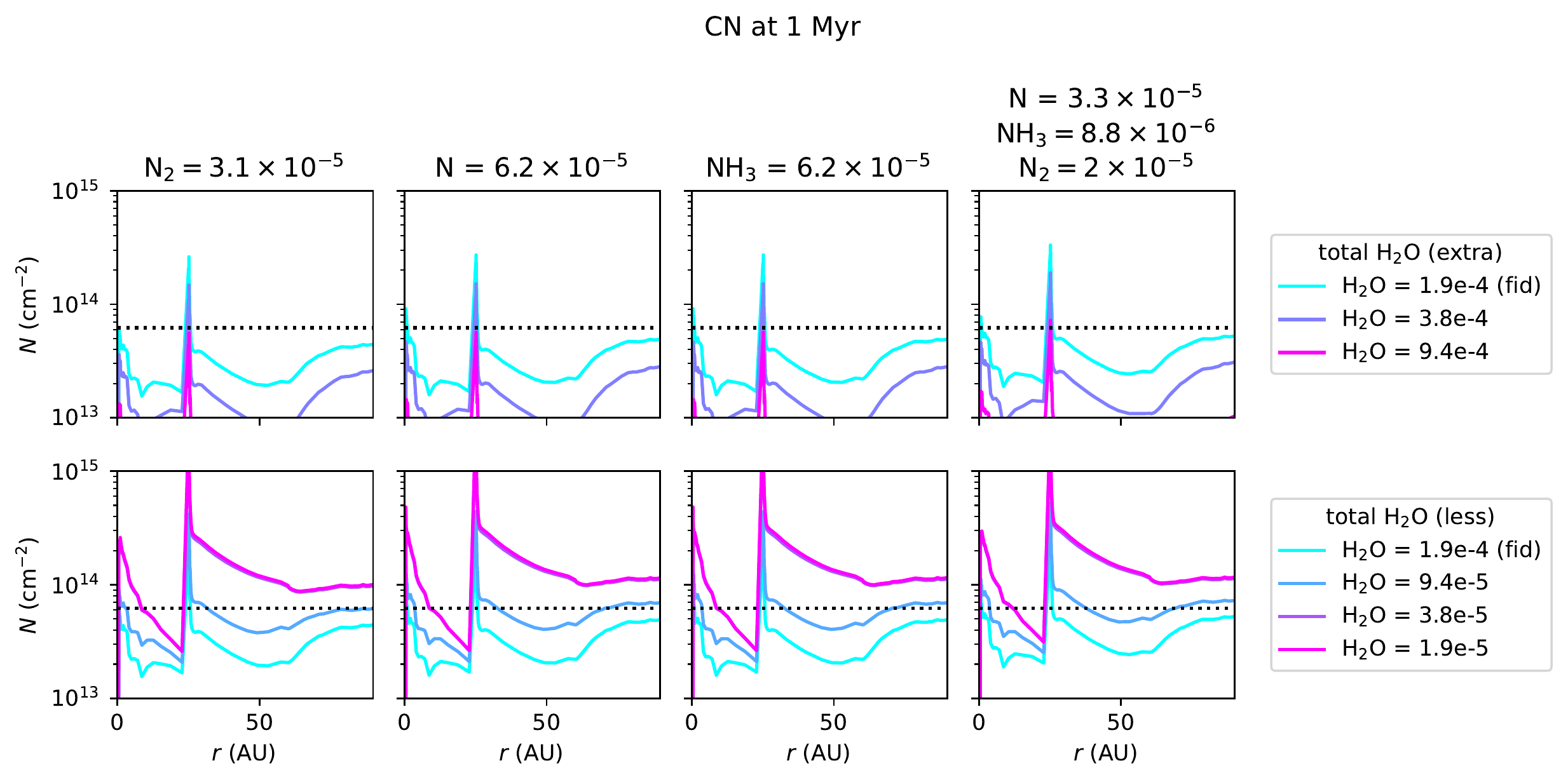}
      \caption{Same as Fig.~\ref{fig:dali_grid_north_NO} but then for CN. Models with an initial \ce{H2O} abundance of $ (3.8-1.9)\times 10^{-5}$ overpredict the upper limit on the CN column density.}
         \label{fig:dali_grid_north_CN}
\end{figure*}

\begin{figure*}
   \centering
  \includegraphics[width=1\linewidth]{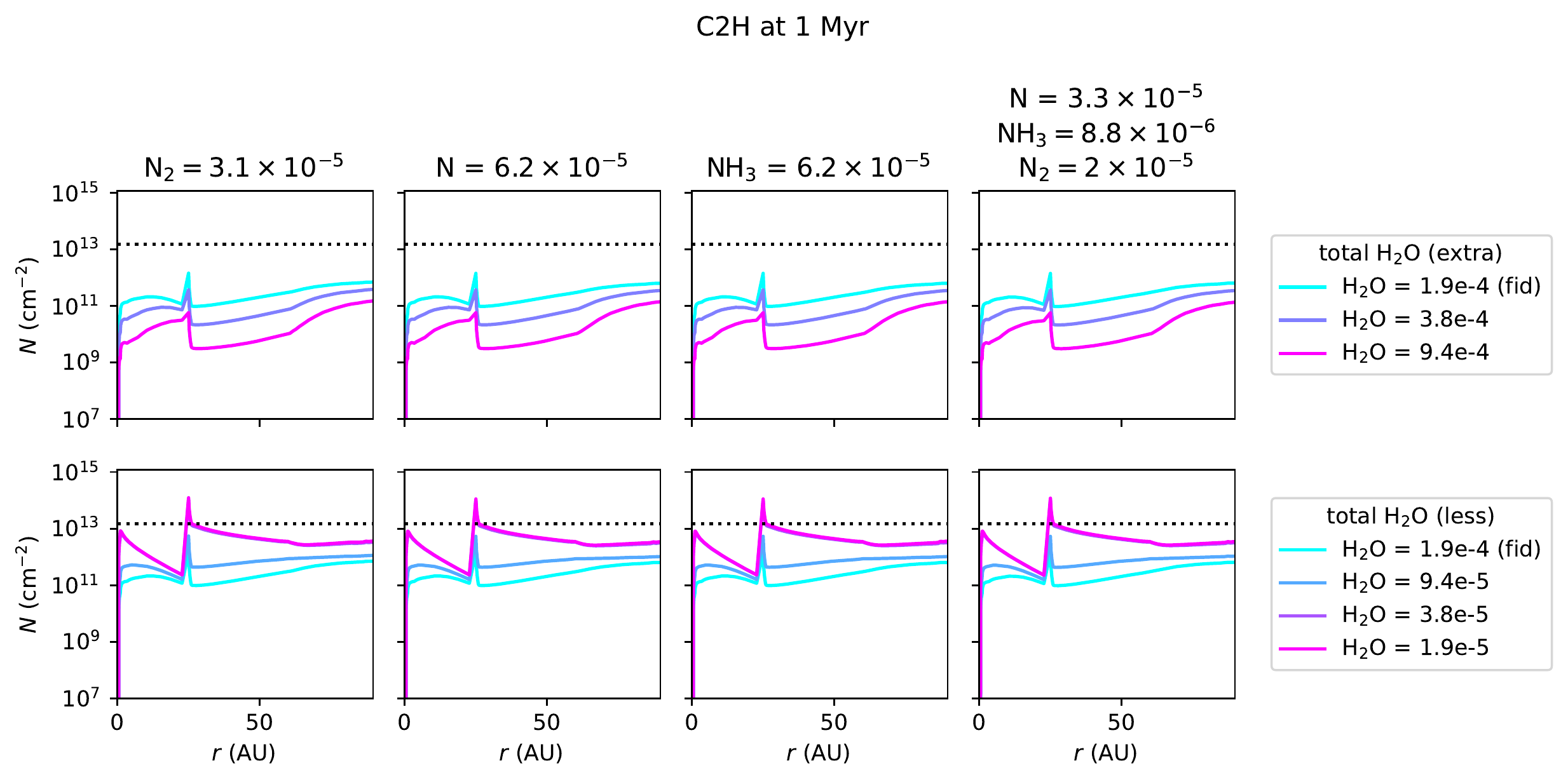}
      \caption{Same as Fig.~\ref{fig:dali_grid_north_NO} but then for \ce{C2H}. The models are consistent with the upper limit on the \ce{C2H} column density derived from observations except for a narrow peak at 25~AU for the models with the lowest initial \ce{H2O} abundance. }
         \label{fig:dali_grid_north_C2H}
\end{figure*}

The chemical composition of the non-dust trap side is modelled by varying the initial abundance of gas-phase water. The resulting NO, CN, and \ce{C2H} column densities are presented in Fig.~\ref{fig:dali_grid_north_NO}, \ref{fig:dali_grid_north_CN}, and \ref{fig:dali_grid_north_C2H}. Similar to the results in the dust trap, the NO and \ce{C2H} emission are not tightly constraining the disk C/O ratio. The CN indicates that the initial abundance of gas-phase water must be $\lesssim 9.4\times 10^{-5}$, setting the C/O ratio to be $\lesssim 0.6$. 

\subsubsection{Time evolution} \label{sec:timeevolution}
\begin{figure*}
   \centering
  \includegraphics[width=1\linewidth]{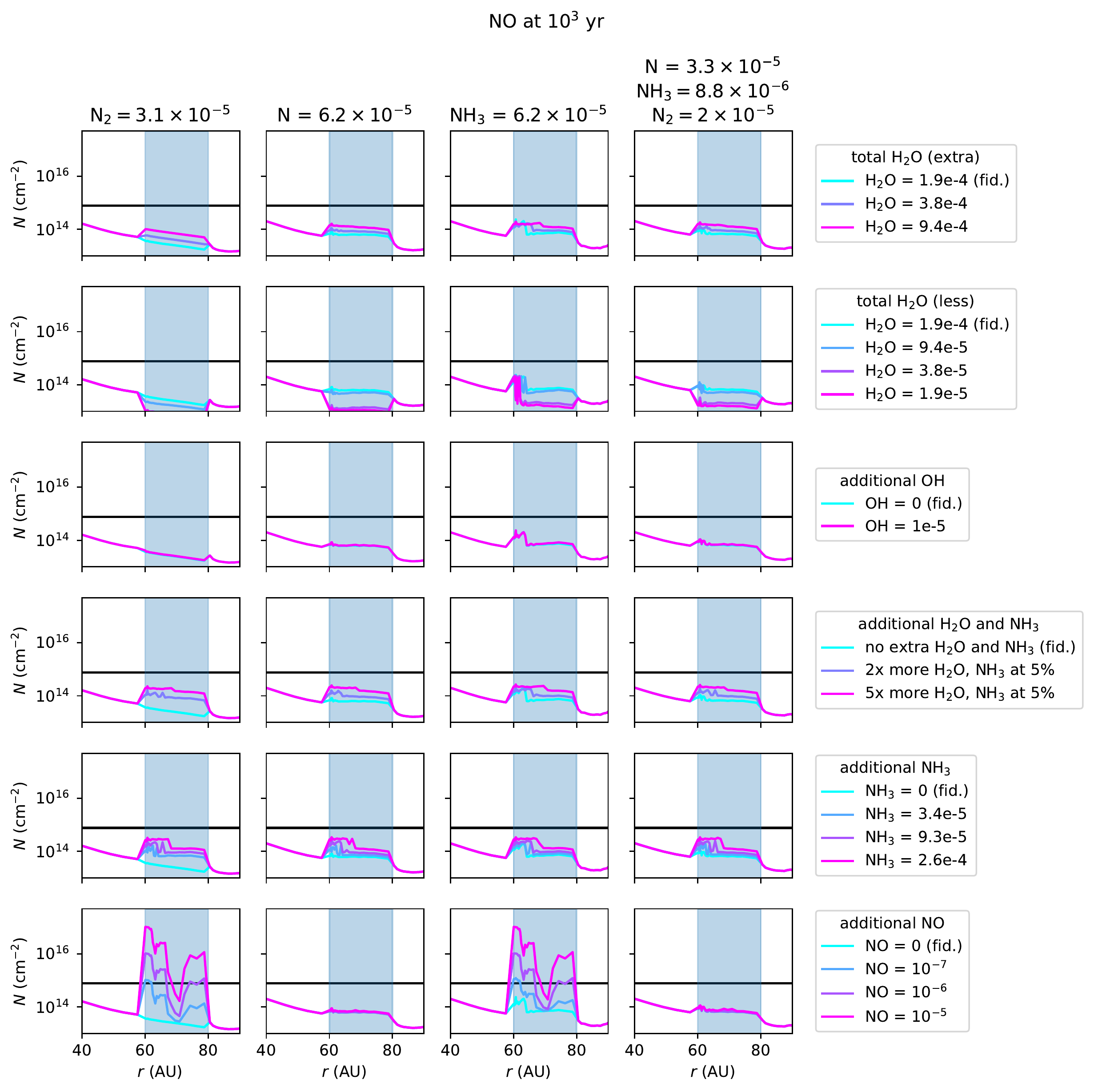}
      \caption{Same as Fig.~\ref{fig:dali_grid_big} but then for NO evolved for $10^3$~yr inside the dust trap.  }
         \label{fig:dali_grid_big_NO_105}
\end{figure*}

The NO column density after $10^3$~yr is presented in \ref{fig:dali_grid_big_NO_105}. The NO column density is relatively stable over a $10^3$~yr time period in the dust trap, except for the models where the initial \ce{NH3} abundance is non-zero. in these cases, the NO column density in the outer parts of the dust trap decreases by a factor of a few up to an order of magnitude in the models where \ce{NH3} is abundant.

\end{appendix}

\end{document}